%% file: paper.tex
\documentclass[11pt,a4paper]{article}
\usepackage{jheppub}

\usepackage{amsmath}
\usepackage{amssymb}
\usepackage{amsfonts}
\usepackage{bm}
\usepackage{bbm}
\usepackage{slashed}
\usepackage{mathrsfs}
\usepackage{graphicx}
\usepackage{wrapfig}

\usepackage[utf8]{inputenc}
\usepackage{verbatim}
\graphicspath{{}{plots/}}

\DeclareMathOperator{\tr}{tr}
\DeclareMathOperator{\sign}{sign}

\DeclareMathOperator{\Pf}{Pf}

\newcommand{\pas}{\partial^{\text{sym}}}
\newcommand{\pasb}{\bar\partial^{\text{sym}}}
\newcommand{\cQ}{\mathcal{Q}}
\newcommand{\cN}{\mathcal{N}}
\newcommand{\ZZ}{\mathcal{Z}}

\newcommand{\abs}[1]{\left| #1 \right|}
\newcommand{\SB}{S_\text{B}}
\newcommand{\vev}[1]{\left\langle #1 \right\rangle}
\newcommand{\bbZ}{\mathbb{Z}}
\newcommand{\trnsp}{\mathsf{T}}

\newcommand{\Nt}{N_\text{t}}
\newcommand{\Ns}{N_\text{s}}
\newcommand{\Ls}{L_\text{s}}
\newcommand{\ord}{\mathcal{O}}

\newcommand{\nN}[1]{\left\langle #1 \right\rangle}

\newcommand{\heco}{\mathrm{h.c.}}
\newcommand{\half}{{\textstyle\frac{1}{2}}}

\newcommand{\bPhi}{\bo{\Phi}}
\newcommand{\be}{\bo{e}}
\newcommand{\bff}{\bo{f}}
\newcommand{\bg}{\bo{g}}
\newcommand{\bU}{\bo{U}}
\newcommand{\bu}{\bo{u}}
\newcommand{\bv}{\bo{v}}
\newcommand{\bw}{\bo{w}}
\newcommand{\bn}{\bo{n}}
\newcommand{\bpsi}{\bo{\psi}}
\newcommand{\blam}{\bo{\lambda}}
\newcommand{\bchi}{\bo{\chi}}
\renewcommand{\Re}{\operatorname{Re}}

\newcommand{\eps}{\varepsilon}
\newcommand{\ii}{\mathrm{i}}
\newcommand{\ee}{\mathrm{e}}
\newcommand{\dd}{\mathrm{d}}
\newcommand{\DD}{\mathrm{D}}
\newcommand{\SF}{S_\text{F}}

\newcommand{\cD}{\mathcal{D}}

\newcommand{\id}{\mathbbm{1}}
\newcommand{\R}{\mathbbm{R}}
\newcommand{\Z}{\mathbbm{Z}}

\newcommand{\cp}[1]{\mathbb{C}\mathrm{P}^{#1}}

\newcommand{\bo}[1]{\boldsymbol{ #1 }}
\newcommand{\Neqtwo}{\cN\!\!=\!\!2}
\newcommand{\Neqone}{\cN\!\!=\!\!1}
\newcommand{\sgn}{\mathop{\rm sgn}}

\newcommand{\FIGURE}[1]{\begin{figure}[htbp] \centering #1 \end{figure}}
\newcommand{\rFIGURE}[1]{\begin{wrapfigure}{r}{0pt} \centering #1 \end{wrapfigure}}

\newcommand{\TABLE}[1]{\begin{table}[htbp] \centering #1 \end{table}}
\newcommand{\rTABLE}[1]{\begin{wraptable}{r}{0pt} \centering #1 \end{wraptable}}

\title{Supersymmetric Nonlinear O(3) Sigma Model on the Lattice}

\author{Raphael Flore,}
\author{Daniel K\"orner,}
\author{Andreas Wipf,}
\author{Christian Wozar}

\affiliation{Theoretisch-Physikalisches Institut, Universit{\"a}t Jena, \\ Fr\"obelstieg 1, D-07743 Jena, Germany}

\emailAdd{Raphael.Flore@uni-jena.de}
\emailAdd{Daniel.Koerner@uni-jena.de}
\emailAdd{wipf@tpi.uni-jena.de}
\emailAdd{christian.wozar@vega.de}

\abstract{A supersymmetric extension of the nonlinear O$(3)$ sigma model in two spacetime dimensions 
is investigated by means of Monte Carlo
simulations. We argue that it is impossible to construct a lattice 
action that implements both the O$(3)$ symmetry as well as at least one 
supersymmetry exactly at finite lattice spacing. It is shown by 
explicit calculations that previously proposed discretizations fail to 
reproduce the exact symmetries of the target manifold in the continuum limit. 
We provide an alternative lattice action with exact O$(3)$ symmetry and compare two 
approaches based on different derivative operators. Using the nonlocal SLAC 
derivative for the quenched model on moderately sized
lattices we extract the value $\sigma(2,u_0)=1.2604(13)$ for the step scaling function 
at $u_0=1.0595$, to be compared with the exact value $1.261210$. 
For the supersymmetric model with SLAC derivative
the discrete chiral symmetry is maintained but we  encounter strong sign 
fluctuations, rendering large lattice simulations ineffective. By applying the Wilson 
prescription, supersymmetry and chiral symmetry are broken explicitly at finite lattice spacing, 
though there is clear evidence that both are restored in the continuum limit by  
fine tuning of a single mass parameter.}


\begin{document}

\maketitle

\section{Introduction}

\input{introduction} \label{ch:introduction}

\section{Symmetries and possible discretizations of the O$(N)$ nonlinear sigma model}

\input{basics} \label{ch:basics}
\input{secondSUSY} \label{ch:secondsusy}
\input{discretization}

\input{derivatives}	\label{ch:derivatives}
\input{exactSUSY} \label{ch:exactsusy}

\section{Numerical results}

\input{catterall} \label{ch:catterall}
\input{o3symmetry} \label{ch:o3symmetry}
\input{Fmasses} \label{ch:fmasses}
\input{chiral}	\label{ch:chiral}
\input{Bmasses} \label{ch:bmasses}
\input{finetuning} \label{ch:finetuning}
\input{ward} \label{ch:ward}

\input{conclusion}

\acknowledgments{
We would like to thankfully mention the numerous discussions with Björn 
Wellegehausen. This work has been supported by the DFG Researach 
Training Group ``Quantum and Gravitational Fields'' GRK 1523 and DFG 
grant Wi 777/11. The simulations have been carried out at the Omega 
cluster of the TPI.}

\appendix
\input{invariance} \label{app:inv}
\input{measure} \label{app:meas}
\input{derivativeEvaluation} \label{ch:eval}
\input{algorithms} \label{app:algo}

\bibliography{literature}
\bibliographystyle{JHEP}

\end{document}

%% file: introduction.tex
Despite its great success in the description of high energy experiments, 
the Standard Model of particle physics still faces serious problems, especially 
with regard to cosmological observations, where yet undiscovered
sources of gravitation are proposed to correct the seemingly unbalanced equation of observed 
matter on the one hand and gravitational attraction on the other hand \cite{ArkaniHamed2008}. 
Another interesting question concerns the existence of far more matter than antimatter in our universe,
and this fact is connected to the strong $C\!P$ problem (see \cite{RevModPhys.82.557} and references listed
therein). Coined as the hierarchy problem is the question of scales that are involved in 
the fundamental theory of particle physics. To address these open issues, extensions of 
the Standard Model have been proposed, among them supersymmetrically
extended versions, in particular the Minimal Supersymmetric Standard Model (MSSM).
It is of great interest to examine the nonperturbative properties of this theory, but the
utilization of the only ab initio method for this purpose, namely a spacetime lattice simulation, is notoriously
difficult because of the many different scales that are involved.\\
In the past, two-dimensional nonlinear sigma models have been applied successfully
to model non-perturbative properties of four-dimensional strongly coupled 
pure gauge theories \cite{Novikov:1984ac}. Similarly one may employ the supersymmetrized version 
of the nonlinear sigma model to effectively describe super-Yang-Mills theories with
a strongly interacting fermionic sector. A first construction of the supersymmetric 
nonlinear sigma model with O$(N)$ target manifold is due to E. Witten \cite{Witten:1977xn} 
and P. Di Vecchia and S. Ferrara \cite{DiVecchia:1977bs}, followed by a number of papers
that established analytical properties, including asymptotic safety, spontaneous breaking of chiral symmetry
and the dynamical generation of particle masses \cite{Shankar1978,Alvarez1978,Evans:1994sy}.
Further studies are particularly concerned with the special case of the O$(3)$ model, which admits 
an extended $\Neqtwo$ supersymmetry algebra since its target manifold is K{\"a}hler 
\cite{Zumino:1979et}.\\
This enhanced supersymmetry has recently raised some interest with regard to the endeavor to
study supersymmetric theories by numerical simulations.
Because it is an extension of the Poincaré symmetry, any lattice discretization of spacetime 
also breaks supersymmetry. This can be traced back to
the failure of the Leibniz rule on the lattice \cite{Dondi:1976tx}. It is therefore necessary 
to take great care in restoring the full symmetry in the continuum limit. In general, one 
needs to introduce appropriate counter terms for each relevant supersymmetry-breaking operator 
that must each be fine tuned such that the desired continuum theory is reached \cite{Montvay:1995rs}.
Depending on the number of parameters, this becomes
practically impossible very fast due to limited computer time. In addition, much information
about the theory is needed prior to numerical investigations. For that reason alternative approaches 
have been proposed in order to reduce the number of fine tuning parameters or even render 
the fine tuning procedure obsolete \cite{Catterall:2009it}. One such approach for theories with
extended supersymmetry aims at the construction of a nilpotent charge composed from the supercharges
such that both the continuum as well as the discretized model are invariant under this charge.
It is then expected that, by preserving a part of the symmetry, invariance under the full supersymmetry 
is restored automatically in the continuum limit, without the need for fine tuning. This procedure 
was applied to the supersymmetric O$(3)$ model
\cite{Catterall:2003uf,Catterall:2006sj} and the authors conclude that the supersymmetric Ward
identities are indeed fulfilled in the continuum limit, hinting at a restoration of the full
supersymmetry. However, the lattice discretization constructed in this way breaks the O$(3)$
symmetry of the target space explicitly at finite lattice spacing and no attempt was made to
show that it is restored in the continuum limit. It is therefore unclear whether this discretization
actually describes the correct model in the continuum and we will address this question
in much detail.\\
Expensive dynamical fermion simulations are needed in order to treat the superpartners
on equal footing and we wish to utilize the SLAC prescription for the fermionic derivative, 
which performed very well in previously examined Wess-Zumino models \cite{Kastner:2008zc},
showing greatly reduced lattice artifacts in comparison to e.g. the Wilson derivative.
Furthermore, the SLAC derivative allows for a discretized action that does not break 
the discrete chiral symmetry explicitly. But this comes at the cost of having a 
nonlocal derivative operator and we will explicitly check whether the SLAC derivative 
is applicable in the present model. In order to compare with an alternative, the model will also 
be studied by simulations based on the Wilson derivative, which is less affected by the sign problem
and enables us to investigate larger lattices.\\
The article is organized as follows: We introduce the supersymmetric version of the 
nonlinear O$(3)$ model along with a derivation of its stereographic projection in the 
superfield formalism. We then discuss the extended supersymmetry of the model and express the 
related transformations in terms of constrained fields. Based on the continuum theory two
alternative discretizations in Euclidean space are presented which resolve the field constraints -- 
one using stereographically projected fields and the other using elements of the orthogonal group SO$(3)$. 
Thus providing the basic setup of our lattice discretization,
we go on to argue that the SLAC derivative is indeed applicable here by performing a high 
precision analysis of the step scaling function in the quenched model. Towards the end of 
the first part we finally answer the question whether a lattice theory exists that
admits both the O$(3)$ symmetry and at least one supersymmetry.\\
In the second part, which 
is primarily dedicated to extensive numerical investigations, we discuss the missing
target space symmetry in a previously given lattice discretization by Catterall et al. 
\cite{Catterall:2003uf} and proceed with results from our own calculations. We examine the 
discrete chiral symmetry as well as supersymmetry
by presenting results for the chiral condensate and the masses of the elementary excitations 
computed from ensembles obtained with both SLAC and Wilson derivative. In the second
case special emphasis is put on the fine
tuning of divergent operators in order to arrive at a supersymmetric continuum limit.
Before coming to our conclusions, we present our results for a simple Ward identity based on the 
expectation value of the bosonic action.

%% file: basics.tex
\subsection{Constrained and unconstrained formulations}
The supersymmetric extension of the nonlinear O$(N)$ sigma model in two-dimensional Euclidean spacetime can be formulated in terms of a real superfield 
\footnote{Conventions concerning the Gamma matrices and the Fierz relations are explained in appendix \ref{app:con}.},
\begin{equation}\label{superfN}
\bPhi=\bn+\ii\bar\theta \bpsi+\tfrac{\ii}{2}\bar\theta
\theta\bff\, ,
\end{equation}
subject to the constraint $\bPhi\bPhi=1$,
where $\bn$ and $\bff$ are $N$-tuples of real scalar fields
and $\bpsi$ is an $N$-tuple of Majorana fields. We shall refer 
to the elements of a tupel as ``flavors''.
The constraint in superspace is equivalent to the following
constraints for the component fields,
\begin{equation}\label{constraints}
\bn^2=1,\quad 
\bn\bpsi = 0\quad\hbox{and}\quad
\bn\bff=\tfrac{\ii}{2}\bar{\bpsi}\bpsi\,.
\end{equation}
The O$(N)$-invariant Lagrangian density is defined in terms of the covariant derivatives 
$D_{\alpha} = \partial_{\bar{\theta}^{\alpha}} + i (\gamma^{\mu} \theta)_{\alpha} \partial_{\mu}$,
\begin{equation} \label{lagr}
\mathcal{L} = \frac{1}{2g^2} \overline{D\bPhi}\, D\bPhi |_{\bar{\theta}\theta} 
= \frac{1}{2g^2} \left(\partial_{\mu}\bn \partial^{\mu}\bn +
\ii\bar{\bpsi}\slashed{\partial} \bpsi - \bff^2\right)\,. 
\end{equation}
The equation of motion for the auxiliary field implies
that $\bff$ and $\bn$ are parallel, $\bff = \tfrac{\ii}{2} (\bar{\bpsi}\bpsi) \bn$, 
and the resulting on-shell Lagrangian density contains a $4$-Fermi term,
\begin{equation}\label{lagron}
\mathcal{L}=  \frac{1}{2g^2} \left(\partial_\mu\bn \partial^\mu\bn 
+ \ii\bar{\bpsi} \slashed{\partial} \bpsi 
+ \tfrac{1}{4} (\bar{\bpsi}\bpsi)^2\right)\,.
\end{equation}
The action and the constraints are both invariant under global O$(N)$ 
``flavor'' transformations and by construction they are also invariant under 
the $\cN=1$ supersymmetry transformations
\begin{equation} \label{trafo}
\delta \bn = \ii \bar{\eps} \bpsi,\quad 
\delta \bpsi = \left(\slashed{\partial} + \tfrac{\ii}{2} \bar{\bpsi}\bpsi\right)\bn\eps\,.
\end{equation}
Besides flavor symmetry and supersymmetry the classical theory admits a further 
$\Z_2$-symmetry generated by the chiral transformation $\bpsi \to i \gamma_* \bpsi$. 
However, quantum fluctuations dynamically generate a mass term and hence
induce spontaneous breaking of the chiral $\bbZ_2$-symmetry.\\
Explicit constraints for the fields are sometimes difficult to handle, e.g. in numerical 
simulations. Therefore, it is useful to construct a formulation of the model in terms 
of an unconstrained real superfield $\bU(x,\theta)
=\bu(x) + \ii\bar{\theta} \blam(x) + \tfrac{\ii}{2} \bar{\theta}\theta \bo{g}(x)$, which
is an $(N-1)$-tupel. It is related to the superfield $\bPhi$ by a stereographic projection
in superspace
\begin{equation} \label{stereo}
\begin{pmatrix}\Phi_1\\ \bPhi_\perp\end{pmatrix}
= \frac{1}{1+\bU^2}\begin{pmatrix}1-\bU^2\\ 2\bU \end{pmatrix}.
\end{equation}
The decomposition of the projection into bosonic and fermionic fields reads:
\begin{equation}
\label{proj}
\bn_\perp = 2 \rho\, \bu,\quad
\bpsi_\perp = 2 \rho \blam - 4\rho^2\, (\bu\blam) \bu\quad\hbox{with}\quad
\rho=\frac{1}{1+\bu^2}\,.
\end{equation}
The expressions for the remaining components $n_1$ and $\psi_1$ can be determined 
either from \eqref{stereo} or from \eqref{proj} and the constraints $\bn^2=1$ and $\bn\bpsi=0$. The inverse
transformation in superspace reads $\bU = \bPhi_\perp/(1+\Phi_1)$ and 
leads to
\begin{equation}
\label{backproj}
\bu = \frac{1}{2\rho}\, \bn_\perp,\quad
\blam = \frac{1}{2\rho}\, \bpsi_\perp - \frac{1}{4\rho^2}\, \psi_1\bn_\perp\quad\hbox{with}\quad
\rho=\frac{1+n_1}{2}\,.
\end{equation}
Applying the stereographic projection (\ref{proj}), the on-shell
Lagrangian density can be written in terms of the unconstrained fields as
\begin{equation} \label{lagrU}
\mathcal{L} = \frac{2}{g^2} \rho^2 
\left(\partial_{\mu} \bu \partial^{\mu} \bu + \ii\bar{\blam} \slashed{\partial} \blam 
+ 4 \ii \rho\,(\bar{\blam}\bu) \gamma^\mu(\partial_\mu \bu\blam) 
+ \rho^2 (\bar{\blam}\blam)^2\right)\,.
\end{equation}
The action is invariant under the supersymmetry transformations
\begin{equation} \label{utrafo}
\delta \bu = \ii \bar{\eps} \blam,\quad
\delta \blam = \left(\slashed{\partial} + \ii 
\rho \,\bar{\blam}\blam\right)\bu\eps 
- 2\ii\rho\, (\bar{\blam}\bu) \blam\eps\,.
\end{equation}

%% file: secondSUSY.tex
\subsection{Extended supersymmetry of the O$(3)$-model}
The fields $\bu$ and $\blam$ are unconstrained, but their target manifold has a
non-trivial metric. In fact, for $N=3$ the target manifold is K\"ahler and the 
corresponding potential can be written in terms of the complex field $u = u_1 + i u_2$ as 
$K(u,\bar{u}) = \ln(1+\bar{u}u)$. As pointed out by Zumino a nonlinear sigma model whose 
target manifold is K\"ahler, possesses an $\Neqtwo$-supersymmetric extension \cite{Zumino:1979et}. 
Hence the O$(3)$ admits an additional supersymmetry besides (\ref{utrafo}) and (\ref{trafo}).\\
In order to derive an explicit expression of this symmetry, we investigate a general ansatz in terms of the
unconstrained fields ($\delta \bu = \ii \bar{\eps}(A_I)\blam$, etc.), and determine the constraints for
the matrices $A_I$, etc., which follow from the supersymmetry algebra and the invariance
of the action\footnote{An example of this approach can be found in \cite{Kirchberg:2004vm}, where it is applied to the 
Wess-Zumino model.}. This approach yields the second pair of symmetry transformations as
\begin{align}
\label{u2trafo}
\delta \bu &= \sigma_2\bar{\eps} \blam\\
\delta \blam &= \ii\sigma_2\left( \slashed{\partial}\bu - 
\ii\rho\, (\bar{\blam}\blam) \bu + 2 \ii \rho\, (\bar{\blam}\bu) \blam\right)\eps\,.
\nonumber
\end{align}
Both supersymmetries (\ref{utrafo},\ref{u2trafo}) can 
also be obtained by deriving the complex supersymmetry 
from the Kähler potential, cf. \cite{Zumino:1979et}, and decomposing the complex 
fields and complex transformation parameters into real ones.\\
Applying (\ref{proj}) and (\ref{backproj}), we can also express these transformations 
in terms of the constrained fields in order to determine the second supersymmetry 
of (\ref{lagron}). One finds the simple transformations
\begin{align}
\label{trafo2}
\delta\bn &= ~\ii\bn\times \bar{\eps}\bpsi\\
\nonumber
\delta \bpsi &= - \bn \times \partial_\mu \bn\, \gamma^\mu\eps
- \ii\bar{\eps}\bpsi \times \bpsi\,,
\end{align}
where $\boldsymbol{a}\times \boldsymbol{b}$ denotes the vector product 
of $\boldsymbol{a}$ and $\boldsymbol{b}$.
A proof that the action (\ref{lagron}) is invariant under these transformations 
is given in appendix \ref{app:inv}. The two on-shell supersymmetries
(\ref{trafo},\ref{trafo2}) are generated by the supercharges
\begin{align}
\cQ_{\rm I} = \ii \int \gamma^\mu\gamma^0 \partial_\mu\bn\bpsi\quad,\quad 
\cQ_{\rm II} = -\ii\int \gamma^\mu\gamma^0 (\bn\times\partial_\mu\bn)\bpsi\,.
\end{align}
These results are in agreement 
with the supercurrents constructed in \cite{Witten:1977xn}.

%% file: discretization.tex
\subsection{Discretization and constraints}
\label{ch:discretization}
So far, sigma models in the continuum have been considered. In order 
to investigate the corresponding lattice models one should try to discretize it
in a way that maintains as many symmetries of the continuum theory as possible. 
This is difficult with respect to supersymmetry, as will be discussed in section
\ref{ch:exactsusy}. But also the flavor symmetry must be treated with care: 
If one starts with an unconstrained formulation of the model and tries to 
discretize it in a straightforward way, one generically breaks the O$(N)$ 
symmetry. For the O$(3)$ model we will illustrate  this in more 
detail in section \ref{ch:catterall}. Simulations demonstrate clearly that 
even in the continuum limit the symmetry is not restored in such cases. 
In order to avoid this problem, we start with a formulation in terms of 
constrained fields, whose discretization is manifestly O$(N)$-invariant:
\begin{equation} \label{disc}
S[\bn,\bpsi] = \frac{1}{2 g^2} \sum_{x,y \in \Lambda}  \left(\bn_{x}^\trnsp K_{xy} \bn_{y} 
+\ii \bar{\bpsi}^\alpha_{x} M^{\alpha \beta}_{xy} \bpsi^{\beta}_{y} + \tfrac{1}{4} 
(\bar{\bpsi}_{x}\delta_{xy}\bpsi_{y})^2\right).
\end{equation}
The subscripts $x,y$ denote lattice sites, while $\alpha$, $\beta$ are spinor indices. 
The lattice derivatives $K_{xy}$ and $M^{\alpha \beta}_{xy}$ are proportional to
the identity in flavor space. The constraints $\bn_x\bn_x = 1$ and 
$\bn_x\bpsi_x = 0$ must be fulfilled at each lattice point $x$ and 
they are implemented as delta-functions in the path integral measure. 
This causes some difficulties in 
numerical simulations.  One can cope with this problem by either applying the 
stereographic projection or by introducing group valued dynamical variables.

\subsubsection{Stereographic projection and measure of path integration}
The stereographic projection \eqref{proj} resolves both constraints and 
leads to an unconstrained but yet O$(N)$-symmetric lattice formulation:
\begin{equation} \label{actiondisc}
\begin{aligned}
S[\bu,\blam] &= S_{\rm B} + S_{\rm 2F} + S_{\rm 4F},\text{~with}\\
&S_{\rm B} = \frac{1}{2 g^2} \sum_{x,y}  4\rho_x \bu^T_{x} K_{xy} \bu_{y} \rho_y 
+ \rho_x (1-\bu_x^2) K_{xy} (1- \bu_y^2) \rho_y, \\
&S_{\rm 2F} = \frac{2\ii}{ g^2} \!\!\sum_{x,y;\alpha,\beta}\!\! 
\bar{\blam}_{x}^{\alpha} \Big[\left(\rho-2\rho^2 \bu\bu^T\right)_x 
\!M_{xy}^{\alpha \beta} \left(\rho -2\rho^2 \bu\bu^T\right)_y 
+4 \left(\rho^2 \bu\right)_x \! M_{xy}^{\alpha \beta} \left(\rho^2\bu^T\right)_y\Big] 
\blam_{y}^{\beta}, \\
&S_{\rm 4F} = \frac{2}{g^2} \sum_{x} 
\rho_x^4 (\bar{\blam}_x \blam_x)^2\,.
\end{aligned}
\end{equation}
The change from the constrained fields $(\bn, \bpsi)$
to the unconstrained fields $(\bu, \blam)$ yields a non-trivial Jacobian
which is computed in appendix \ref{app:meas}. The result is\footnote{One obtains 
the same result if one does not eliminate the auxiliary field $f$ at the 
beginning but projects it in accordance to (\ref{stereo}) and integrates out the 
unconstrained auxiliary field $g$ afterwards. Following this approach, 
the superdeterminant yields only a trival factor, while the density 
$1/\rho^{N-1}$ stems from integrating w.r.t. $g$.}:
\begin{equation}
\label{eq:discretization:measure}
\prod_{x\in \Lambda} \dd\bn_{x} ~\dd\bpsi^1_x\,\dd\bpsi^2_x
~ \delta(\bn_x^2-1) \delta(\bn\bpsi_x^1) \delta(\bn\bpsi_x^2) 
\longrightarrow \prod_{x\in \Lambda} \dd\bu_{x}~ \dd\blam^1_x\,\dd \blam^1_x 
~\left(1+\bu_x^2\right)^{N-1}\,.
\end{equation}
The four-fermion interaction can be eliminated by employing the usual 
Hubbard-Stratonovich transformation \cite{Hubbard:1959ub}, which introduces 
an auxiliary bosonic field $\sigma$:
\begin{equation} 
\label{stereoaction}
  S[\bu,\blam] = S_{\rm B} + S_{\rm 2F} 
+ \frac{1}{2 g^2} \sum_{x\in \Lambda} \left(\sigma_x^2 
+ 4\ii  \sigma_x \rho_x^2~ \bar{\blam}_x\blam_x\right)
\end{equation}


\subsubsection{Coset formulation}
The constrained field $\bn$ propagates on the unit sphere $S^{N'}$
with $N'=N-1$ which can be viewed as coset space SO$(N)$/SO$(N')$.
To relate the constrained field to an element of the orthogonal
group SO$(N)$ we supplement at each lattice site the unit vector $\bn$ by
orthonormal vectors $\be_1,\dots,\be_{N'}$ such that $\{\bn,\be_1,\dots,\be_{N'}\}$
forms an oriented orthonormal basis of $\R^N$. The $N$-component Majorana spinor 
orthogonal to $\bn$ is a linear combination of the $\be_i$, i.e. 
\begin{equation}
\bpsi=\chi_1\be_1+\dots+\chi_{N'}\be_{N'}\,.\label{cosetfermion}
\end{equation}
These $N$ basis vectors may be viewed as columns of a rotation 
matrix $R=(\bn,\be_1,\dots,\be_{N'})$ in $N$ dimensions. Therefore 
the $x$-dependent orthonormal basis is given by a spacetime-dependent 
rotation $R\in {\rm SO}(N)$ acting on a constant ($x$-independent) orthonormal
frame $\{\bn_0,\bg_1,\dots,\bg_{N'}\}$:
\begin{equation}
\bn(x)=R(x) \bn_0,\quad \be_i(x) = R(x)\bg_i\,,
\end{equation}
and the path integral in terms of new variables $R,\,\bchi$ and $\sigma$ reads
\begin{equation}
\label{eq:sigma:pathIntegralInRvars}
\ZZ = \int \cD R\, \cD\sigma\, \cD \bchi\, \ee^{-\SB-\SF}
\end{equation}
with actions 
\begin{equation}
\SB = \frac{1}{2g^2} \sum_{x,y} \bigl( \bn_0^\trnsp R_x^\trnsp K_{xy} R_y \bn_0 + 
\sigma_x\delta_{xy} \sigma_y\bigr),\quad
\SF=\frac{\ii}{2g^2}\sum_{x,y}\bchi_x^\trnsp Q_{xy} \bchi_y\,.\label{rotaction}
\end{equation}
The action for the fermions contains the real and antisymmetric matrix
\begin{equation}
\label{diracoperator}
Q_{xy,ij}=\bg^\trnsp_i R_x^\trnsp C M_{xy}R_y\bg_j+\delta_{xy}\delta_{ij}C\sigma_x\,.
\end{equation}
The base vectors $\{\bg_i\}$ (or $\{\be_i\}$) are not unique since any rotation
in the plane orthogonal to $\bn_0$ transforms an admissible
set of base vectors into another admissible set. More precisely,
$$
R\longrightarrow R'=RS,\;\;\; \chi_i\longrightarrow
\chi'_i=\chi_k S_{ki},\quad\hbox{with}\quad
S\bn_0=\bn_0,\quad S_{ij}=\bg_i^\trnsp S\bg_j\,
$$
are local SO$(N')$ symmetries of the action (\ref{rotaction}).
The measure of the path integral is not affected by the change of the 
dynamical fields from $\bn$ to $R$, i.e. a distribution of $R$ according
to the Haar measure on SO$(N)$ will lead to a uniform distribution 
of $\bn=R \bn_0$ on the sphere. This renders the structure of the
lattice action rather simple. The price we pay is a local SO$(N')$
symmetry in the choice of the basis vectors $\be_i$.

\subsubsection{Fermion determinant}
The treatment of the fermion operator is very similar for stereographically projected or group valued variables and we will only depict the second case here. Performing the Grassmannian integral in (\ref{eq:sigma:pathIntegralInRvars}) 
leads to the bosonic path integral
\begin{equation}
\label{eq:sigma:bosonicActionInRvars}
\ZZ = \int \cD R\,\cD\sigma\,\sign \Pf (Q)\, \ee^{-\SB+\ln \abs{\Pf (Q)}}\,,
\end{equation}
and simulations are performed in the sign quenched ensemble.
This means that the configurations are sampled according to the
probability distribution defined by the exponential function. 
The sign of the Pfaffian is handled by reweighting. 
In this formulation the invariance of the path integral under the
local SO$(N')$ transformation is obvious: The bosonic action and
Haar measure are both invariant under the change of variables $R\to R S$ 
with $S\bn_0=\bn_0$. The operator $Q_{xy}$ 
transforms as $S_x^\trnsp Q_{xy} S_y$, which leaves the 
Pfaffian invariant since $S=\otimes_x S_x$ cancels
in the general relation $\Pf(S^\trnsp QS)=\det(S)\Pf(Q)$.
In the simulations the effective fermionic action is rewritten according 
to  $\ln \abs{\Pf (Q)} = \half \ln \det(Q)$ and the hybrid Monte Carlo algorithm 
is used, such that the Hamiltonian evolution 
of the group valued $R$ field is similar to the quenched case.\\


%% file: derivatives.tex
\subsection{Lattice derivatives}

Now we specify the lattice derivative for the fermions $M_{xy}^{\alpha \beta}$ and 
bosons $K_{xy}$ in the $O(N)$ invariant formulation. We will use two 
different implementations, the SLAC derivative and 
the Wilson derivative, so that we can compare results. 

\subsubsection{SLAC derivative}

With the help of discrete Fourier transformation on
a finite $\Nt\times \Ns$ lattice $\Lambda$ with $N$
lattice sites
one may motivate the SLAC derivative
\cite{Drell:1976bq},
\begin{equation}
\frac{1}{\sqrt{N}}\sum_{p\in\Lambda^{\!*}}
\ii p_\mu\ee^{\ii p x}\tilde f(p) 
= \sum_{y\in \Lambda} f(y) \Big(\frac{1}{N}\sum_{p\in
\Lambda^{\!*}} \ii p_\mu\ee^{ip (x-y)}\Big) \equiv \sum_{y\in
\Lambda}\big(\partial^\text{SLAC}_\mu\big)_{xy}f(y)\, ,
\end{equation}
where $p$ is from the dual Lattice $\Lambda^{\!*}$. This non-local derivative has
been proven to be useful in the context of Wess-Zumino models \cite{Kastner:2008zc} 
and supersymmetric quantum mechanics \cite{Bergner:2007pu,Wozar:2011gu}.
Because the formulations based on the SLAC derivative show only 
small lattice artifacts and do \emph{not break} the $\bbZ_2$ chiral symmetry 
explicitly, we use this derivative for the supersymmetric nonlinear 
sigma model:
\begin{equation}
\label{eq:sigma:operatorChoice}
K_{xy} =
-\sum_{\mu}\big(\partial_\mu^\text{SLAC}\big)^2_{xy},\quad
M_{xy} = \big(\gamma^\mu \partial^\text{SLAC}_\mu\big)_{xy}\,.
\end{equation}
The SLAC-derivative leads to lattice models without doublers.
\subsubsection{Wilson derivative}

Another suitable choice that avoids fermion doublers is the 
Wilson derivative which introduces an additional momentum-dependent mass term 
that vanishes in the naive continuum limit,
\begin{equation}
M_{xy}^{\alpha\beta} = \gamma_\mu^{\alpha\beta}\big(\partial_{\mu}^\text{sym}\big)_{xy} 
+ \delta^{\alpha\beta}\frac{ra}{2}\Delta_{xy}\,,
\end{equation}
where $\partial_\mu^\text{sym}$ is the symmetric lattice derivative, 
$a$ the lattice spacing and $\Delta_{xy}$ the lattice Laplacian. We 
choose the bosonic derivative in a particular form that has demonstrated 
superior results in Wess-Zumino models \cite{Kastner:2008zc},
\begin{equation}
K_{xy} = -\sum_{\mu} \big(\partial_\mu^\text{sym}
\big)^2_{xy} + \Big( \frac{ra}{2}\Delta_{xy} \Big)^2\,.
\end{equation}
The Wilson prescription breaks chiral symmetry explicitly, but is ultralocal in
contrast to the SLAC derivative.

\subsubsection{Universality and the SLAC derivative}

It is not obvious that the SLAC derivative can be used for models with 
curved target space. To justify its use we will first study the purely bosonic 
O$(3)$-model with this derivative and test if the known continuum result 
with scaling of the finite volume mass gap $m$ can be reproduced.
A quantity accessible at finite volumes is the step scaling function
introduced by L\"uscher, Weisz and Wolff \cite{Luscher:1991wu}. At finite 
spatial volume  $L=a\Ns$, the renormalization group invariant
variable $m(\Ns)L$ 
\rTABLE{
\begin{tabular}{ccc} \hline \hline
$\Ns$  & $\Sigma(2,u_0,\Ns^{-1})$ & $g^{-2}$ \\ \hline
$4$   & $1.28914(19)$ & $1.22906(6)$\\
$6$   & $1.27938(18)$ & $1.31071(5)$ \\
$8$   & $1.27368(34)$ & $1.36526(9)$ \\
$10$  & $1.27049(31)$ & $1.40622(8)$ \\
$12$  & $1.26742(33)$ & $1.43903(8)$ \\
$16$  & $1.26587(35)$ & $1.48986(8)$ \\
$20$  & $1.26416(70)$ & $1.52870(12)$ \\ \hline \hline
\end{tabular}
\caption{\label{tab:sigma:stepScalingInterpolation} 
Value of the step scaling function $\Sigma(2,u_0,\Ns^{-1})$ after interpolating to $u_0=1.0595$ for
different spatial lattice sizes $\Ns$ and corresponding $g^{-2}$ at which
$m(\Ns)L=u_0$ is reached.}}
can be computed at large temporal extent. 
On every finite lattice the step scaling function $\Sigma$ is then 
determined according to
\begin{equation}
\Sigma(2,u,\Ns^{-1}) = \left. m(2\Ns)2L\right\vert_{m(\Ns)L=u}\, ,
\end{equation}
where the condition $m(\Ns)L =u$ determines the bare coupling $g^{-2}$ that
is used on both spatial volumes $\Ns$ and $2\Ns$.\\
One has to measure the finite volume mass gap $m$ twice
in order to determine $\Sigma$ at a given $\Ns$.
First one has to tune the bare coupling until $mL$ is $u$ on
the small lattice with $\Ns$ spatial sites. Then one goes to the 
larger lattice with $2\Ns$ sites and measures $m(2\Ns)2\Ls$
with the previously determined bare coupling.
The lattice step scaling function is
expected to have a universal continuum limit 
$$\sigma(2,u) = \Sigma(2,u,0) =
\lim_{\Ns\to \infty} \Sigma(2,u,\Ns^{-1})
$$ 
and the high-precision results in
\cite{Balog:2009np,Wolff:2009kp} demonstrate
this universal behaviour. Here, the step scaling function
is computed using the SLAC derivative.
Following \cite{Luscher:1991wu} we use 
$u_0=1.0595$ which, according to  \cite{Balog:2003yr},
leads to the continuum value
\begin{equation}
\label{eq:sigma:pointOfInterest}
\sigma(2,u_0)=1.261210\,.
\end{equation}
Using the SLAC derivative the regularized action is given by
\begin{equation}
S = -\frac{1}{2g^2} \sum_{x,y} \bn^\trnsp_0 
R_x^\trnsp (\partial^\text{SLAC}_\mu)^2_{xy} R_y 
\bn_0,
\end{equation}
where $R$ is the group valued dynamical field. Since the action is
not given by nearest neighbour interactions, a cluster algorithm is not
applicable and the hybrid Monte-Carlo algorithm will be used instead. 
The momenta of the `Hamiltonian' used for the hybrid Monte-Carlo algorithm 
are elements of the Lie algebra $so(3)$.\\
The O$(3)$ invariant correlator is naturally defined as 
\begin{equation}\label{eq:sigma:bosonCorrelator}
C(t) = \Ns^{-2}\sum_{x,y} \bigl\langle\bn_{(t,x)} \bn_{(0,y)}\bigr\rangle
\end{equation}
and the mass is extracted via a fit to
\begin{equation}
C(t) \propto \cosh\left(ma\left(t-\Nt/2\right)\right)
\end{equation}
on a logarithmic scale, so that contributions from $t$'s
near $\Nt/2$ are taken into account, where the contributions of the higher excited states
are suppressed. However, reliable high-precision results can only be
obtained if systematic errors are under control. To actually 
see the contribution of higher excited states, one considers
the extracted mass as a function of $t_0$ where the fit is performed 
over a range $t\in[t_0,\Nt-t_0]$ for fixed $\Nt$ and $\Ns$. 
This effect is analysed for $\Ns\in\{6,12\}$ and $\Nt=6\Ns$ for a
coupling $g^{-2}=1.309$, which is quite close to the point of interest
\eqref{eq:sigma:pointOfInterest}, with extremely large statistics of about
$5\cdot 10^{9}$ configurations, distributed over $1\,000$ replica. The results
depicted in Fig.~\ref{fig:sigma:massgapSkip} show that for
larger lattices the contribution of higher excited states is well 
below the usual statistical accuracy that is used for most of the 
results given below. 
For small lattices there are two competing effects: For \emph{small} $t_0$ 
the contribution of higher states is well visible, whereas for \emph{large} $t_0$
the well-known fluctuations arising from the non-locality of the 
SLAC derivative begin to grow. Therefore the optimal choice leading 
to systematic errors of the same order as the statistical ones, 
is given by $t_0=\Ns$ and will be used in the following.
The second possible systematic error is due to the finite $\Nt$. 
For $\Nt$ too small a thermal contribution to the mass gap will 
show up. This effect is investigated by keeping $t_0=\Ns$ fixed but varying $\Nt$, see
Fig.~\ref{fig:sigma:massgapNt} for $\Ns\in\{6,12\}$. For the smaller lattice the
contributions at small $\Nt$ are more pronounced and become negligible for
$\Nt>6\Ns$ while for the larger lattice $\Nt>5\Ns$ is sufficient. To
suppress these systematic errors $\Nt=8\Ns$ ($\Nt=6\Ns$) is used 
on the smaller (larger) lattice of each step scaling computation.
\FIGURE{\hfill
\input{massOverSkip6}\hfill
\input{massOverSkip12}\hfill\hfill\phantom.
\caption{\label{fig:sigma:massgapSkip}Mass gap extracted from a logarithmic
$\cosh$ fit of the correlator
\eqref{eq:sigma:bosonCorrelator}
in a range
$t\in[t_0,\Nt-t_0]$ for $\Ns=6$ (left panel) and $\Ns=12$ (right panel) at
coupling $g^{-2}=1.309$ for fixed $\Nt=6\Ns$. The shaded area denotes the usual accuracy of results at other
$g^{-2}$.}}
\FIGURE{\hfill
\input{massOverNt6}\hfill
\input{massOverNt12}\hfill\hfill\phantom.
\caption{\label{fig:sigma:massgapNt}
Mass gap extracted from a logarithmic $\cosh$ fit of the correlator \eqref{eq:sigma:bosonCorrelator} in a range
$t\in[\Ns,\Nt-\Ns]$ for $\Ns=6$ (left panel) and $\Ns=12$ (right panel) at
coupling $\beta=1.309$ for different $\Nt/\Ns$.}}
With the systematic errors under control it is possible to extrapolate
the lattice step scaling function to the continuum limit. The
bare coupling is tuned such that $u=m(\Ns)L$ is near
$u_0=1.0595$ on lattices with sizes $\Ns\in\{4,6,8,10,12,16,20\}$. 
The corresponding $\Sigma(2,u,\Ns^{-1})$ are
plotted over $u$ for a subset of these $\Ns$ in
Fig.~\ref{fig:sigma:stepScalingFunctionContinuum} (left panel), and a linear
interpolation based on seven different coupling $g^{-2}$ allows for the
extraction of $\Sigma(2,u_0,\Ns^{-1})$ at the point $u_0=1.0595$. The
explicit values are collected in Tab.~\ref{tab:sigma:stepScalingInterpolation}. 
With Symanzik's theory of lattice artefacts it has been argued in
\cite{Balog:2009np} that finite $a$ corrections are of 
order $\ord(a^2 (\ln a)^3)$ and appear \emph{nearly linear} for 
a large range of computationally accessible lattice sizes \cite{Hasenbusch:2001ht}. 
For that reason an extrapolation to $a=0$ based on the formula
\begin{equation}
\label{eq:sigma:continuumExtrapolation}
\Sigma(2,u_0,\Ns^{-1}) = \sigma(2,u_0)+ A  \left(\frac{B}{\Ns}\right)^2
\left(\ln \frac{B}{\Ns}\right)^3
\end{equation}
is used. The results for $\Nt=4$ have been omitted because
of the large systematic errors introduced by the fluctuations arising from the
SLAC derivative for large lattice spacings. The extrapolation is shown in
Fig.~\ref{fig:sigma:stepScalingFunctionContinuum} (right panel) and a value of 
$\sigma(2,u_0)=1.2604(13)$ is extracted.
This value is in complete agreement with the exactly
known result in the continuum limit, see eq.~\eqref{eq:sigma:pointOfInterest}.
Therefore a discretisation of the (bosonic) O$(3)$ nonlinear sigma model with
the SLAC derivative is feasible and may also be used for the
supersymmetric model.
\FIGURE{\hfill
\input{stepScalingFunction}\hfill
\input{stepScalingContinuum}\hfill\hfill\phantom.
\caption{\label{fig:sigma:stepScalingFunctionContinuum}Left panel: Step scaling
function for lattices with spatial size $\Ns\in\{4,6,8,12\}$. Shaded regions indicate error
bounds of a linear interpolation.
Right panel: Continuum limit of the step
scaling function for $u_0=1.0595$. The shaded area indicates the
error bounds of a fit according to Eq.~\eqref{eq:sigma:continuumExtrapolation}, where the
value for $\Ns^{-1}=0.25$ has been omitted. The black dot marks the continuum
value given in Eq.~\eqref{eq:sigma:pointOfInterest}.}}

%% file: massOverSkip6.tex
\begingroup
\footnotesize
  \makeatletter
  \providecommand\color[2][]{%
    \GenericError{(gnuplot) \space\space\space\@spaces}{%
      Package color not loaded in conjunction with
      terminal option `colourtext'%
    }{See the gnuplot documentation for explanation.%
    }{Either use 'blacktext' in gnuplot or load the package
      color.sty in LaTeX.}%
    \renewcommand\color[2][]{}%
  }%
  \providecommand\includegraphics[2][]{%
    \GenericError{(gnuplot) \space\space\space\@spaces}{%
      Package graphicx or graphics not loaded%
    }{See the gnuplot documentation for explanation.%
    }{The gnuplot epslatex terminal needs graphicx.sty or graphics.sty.}%
    \renewcommand\includegraphics[2][]{}%
  }%
  \providecommand\rotatebox[2]{#2}%
  \@ifundefined{ifGPcolor}{%
    \newif\ifGPcolor
    \GPcolortrue
  }{}%
  \@ifundefined{ifGPblacktext}{%
    \newif\ifGPblacktext
    \GPblacktexttrue
  }{}%
  \let\gplgaddtomacro\g@addto@macro
  \gdef\gplbacktext{}%
  \gdef\gplfronttext{}%
  \makeatother
  \ifGPblacktext
    \def\colorrgb#1{}%
    \def\colorgray#1{}%
  \else
    \ifGPcolor
      \def\colorrgb#1{\color[rgb]{#1}}%
      \def\colorgray#1{\color[gray]{#1}}%
      \expandafter\def\csname LTw\endcsname{\color{white}}%
      \expandafter\def\csname LTb\endcsname{\color{black}}%
      \expandafter\def\csname LTa\endcsname{\color{black}}%
      \expandafter\def\csname LT0\endcsname{\color[rgb]{1,0,0}}%
      \expandafter\def\csname LT1\endcsname{\color[rgb]{0,1,0}}%
      \expandafter\def\csname LT2\endcsname{\color[rgb]{0,0,1}}%
      \expandafter\def\csname LT3\endcsname{\color[rgb]{1,0,1}}%
      \expandafter\def\csname LT4\endcsname{\color[rgb]{0,1,1}}%
      \expandafter\def\csname LT5\endcsname{\color[rgb]{1,1,0}}%
      \expandafter\def\csname LT6\endcsname{\color[rgb]{0,0,0}}%
      \expandafter\def\csname LT7\endcsname{\color[rgb]{1,0.3,0}}%
      \expandafter\def\csname LT8\endcsname{\color[rgb]{0.5,0.5,0.5}}%
    \else
      \def\colorrgb#1{\color{black}}%
      \def\colorgray#1{\color[gray]{#1}}%
      \expandafter\def\csname LTw\endcsname{\color{white}}%
      \expandafter\def\csname LTb\endcsname{\color{black}}%
      \expandafter\def\csname LTa\endcsname{\color{black}}%
      \expandafter\def\csname LT0\endcsname{\color{black}}%
      \expandafter\def\csname LT1\endcsname{\color{black}}%
      \expandafter\def\csname LT2\endcsname{\color{black}}%
      \expandafter\def\csname LT3\endcsname{\color{black}}%
      \expandafter\def\csname LT4\endcsname{\color{black}}%
      \expandafter\def\csname LT5\endcsname{\color{black}}%
      \expandafter\def\csname LT6\endcsname{\color{black}}%
      \expandafter\def\csname LT7\endcsname{\color{black}}%
      \expandafter\def\csname LT8\endcsname{\color{black}}%
    \fi
  \fi
  \setlength{\unitlength}{0.0500bp}%
  \begin{picture}(3968.00,2976.00)%
    \gplgaddtomacro\gplbacktext{%
      \csname LTb\endcsname%
      \put(768,480){\makebox(0,0)[r]{\strut{}$1.0610$}}%
      \put(768,1092){\makebox(0,0)[r]{\strut{}$1.0615$}}%
      \put(768,1703){\makebox(0,0)[r]{\strut{}$1.0620$}}%
      \put(768,2315){\makebox(0,0)[r]{\strut{}$1.0625$}}%
      \put(768,2927){\makebox(0,0)[r]{\strut{}$1.0630$}}%
      \put(864,320){\makebox(0,0){\strut{}$0$}}%
      \put(1303,320){\makebox(0,0){\strut{}$2$}}%
      \put(1743,320){\makebox(0,0){\strut{}$4$}}%
      \put(2182,320){\makebox(0,0){\strut{}$6$}}%
      \put(2621,320){\makebox(0,0){\strut{}$8$}}%
      \put(3060,320){\makebox(0,0){\strut{}$10$}}%
      \put(3500,320){\makebox(0,0){\strut{}$12$}}%
      \put(3939,320){\makebox(0,0){\strut{}$14$}}%
      \put(64,2023){\makebox(0,0){\strut{}$mL$}}%
      \put(2401,80){\makebox(0,0){\strut{}$t_0$}}%
    }%
    \gplgaddtomacro\gplfronttext{%
    }%
    \gplbacktext
    \put(0,0){\includegraphics{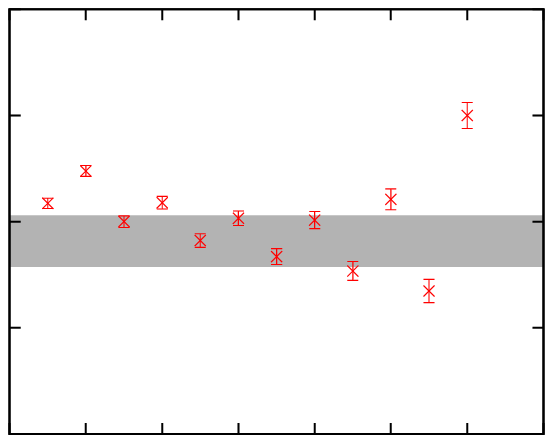}}%
    \gplfronttext
  \end{picture}%
\endgroup

%% file: massOverSkip12.tex
\begingroup
\footnotesize
  \makeatletter
  \providecommand\color[2][]{%
    \GenericError{(gnuplot) \space\space\space\@spaces}{%
      Package color not loaded in conjunction with
      terminal option `colourtext'%
    }{See the gnuplot documentation for explanation.%
    }{Either use 'blacktext' in gnuplot or load the package
      color.sty in LaTeX.}%
    \renewcommand\color[2][]{}%
  }%
  \providecommand\includegraphics[2][]{%
    \GenericError{(gnuplot) \space\space\space\@spaces}{%
      Package graphicx or graphics not loaded%
    }{See the gnuplot documentation for explanation.%
    }{The gnuplot epslatex terminal needs graphicx.sty or graphics.sty.}%
    \renewcommand\includegraphics[2][]{}%
  }%
  \providecommand\rotatebox[2]{#2}%
  \@ifundefined{ifGPcolor}{%
    \newif\ifGPcolor
    \GPcolortrue
  }{}%
  \@ifundefined{ifGPblacktext}{%
    \newif\ifGPblacktext
    \GPblacktexttrue
  }{}%
  \let\gplgaddtomacro\g@addto@macro
  \gdef\gplbacktext{}%
  \gdef\gplfronttext{}%
  \makeatother
  \ifGPblacktext
    \def\colorrgb#1{}%
    \def\colorgray#1{}%
  \else
    \ifGPcolor
      \def\colorrgb#1{\color[rgb]{#1}}%
      \def\colorgray#1{\color[gray]{#1}}%
      \expandafter\def\csname LTw\endcsname{\color{white}}%
      \expandafter\def\csname LTb\endcsname{\color{black}}%
      \expandafter\def\csname LTa\endcsname{\color{black}}%
      \expandafter\def\csname LT0\endcsname{\color[rgb]{1,0,0}}%
      \expandafter\def\csname LT1\endcsname{\color[rgb]{0,1,0}}%
      \expandafter\def\csname LT2\endcsname{\color[rgb]{0,0,1}}%
      \expandafter\def\csname LT3\endcsname{\color[rgb]{1,0,1}}%
      \expandafter\def\csname LT4\endcsname{\color[rgb]{0,1,1}}%
      \expandafter\def\csname LT5\endcsname{\color[rgb]{1,1,0}}%
      \expandafter\def\csname LT6\endcsname{\color[rgb]{0,0,0}}%
      \expandafter\def\csname LT7\endcsname{\color[rgb]{1,0.3,0}}%
      \expandafter\def\csname LT8\endcsname{\color[rgb]{0.5,0.5,0.5}}%
    \else
      \def\colorrgb#1{\color{black}}%
      \def\colorgray#1{\color[gray]{#1}}%
      \expandafter\def\csname LTw\endcsname{\color{white}}%
      \expandafter\def\csname LTb\endcsname{\color{black}}%
      \expandafter\def\csname LTa\endcsname{\color{black}}%
      \expandafter\def\csname LT0\endcsname{\color{black}}%
      \expandafter\def\csname LT1\endcsname{\color{black}}%
      \expandafter\def\csname LT2\endcsname{\color{black}}%
      \expandafter\def\csname LT3\endcsname{\color{black}}%
      \expandafter\def\csname LT4\endcsname{\color{black}}%
      \expandafter\def\csname LT5\endcsname{\color{black}}%
      \expandafter\def\csname LT6\endcsname{\color{black}}%
      \expandafter\def\csname LT7\endcsname{\color{black}}%
      \expandafter\def\csname LT8\endcsname{\color{black}}%
    \fi
  \fi
  \setlength{\unitlength}{0.0500bp}%
  \begin{picture}(3968.00,2976.00)%
    \gplgaddtomacro\gplbacktext{%
      \csname LTb\endcsname%
      \put(768,480){\makebox(0,0)[r]{\strut{}$1.280$}}%
      \put(768,888){\makebox(0,0)[r]{\strut{}$1.281$}}%
      \put(768,1296){\makebox(0,0)[r]{\strut{}$1.282$}}%
      \put(768,1703){\makebox(0,0)[r]{\strut{}$1.283$}}%
      \put(768,2111){\makebox(0,0)[r]{\strut{}$1.284$}}%
      \put(768,2519){\makebox(0,0)[r]{\strut{}$1.285$}}%
      \put(768,2927){\makebox(0,0)[r]{\strut{}$1.286$}}%
      \put(864,320){\makebox(0,0){\strut{}$0$}}%
      \put(1479,320){\makebox(0,0){\strut{}$5$}}%
      \put(2094,320){\makebox(0,0){\strut{}$10$}}%
      \put(2709,320){\makebox(0,0){\strut{}$15$}}%
      \put(3324,320){\makebox(0,0){\strut{}$20$}}%
      \put(3939,320){\makebox(0,0){\strut{}$25$}}%
      \put(160,1911){\makebox(0,0){\strut{}$mL$}}%
      \put(2401,80){\makebox(0,0){\strut{}$t_0$}}%
    }%
    \gplgaddtomacro\gplfronttext{%
    }%
    \gplbacktext
    \put(0,0){\includegraphics{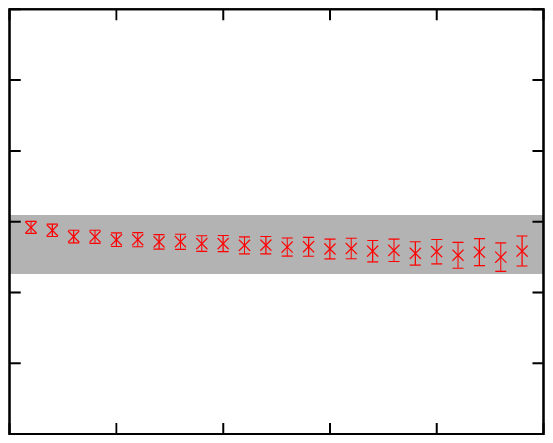}}%
    \gplfronttext
  \end{picture}%
\endgroup

%% file: massOverNt6.tex
\begingroup
\footnotesize
  \makeatletter
  \providecommand\color[2][]{%
    \GenericError{(gnuplot) \space\space\space\@spaces}{%
      Package color not loaded in conjunction with
      terminal option `colourtext'%
    }{See the gnuplot documentation for explanation.%
    }{Either use 'blacktext' in gnuplot or load the package
      color.sty in LaTeX.}%
    \renewcommand\color[2][]{}%
  }%
  \providecommand\includegraphics[2][]{%
    \GenericError{(gnuplot) \space\space\space\@spaces}{%
      Package graphicx or graphics not loaded%
    }{See the gnuplot documentation for explanation.%
    }{The gnuplot epslatex terminal needs graphicx.sty or graphics.sty.}%
    \renewcommand\includegraphics[2][]{}%
  }%
  \providecommand\rotatebox[2]{#2}%
  \@ifundefined{ifGPcolor}{%
    \newif\ifGPcolor
    \GPcolortrue
  }{}%
  \@ifundefined{ifGPblacktext}{%
    \newif\ifGPblacktext
    \GPblacktexttrue
  }{}%
  \let\gplgaddtomacro\g@addto@macro
  \gdef\gplbacktext{}%
  \gdef\gplfronttext{}%
  \makeatother
  \ifGPblacktext
    \def\colorrgb#1{}%
    \def\colorgray#1{}%
  \else
    \ifGPcolor
      \def\colorrgb#1{\color[rgb]{#1}}%
      \def\colorgray#1{\color[gray]{#1}}%
      \expandafter\def\csname LTw\endcsname{\color{white}}%
      \expandafter\def\csname LTb\endcsname{\color{black}}%
      \expandafter\def\csname LTa\endcsname{\color{black}}%
      \expandafter\def\csname LT0\endcsname{\color[rgb]{1,0,0}}%
      \expandafter\def\csname LT1\endcsname{\color[rgb]{0,1,0}}%
      \expandafter\def\csname LT2\endcsname{\color[rgb]{0,0,1}}%
      \expandafter\def\csname LT3\endcsname{\color[rgb]{1,0,1}}%
      \expandafter\def\csname LT4\endcsname{\color[rgb]{0,1,1}}%
      \expandafter\def\csname LT5\endcsname{\color[rgb]{1,1,0}}%
      \expandafter\def\csname LT6\endcsname{\color[rgb]{0,0,0}}%
      \expandafter\def\csname LT7\endcsname{\color[rgb]{1,0.3,0}}%
      \expandafter\def\csname LT8\endcsname{\color[rgb]{0.5,0.5,0.5}}%
    \else
      \def\colorrgb#1{\color{black}}%
      \def\colorgray#1{\color[gray]{#1}}%
      \expandafter\def\csname LTw\endcsname{\color{white}}%
      \expandafter\def\csname LTb\endcsname{\color{black}}%
      \expandafter\def\csname LTa\endcsname{\color{black}}%
      \expandafter\def\csname LT0\endcsname{\color{black}}%
      \expandafter\def\csname LT1\endcsname{\color{black}}%
      \expandafter\def\csname LT2\endcsname{\color{black}}%
      \expandafter\def\csname LT3\endcsname{\color{black}}%
      \expandafter\def\csname LT4\endcsname{\color{black}}%
      \expandafter\def\csname LT5\endcsname{\color{black}}%
      \expandafter\def\csname LT6\endcsname{\color{black}}%
      \expandafter\def\csname LT7\endcsname{\color{black}}%
      \expandafter\def\csname LT8\endcsname{\color{black}}%
    \fi
  \fi
  \setlength{\unitlength}{0.0500bp}%
  \begin{picture}(3968.00,2976.00)%
    \gplgaddtomacro\gplbacktext{%
      \csname LTb\endcsname%
      \put(768,480){\makebox(0,0)[r]{\strut{}$1.060$}}%
      \put(768,830){\makebox(0,0)[r]{\strut{}$1.065$}}%
      \put(768,1179){\makebox(0,0)[r]{\strut{}$1.070$}}%
      \put(768,1529){\makebox(0,0)[r]{\strut{}$1.075$}}%
      \put(768,1878){\makebox(0,0)[r]{\strut{}$1.080$}}%
      \put(768,2228){\makebox(0,0)[r]{\strut{}$1.085$}}%
      \put(768,2577){\makebox(0,0)[r]{\strut{}$1.090$}}%
      \put(768,2927){\makebox(0,0)[r]{\strut{}$1.095$}}%
      \put(864,320){\makebox(0,0){\strut{}3}}%
      \put(1479,320){\makebox(0,0){\strut{}4}}%
      \put(2094,320){\makebox(0,0){\strut{}5}}%
      \put(2709,320){\makebox(0,0){\strut{}6}}%
      \put(3324,320){\makebox(0,0){\strut{}7}}%
      \put(3939,320){\makebox(0,0){\strut{}8}}%
      \put(112,1703){\makebox(0,0){\strut{}$mL$}}%
      \put(2401,80){\makebox(0,0){\strut{}$\Nt/\Ns$}}%
    }%
    \gplgaddtomacro\gplfronttext{%
    }%
    \gplbacktext
    \put(0,0){\includegraphics{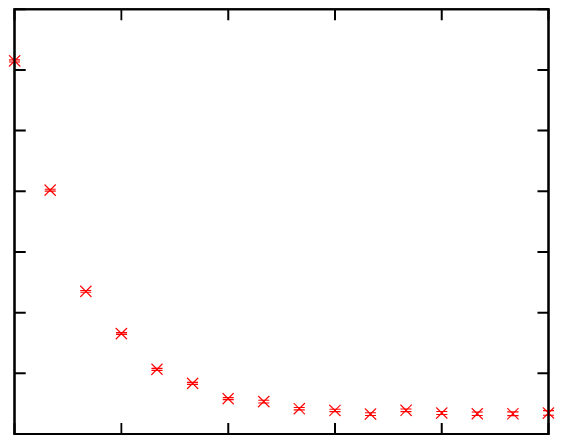}}%
    \gplfronttext
  \end{picture}%
\endgroup

%% file: massOverNt12.tex
\begingroup
\footnotesize
  \makeatletter
  \providecommand\color[2][]{%
    \GenericError{(gnuplot) \space\space\space\@spaces}{%
      Package color not loaded in conjunction with
      terminal option `colourtext'%
    }{See the gnuplot documentation for explanation.%
    }{Either use 'blacktext' in gnuplot or load the package
      color.sty in LaTeX.}%
    \renewcommand\color[2][]{}%
  }%
  \providecommand\includegraphics[2][]{%
    \GenericError{(gnuplot) \space\space\space\@spaces}{%
      Package graphicx or graphics not loaded%
    }{See the gnuplot documentation for explanation.%
    }{The gnuplot epslatex terminal needs graphicx.sty or graphics.sty.}%
    \renewcommand\includegraphics[2][]{}%
  }%
  \providecommand\rotatebox[2]{#2}%
  \@ifundefined{ifGPcolor}{%
    \newif\ifGPcolor
    \GPcolortrue
  }{}%
  \@ifundefined{ifGPblacktext}{%
    \newif\ifGPblacktext
    \GPblacktexttrue
  }{}%
  \let\gplgaddtomacro\g@addto@macro
  \gdef\gplbacktext{}%
  \gdef\gplfronttext{}%
  \makeatother
  \ifGPblacktext
    \def\colorrgb#1{}%
    \def\colorgray#1{}%
  \else
    \ifGPcolor
      \def\colorrgb#1{\color[rgb]{#1}}%
      \def\colorgray#1{\color[gray]{#1}}%
      \expandafter\def\csname LTw\endcsname{\color{white}}%
      \expandafter\def\csname LTb\endcsname{\color{black}}%
      \expandafter\def\csname LTa\endcsname{\color{black}}%
      \expandafter\def\csname LT0\endcsname{\color[rgb]{1,0,0}}%
      \expandafter\def\csname LT1\endcsname{\color[rgb]{0,1,0}}%
      \expandafter\def\csname LT2\endcsname{\color[rgb]{0,0,1}}%
      \expandafter\def\csname LT3\endcsname{\color[rgb]{1,0,1}}%
      \expandafter\def\csname LT4\endcsname{\color[rgb]{0,1,1}}%
      \expandafter\def\csname LT5\endcsname{\color[rgb]{1,1,0}}%
      \expandafter\def\csname LT6\endcsname{\color[rgb]{0,0,0}}%
      \expandafter\def\csname LT7\endcsname{\color[rgb]{1,0.3,0}}%
      \expandafter\def\csname LT8\endcsname{\color[rgb]{0.5,0.5,0.5}}%
    \else
      \def\colorrgb#1{\color{black}}%
      \def\colorgray#1{\color[gray]{#1}}%
      \expandafter\def\csname LTw\endcsname{\color{white}}%
      \expandafter\def\csname LTb\endcsname{\color{black}}%
      \expandafter\def\csname LTa\endcsname{\color{black}}%
      \expandafter\def\csname LT0\endcsname{\color{black}}%
      \expandafter\def\csname LT1\endcsname{\color{black}}%
      \expandafter\def\csname LT2\endcsname{\color{black}}%
      \expandafter\def\csname LT3\endcsname{\color{black}}%
      \expandafter\def\csname LT4\endcsname{\color{black}}%
      \expandafter\def\csname LT5\endcsname{\color{black}}%
      \expandafter\def\csname LT6\endcsname{\color{black}}%
      \expandafter\def\csname LT7\endcsname{\color{black}}%
      \expandafter\def\csname LT8\endcsname{\color{black}}%
    \fi
  \fi
  \setlength{\unitlength}{0.0500bp}%
  \begin{picture}(3968.00,2976.00)%
    \gplgaddtomacro\gplbacktext{%
      \csname LTb\endcsname%
      \put(768,480){\makebox(0,0)[r]{\strut{}$1.282$}}%
      \put(768,786){\makebox(0,0)[r]{\strut{}$1.284$}}%
      \put(768,1092){\makebox(0,0)[r]{\strut{}$1.286$}}%
      \put(768,1398){\makebox(0,0)[r]{\strut{}$1.288$}}%
      \put(768,1704){\makebox(0,0)[r]{\strut{}$1.290$}}%
      \put(768,2009){\makebox(0,0)[r]{\strut{}$1.292$}}%
      \put(768,2315){\makebox(0,0)[r]{\strut{}$1.294$}}%
      \put(768,2621){\makebox(0,0)[r]{\strut{}$1.296$}}%
      \put(768,2927){\makebox(0,0)[r]{\strut{}$1.298$}}%
      \put(864,320){\makebox(0,0){\strut{}$3.0$}}%
      \put(1377,320){\makebox(0,0){\strut{}$3.5$}}%
      \put(1889,320){\makebox(0,0){\strut{}$4.0$}}%
      \put(2402,320){\makebox(0,0){\strut{}$4.5$}}%
      \put(2914,320){\makebox(0,0){\strut{}$5.0$}}%
      \put(3427,320){\makebox(0,0){\strut{}$5.5$}}%
      \put(3939,320){\makebox(0,0){\strut{}$6.0$}}%
      \put(112,1863){\makebox(0,0){\strut{}$mL$}}%
      \put(2401,80){\makebox(0,0){\strut{}$\Nt/\Ns$}}%
    }%
    \gplgaddtomacro\gplfronttext{%
    }%
    \gplbacktext
    \put(0,0){\includegraphics{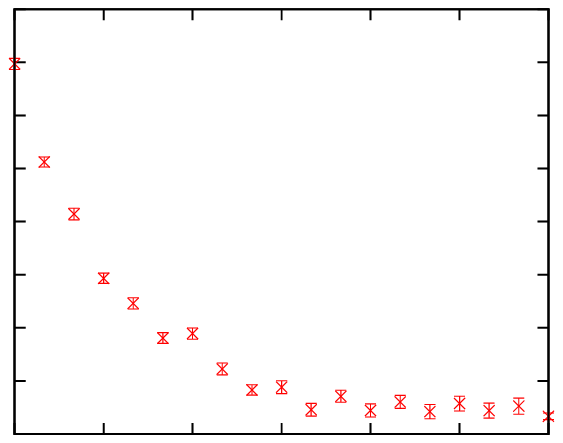}}%
    \gplfronttext
  \end{picture}%
\endgroup

%% file: stepScalingFunction.tex
\begingroup
\footnotesize
  \makeatletter
  \providecommand\color[2][]{%
    \GenericError{(gnuplot) \space\space\space\@spaces}{%
      Package color not loaded in conjunction with
      terminal option `colourtext'%
    }{See the gnuplot documentation for explanation.%
    }{Either use 'blacktext' in gnuplot or load the package
      color.sty in LaTeX.}%
    \renewcommand\color[2][]{}%
  }%
  \providecommand\includegraphics[2][]{%
    \GenericError{(gnuplot) \space\space\space\@spaces}{%
      Package graphicx or graphics not loaded%
    }{See the gnuplot documentation for explanation.%
    }{The gnuplot epslatex terminal needs graphicx.sty or graphics.sty.}%
    \renewcommand\includegraphics[2][]{}%
  }%
  \providecommand\rotatebox[2]{#2}%
  \@ifundefined{ifGPcolor}{%
    \newif\ifGPcolor
    \GPcolortrue
  }{}%
  \@ifundefined{ifGPblacktext}{%
    \newif\ifGPblacktext
    \GPblacktexttrue
  }{}%
  \let\gplgaddtomacro\g@addto@macro
  \gdef\gplbacktext{}%
  \gdef\gplfronttext{}%
  \makeatother
  \ifGPblacktext
    \def\colorrgb#1{}%
    \def\colorgray#1{}%
  \else
    \ifGPcolor
      \def\colorrgb#1{\color[rgb]{#1}}%
      \def\colorgray#1{\color[gray]{#1}}%
      \expandafter\def\csname LTw\endcsname{\color{white}}%
      \expandafter\def\csname LTb\endcsname{\color{black}}%
      \expandafter\def\csname LTa\endcsname{\color{black}}%
      \expandafter\def\csname LT0\endcsname{\color[rgb]{1,0,0}}%
      \expandafter\def\csname LT1\endcsname{\color[rgb]{0,1,0}}%
      \expandafter\def\csname LT2\endcsname{\color[rgb]{0,0,1}}%
      \expandafter\def\csname LT3\endcsname{\color[rgb]{1,0,1}}%
      \expandafter\def\csname LT4\endcsname{\color[rgb]{0,1,1}}%
      \expandafter\def\csname LT5\endcsname{\color[rgb]{1,1,0}}%
      \expandafter\def\csname LT6\endcsname{\color[rgb]{0,0,0}}%
      \expandafter\def\csname LT7\endcsname{\color[rgb]{1,0.3,0}}%
      \expandafter\def\csname LT8\endcsname{\color[rgb]{0.5,0.5,0.5}}%
    \else
      \def\colorrgb#1{\color{black}}%
      \def\colorgray#1{\color[gray]{#1}}%
      \expandafter\def\csname LTw\endcsname{\color{white}}%
      \expandafter\def\csname LTb\endcsname{\color{black}}%
      \expandafter\def\csname LTa\endcsname{\color{black}}%
      \expandafter\def\csname LT0\endcsname{\color{black}}%
      \expandafter\def\csname LT1\endcsname{\color{black}}%
      \expandafter\def\csname LT2\endcsname{\color{black}}%
      \expandafter\def\csname LT3\endcsname{\color{black}}%
      \expandafter\def\csname LT4\endcsname{\color{black}}%
      \expandafter\def\csname LT5\endcsname{\color{black}}%
      \expandafter\def\csname LT6\endcsname{\color{black}}%
      \expandafter\def\csname LT7\endcsname{\color{black}}%
      \expandafter\def\csname LT8\endcsname{\color{black}}%
    \fi
  \fi
  \setlength{\unitlength}{0.0500bp}%
  \begin{picture}(3968.00,2976.00)%
    \gplgaddtomacro\gplbacktext{%
      \csname LTb\endcsname%
      \put(768,480){\makebox(0,0)[r]{\strut{}$1.23$}}%
      \put(768,725){\makebox(0,0)[r]{\strut{}$1.24$}}%
      \put(768,969){\makebox(0,0)[r]{\strut{}$1.25$}}%
      \put(768,1214){\makebox(0,0)[r]{\strut{}$1.26$}}%
      \put(768,1459){\makebox(0,0)[r]{\strut{}$1.27$}}%
      \put(768,1704){\makebox(0,0)[r]{\strut{}$1.28$}}%
      \put(768,1948){\makebox(0,0)[r]{\strut{}$1.29$}}%
      \put(768,2193){\makebox(0,0)[r]{\strut{}$1.30$}}%
      \put(768,2438){\makebox(0,0)[r]{\strut{}$1.31$}}%
      \put(768,2682){\makebox(0,0)[r]{\strut{}$1.32$}}%
      \put(768,2927){\makebox(0,0)[r]{\strut{}$1.33$}}%
      \put(864,320){\makebox(0,0){\strut{}$1.04$}}%
      \put(1633,320){\makebox(0,0){\strut{}$1.05$}}%
      \put(2402,320){\makebox(0,0){\strut{}$1.06$}}%
      \put(3170,320){\makebox(0,0){\strut{}$1.07$}}%
      \put(3939,320){\makebox(0,0){\strut{}$1.08$}}%
      \put(112,1703){\rotatebox{90}{\makebox(0,0){\strut{}$\Sigma(2,u,\Ns^{-1})$}}}%
      \put(2401,80){\makebox(0,0){\strut{}$u$}}%
    }%
    \gplgaddtomacro\gplfronttext{%
      \csname LTb\endcsname%
      \put(1728,2776){\makebox(0,0)[r]{\strut{}$\Ns=4$}}%
      \csname LTb\endcsname%
      \put(1728,2600){\makebox(0,0)[r]{\strut{}$\Ns=6$}}%
      \csname LTb\endcsname%
      \put(1728,2424){\makebox(0,0)[r]{\strut{}$\Ns=8$}}%
      \csname LTb\endcsname%
      \put(1728,2248){\makebox(0,0)[r]{\strut{}$\Ns=12$}}%
    }%
    \gplbacktext
    \put(0,0){\includegraphics{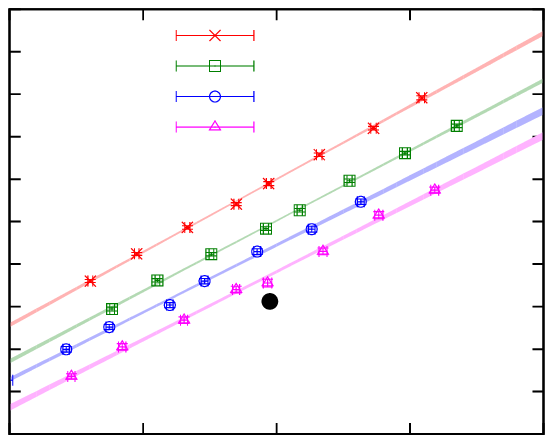}}%
    \gplfronttext
  \end{picture}%
\endgroup

%% file: stepScalingContinuum.tex
\begingroup
\footnotesize
  \makeatletter
  \providecommand\color[2][]{%
    \GenericError{(gnuplot) \space\space\space\@spaces}{%
      Package color not loaded in conjunction with
      terminal option `colourtext'%
    }{See the gnuplot documentation for explanation.%
    }{Either use 'blacktext' in gnuplot or load the package
      color.sty in LaTeX.}%
    \renewcommand\color[2][]{}%
  }%
  \providecommand\includegraphics[2][]{%
    \GenericError{(gnuplot) \space\space\space\@spaces}{%
      Package graphicx or graphics not loaded%
    }{See the gnuplot documentation for explanation.%
    }{The gnuplot epslatex terminal needs graphicx.sty or graphics.sty.}%
    \renewcommand\includegraphics[2][]{}%
  }%
  \providecommand\rotatebox[2]{#2}%
  \@ifundefined{ifGPcolor}{%
    \newif\ifGPcolor
    \GPcolortrue
  }{}%
  \@ifundefined{ifGPblacktext}{%
    \newif\ifGPblacktext
    \GPblacktexttrue
  }{}%
  \let\gplgaddtomacro\g@addto@macro
  \gdef\gplbacktext{}%
  \gdef\gplfronttext{}%
  \makeatother
  \ifGPblacktext
    \def\colorrgb#1{}%
    \def\colorgray#1{}%
  \else
    \ifGPcolor
      \def\colorrgb#1{\color[rgb]{#1}}%
      \def\colorgray#1{\color[gray]{#1}}%
      \expandafter\def\csname LTw\endcsname{\color{white}}%
      \expandafter\def\csname LTb\endcsname{\color{black}}%
      \expandafter\def\csname LTa\endcsname{\color{black}}%
      \expandafter\def\csname LT0\endcsname{\color[rgb]{1,0,0}}%
      \expandafter\def\csname LT1\endcsname{\color[rgb]{0,1,0}}%
      \expandafter\def\csname LT2\endcsname{\color[rgb]{0,0,1}}%
      \expandafter\def\csname LT3\endcsname{\color[rgb]{1,0,1}}%
      \expandafter\def\csname LT4\endcsname{\color[rgb]{0,1,1}}%
      \expandafter\def\csname LT5\endcsname{\color[rgb]{1,1,0}}%
      \expandafter\def\csname LT6\endcsname{\color[rgb]{0,0,0}}%
      \expandafter\def\csname LT7\endcsname{\color[rgb]{1,0.3,0}}%
      \expandafter\def\csname LT8\endcsname{\color[rgb]{0.5,0.5,0.5}}%
    \else
      \def\colorrgb#1{\color{black}}%
      \def\colorgray#1{\color[gray]{#1}}%
      \expandafter\def\csname LTw\endcsname{\color{white}}%
      \expandafter\def\csname LTb\endcsname{\color{black}}%
      \expandafter\def\csname LTa\endcsname{\color{black}}%
      \expandafter\def\csname LT0\endcsname{\color{black}}%
      \expandafter\def\csname LT1\endcsname{\color{black}}%
      \expandafter\def\csname LT2\endcsname{\color{black}}%
      \expandafter\def\csname LT3\endcsname{\color{black}}%
      \expandafter\def\csname LT4\endcsname{\color{black}}%
      \expandafter\def\csname LT5\endcsname{\color{black}}%
      \expandafter\def\csname LT6\endcsname{\color{black}}%
      \expandafter\def\csname LT7\endcsname{\color{black}}%
      \expandafter\def\csname LT8\endcsname{\color{black}}%
    \fi
  \fi
  \setlength{\unitlength}{0.0500bp}%
  \begin{picture}(3968.00,2976.00)%
    \gplgaddtomacro\gplbacktext{%
      \csname LTb\endcsname%
      \put(768,480){\makebox(0,0)[r]{\strut{}$1.255$}}%
      \put(768,786){\makebox(0,0)[r]{\strut{}$1.260$}}%
      \put(768,1092){\makebox(0,0)[r]{\strut{}$1.265$}}%
      \put(768,1398){\makebox(0,0)[r]{\strut{}$1.270$}}%
      \put(768,1703){\makebox(0,0)[r]{\strut{}$1.275$}}%
      \put(768,2009){\makebox(0,0)[r]{\strut{}$1.280$}}%
      \put(768,2315){\makebox(0,0)[r]{\strut{}$1.285$}}%
      \put(768,2621){\makebox(0,0)[r]{\strut{}$1.290$}}%
      \put(768,2927){\makebox(0,0)[r]{\strut{}$1.295$}}%
      \put(864,320){\makebox(0,0){\strut{}$0.00$}}%
      \put(1377,320){\makebox(0,0){\strut{}$0.05$}}%
      \put(1889,320){\makebox(0,0){\strut{}$0.10$}}%
      \put(2402,320){\makebox(0,0){\strut{}$0.15$}}%
      \put(2914,320){\makebox(0,0){\strut{}$0.20$}}%
      \put(3427,320){\makebox(0,0){\strut{}$0.25$}}%
      \put(3939,320){\makebox(0,0){\strut{}$0.30$}}%
      \put(16,1703){\rotatebox{90}{\makebox(0,0){\strut{}$\Sigma(2,u_0,\Ns^{-1})$}}}%
      \put(2401,80){\makebox(0,0){\strut{}$\Ns^{-1}$}}%
    }%
    \gplgaddtomacro\gplfronttext{%
    }%
    \gplbacktext
    \put(0,0){\includegraphics{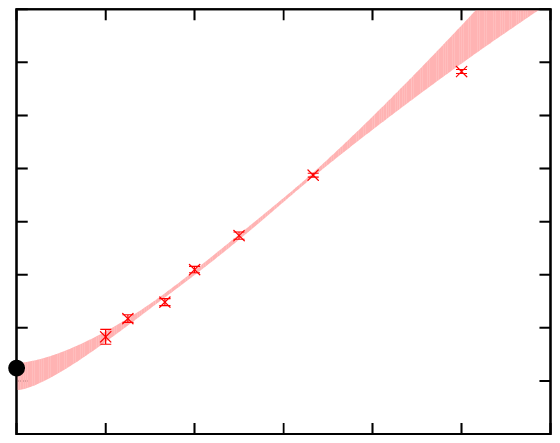}}%
    \gplfronttext
  \end{picture}%
\endgroup

%% file: exactSUSY.tex
\subsection{Supersymmetries on the lattice}
Is it possible to discretize the supersymmetric nonlinear O$(3)$-model such
that its characteristic symmetries are maintained? We have seen
that it is straightforward to find a manifestly O$(3)$-symmetric 
discretization. The treatment of
supersymmetry on the lattice is more difficult, because it is an extension of the
Poincar\'e symmetry, which is broken by any discretization of spacetime. However, 
there are some approaches to maintain at least a part of the supersymmetry on the 
lattice.\\
One of these approaches relies on a twisting of the supersymmetry in a way that provides a 
\emph{nilpotent} supercharge $\cQ$ which is used to find a $\cQ$-exact formulation $S = \cQ \Lambda$
of the lattice action. By construction this action is invariant under 
supersymmetries generated by $\cQ$. This approach has been applied to the O$(3)$-NLSM in
\cite{Catterall:2003uf,Catterall:2006sj}. However, the authors employ a 
formulation of the model in terms of unconstrained fields and its discretization 
breaks the O$(3)$ symmetry in such a way that it is not restored in the 
continuum limit. We will demonstrate this explicitly in section \ref{ch:catterall}. 
We must thus conclude that the continuum limits of these lattice constructions
cannot be identified with the two-dimensional non-linear O$(3)$ sigma model.\\
Are there other ways to find a partly supersymmetric but still O$(3)$ 
symmetric discretization? A symmetry of the model has to be a symmetry of the action \eqref{lagr}, 
but is also has to be compatible with the constraints $\bn^2= 1$  and $\bn\bpsi =0$.
Any supersymmetry has to be a combination of the transformations
given in (\ref{trafo}) and (\ref{trafo2}). If we consider the 
discretization of these, we notice that the first transformation (\ref{trafo}) 
breaks the constraint $\bn\bpsi = 0$ on the lattice, because
\begin{equation}
\delta_{\rm I}(\bn_x\bpsi_x^{\alpha}) = \ii\bar{\eps} \bpsi_x \bpsi_x^{\alpha} + \sum_{y\in\Lambda} \bn_x D_{xy}^{\alpha \beta} \bn_y \eps^{\beta} + \tfrac{\ii}{2}(\bar{\bpsi}_x \bpsi_x) \bn_x^2 \eps^{\alpha} \stackrel{(\ref{cf7})}{=}
\sum_{y\in\Lambda} \bn_x D_{xy}^{\alpha \beta} \bn_y \eps^{\beta}
\label{lconst1}
\end{equation}
does not vanish for arbitrary $\bn_x$, no matter which derivative 
$D_{xy}$ we use\footnote{Actually, it is also not a symmetry of the discretized action, 
but the breaking of the constraints is more severe.}. In contrast, 
the second transformation respects the constraints 
at each point:
\begin{eqnarray}
\label{constSecond}
\delta_{\rm II}(\bn_x \bpsi_x^{\alpha}) 
&=& \ii (\bn_x \times \bar{\eps}\bpsi_x)  \bpsi_x^{\alpha} - \sum_{y\in\Lambda} \bn_x  ( \bn_x \times D_{xy} \bn_y \eps^{\alpha})
- \ii\bn_x  (\bar{\eps}\bpsi_x \times \bpsi_x^{\alpha})=0 \nonumber\\
\delta_{\rm II}(\bn^2_x) &=& 2\ii \bn_x (\bn_x \times \bar{\eps}\bpsi_x) = 0\,.
\end{eqnarray}
We conclude that no nontrivial combination of the two transformations
$\delta_{\rm I}$ and $\delta_{\rm II}$ can be a symmetry of the lattice theory, since 
the second transformation cannot restore the violation of the constraints caused by 
the first one. The second transformation on its own, however, can also not be a 
symmetry of the action because of $\{\cQ^{\rm II}_{\alpha},\bar{\cQ}^{\rm II}_{\beta}\} = 
2i\gamma^{\mu}_{\alpha\beta} \partial_{\mu}$. The superalgebra furthermore tells us that an 
approach based on a nilpotent supercharge is not possible, either, because such a nilpotent
charge would have to be a combination of both supercharges and would hence violate 
the constraints.\\
Could we circumvent this restriction by improving the 
discretiziation in some way? Comparing our formulation with the one investigated in
\cite{Catterall:2003uf,Catterall:2006sj}, we see that the latter one contains 
an additional topological term. However, such a term does not affect
the supersymmetry transformations (\ref{trafo}) and (\ref{trafo2}) and
hence cannot solve the problem.
From a systematic point of view, there are only two modifications possible which are compatible
with an O$(3)$-invariant continuum limit. 
The first possibility is to modify the terms that are already present in the action. For example,
one could introduce non-local interaction 
terms like $\sum_{x,y,z,w} C_{xyzw}(\bar\bpsi_x\bpsi_y)(\bar\bpsi_z\bpsi_w)$ 
instead of $\sum_x(\bar\bpsi_x\bpsi_x)^2$ \cite{Bergner:2009vg}.
The second possibility could be an inclusion of additional terms
in the lattice action which vanish in the continuum limit.  Any change 
of the action, however, does not have an impact on the constraints
and hence cannot prevent their breaking. A modification of the 
constraints, by contrast, would directly alter the geometry of the 
target manifold and is thus no alternative. It follows that an 
improvement of the discretization could only maintain a part of 
supersymmetry by rendering the lattice action invariant under the 
second transformations. But this is not possible due to the structure 
of the superalgebra.\\
Even though these arguments were developed for a specific choice of 
coordinates, on finite lattices they also hold true for any reparametrization 
$(\bn,\bpsi) \rightarrow (\bn^\prime,\bpsi^\prime)$, because such a 
transformation is a bijective mapping between field values at a certain 
point $x$ in spacetime, which commutes with discretization. As a 
consequence, one would observe the same pattern of symmetry breaking as 
depicted in \eqref{lconst1} and \eqref{constSecond}. The single 
ambiguity that could arise here from the discretized derivative of the 
bosonic field is irrelevant since the presented arguments do 
not depend on the details of the lattice derivatives.\\
We conclude that it is just not possible to construct a discretization of 
the nonlinear O$(3)$ model which maintains O$(3)$ invariance as well as an exact supersymmetry.
From this point of view, the symmetry breaking in the ansatz \cite{Catterall:2003uf,Catterall:2006sj} was inevitable.
In this article we will work with the lattice formulation introduced in the 
previous section. It maintains the important O$(3)$ symmetry of the model 
while it breaks all supersymmetries. The latter should be restored in
the continuum limit.

%% file: catterall.tex
Before we present the results which are obtained by simulations for the 
O$(3)$-symmetric lattice models described above, we want to 
demonstrate how a violation of the O$(3)$ symmetry arises in
generic lattice formulations given in terms of unconstrained fields.
We shall see that the symmetry of the continuum model is 
not recovered in the continuum limit.
\subsection{Drawbacks of a $\cQ$-exact lattice formulation}
\label{sssec:sigma:qExact}
This problem can already be illustrated by investigating the bosonic 
O$(3)$-NLSM. If we use a lattice derivative that is based on 
nearest neighbour interactions, the discretization of the bosonic part 
of \eqref{lagrU} reads
\begin{equation}
S = \frac{2}{g^2} \sum_{\nN{xy}} \rho_{xy}^2 |\bu_x -\bu_y|^2\,.
\end{equation}
At this point, an ambiguity in defining the density $\rho^2_{xy}$ 
arises. It must interpolate between $\rho^2_x$ and $\rho^2_y$, 
which coincide in the continuum limit. Naively, 
many interpolations are feasible, e.g. one could
use the arithmetic mean, $\rho^2_{xy} = \half(\rho_x^2+\rho_y^2)$, or the
geometric mean $\rho^2_{xy} = \rho_x \rho_y$. In order to determine the appropriate 
interpolation one can apply the stereographic projection on the manifestly 
O$(3)$ invariant discretization in terms of the constrained $\bn$-field and finds that 
$\rho^2_{xy}$ has to be implemented as the geometric mean.\\
To analyze a possible symmetry breaking in the formulation based on
the arithmetic mean, simulations for both lattice prescriptions have been
performed\footnote{In the hybrid MC algorithm the factor $\rho^2$ 
of the path integral measure for the bosonic model, see (\ref{measure_bos}), 
is absorbed in the action.} and it was tested if $\vev{\bar\bn}=0$ 
for $\bar\bn = N^{-1} \sum_x \bn_{x}$ in accordance with the global O$(3)$ symmetry, 
which cannot be spontaneously broken in two 
spacetime dimensions \cite{Mermin:1966fe}.
\FIGURE{
\hfill
\input{sigmaMeanZquenchedAM}
\hfill
\input{sigmaMeanZquenchedGM}
\hfill
\phantom.
\caption{\label{fig:sigma:averageFieldDistribution} Scatter plot (projected to
the $\bar n_1$--$\bar n_3$ plane) of the averaged field $\bar n$ for a
lattice discretisation based on the arithmetic mean (left panel) and geometric
mean (right panel) at $g^{-2}=1$ and lattice size $N=10\times10$. } 
} 

\noindent Fig.~\ref{fig:sigma:averageFieldDistribution} shows that the
discretization based on the geometric mean indeed yields
O$(3)$symmetric results whereas for the arithmetic mean only a 
O$(2)$ symmetry around the $1$-axis remains. This
is a direct consequence of the global $U(1)$ symmetry 
$u\to\ee^{\ii\phi} u$, where $u=u_1+\ii u_2$.
In the naive continuum limit both prescriptions are expected to coincide. 
In order to investigate this issue, simulations of the model 
based on the arithmetic mean have been carried out with different 
lattice sizes $N$ and at different couplings $g^{-2}$. A restoration of the
O$(3)$ symmetry implies a vanishing $\vev{\bar n_3}$ in the continuum limit. 
For a wide range of couplings $\vev{\bar n_3}$ is independent of the 
lattice sizes such that the vacuum expectation value is assumed 
to be free of finite size effects  (see Fig.~\ref{fig:sigma:arithmeticMeanLimits}, 
left panel). By fitting  the correlator
\begin{equation}
C_\text{B}(t) = \Ns^{-2}\Re \sum_{x,y}\vev{ u_{(t,y)} \bar
u_{(0,x)}}
\end{equation}
to $\cosh (ma (t-\Nt/2))$ the mass $mL=ma\Ns$
measured in units of the physical box length can be extracted\footnote{Due to the 
U$(1)$ symmetry $\vev{u}=0$, and $C_\text{B}(t)$ is the connected
$2$-point function.}. The analysis of $\vev{\bar n_3}$ at fixed physical
box size $mL$ in the continuum limit (see Fig.~\ref{fig:sigma:arithmeticMeanLimits}, 
right panel) clearly shows that $\vev{\bar n_3}$ grows to a value 
close to $1$ for small lattice spacings, i.e.\ for 
large $N$ and large $g^{-2}$. Therefore it is
\emph{impossible} to reach the correct O$(3)$ symmetric continuum limit 
for a regularization based on the arithmetic mean.\\
\FIGURE{
\hfill
\input{sigmaArithmeticMeanLimitsBeta}
\hfill
\input{sigmaArithmeticMeanLimitsML}
\hfill
\phantom.
\caption{\label{fig:sigma:arithmeticMeanLimits} The value of $\vev{\bar n_3}$
as indicator for a broken global O$(3)$ symmetry for three lattice volumes
plotted over the coupling $g^{-2}$ (left panel) and physical box size (right
panel).}}

\noindent Having encountered these problems in the simple bosonic case, we ought to be very 
careful with a discretization of the superymmetric model and should always check the 
restoration of the O$(3)$ symmetry. Unfortunately this was not done in
\cite{Catterall:2003uf,Catterall:2006sj} and we shall see below that
this symmetry is actually broken in the lattice models constructed
by Catterall and Ghadab.\\
The formulation in \cite{Catterall:2003uf,Catterall:2006sj} is based on
a $\cQ$-exact deformation of the O$(3)$ sigma model in the 
unconstrained $\cp{1}$ formulation. The authors used Wilson fermions
which break supersymmetry softly and showed that the $\cQ$-based supersymmetry
is restored in the continuum limit by studying the associated Ward identities.
After performing a Hubbard-Stratonovich transformation to elimate four-fermi 
terms the model contains two complex scalars $u,\; \sigma$ and 
Dirac fermions $\Psi$. The path integral is
\begin{align}
\label{eq:sigma:qExactModel}
\ZZ &= \int \cD u\,\cD\sigma\, \cD
\Psi\,\rho^{-2}\,\ee^{-\SB[u,\sigma]-\SF[u,\sigma,\Psi]}\quad \text{with}\\
\SB &= \frac{2}{g^2}\sum_x\left(
\rho^2_x (\pasb u)_x (\pas \bar u)_x
+ \rho^2_x (\Delta u)_x(\Delta \bar u)_x + \half
\sigma_x\bar\sigma_x\right)\;,
\end{align}
where the symmetric derivative in direction $\mu$ is the 
arithmetic mean of the forward and backward derivates, 
$\partial_\mu^{\text{sym}} =\half (\partial_\mu^+ + \partial_\mu^-)$.
The derivative operators without lower index denote the complex
lattice derivatives $\partial=\partial_1-\ii\partial_2$ and
$\bar\partial=\partial_1+\ii\partial_2$. The lattice
Laplacian $\Delta=\sum_\mu\partial_\mu^+\partial_\mu^-$
originates from the Wilson mass term in the fermionic
action
\begin{equation}
 \SF=\frac{2}{g^2}\bar\Psi M[u,\sigma]\Psi
\end{equation}
with fermion matrix
\begin{equation}
M[u,\sigma] = \rho^2\begin{pmatrix}
\frac{1}{2}\Delta - \rho \bar u (\Delta u) + \heco &
\pasb - 2\rho \bar u (\pasb u)+\sigma\\ 
\pas + 2\rho  u (\pas \bar u)-\bar\sigma & 
\frac{1}{2}\Delta - \rho \bar u (\Delta u) + \heco 
\end{pmatrix}.
\end{equation}
Periodic boundary conditions are assumed for \emph{all fields}
such that supersymmetry is not broken by the boundary conditions.
The difference to a straightforward discretization 
is given by an \emph{improvement term}
\begin{equation}
\Delta S = \frac{4}{g^2} \sum_x \rho_x^2 \left(\pas_1 u_2\,
\pas_2 u_1 -\pas_1 u_1\,\pas_2
u_2\right)
\end{equation}
which corresponds to the topological winding number and becomes irrelevant in the naive continuum limit, similar to the improvement
term in the $\mathcal{N}=2$ Wess-Zumino model \cite{Kastner:2008zc}.
For the simulation of this model on smaller lattices the HMC algorithm 
with an explicit calculation of the fermionic determinant is used.
This has the advantage that no instabilities arise from 
introducing pseudo-fermions. Such instabilities may hide potential
shortcomings of the lattice formulation.
\FIGURE{
\hfill 
\input{catterallImprovementNZ}
\hfill
\input{catterallImprovementDS}
\hfill
\phantom.
\caption{\label{fig:sigma:catterallImprovement} MC history of 
$\bar n_3$ (left panel) and $\Delta S$ (right panel) for a simulation of the lattice
model \eqref{eq:sigma:qExactModel} at coupling $g^{-2}=1.5$ on an $8\times 8$
lattice. }
}

\noindent The improvement term is analyzed for a
lattice size of $N=8\times 8$ at coupling $g^{-2}=1.5$. In our
simulations a value of $\SB\approx 2 N$ is found
(with statistical fluctuations of about $10\%$) in
agreement with the simplest Ward identity \cite{Catterall:2006sj}, which is a
consequence of the (nearly, up to the Wilson mass) lattice supersymmetry.
In Fig.~\ref{fig:sigma:catterallImprovement} the MC histories of $\bar n_3$ 
and $\Delta S$ are plotted and  they shed light on the influence of the 
improvement term on the dynamics. At a certain point in the simulation
the value of $\bar n_3$ freezes out and the improvement term starts growing
largely negative. Just as for the Nicolai improved Wess-Zumino model 
\cite{Kastner:2008zc} the lattice system is driven away from the continuum physics, 
where $\Delta S$ must vanish, into an unphysical phase\footnote{The 
normalized improvement term $\Delta S/N$ fluctuates around zero 
with a width of about $0.002$ in the physical phase.}. 
We conclude that the known problems of improved actions 
must be taken into account also in simulations of supersymmetric
sigma models -- only configurations from the physical phase with
nearly vanishing improvement term should be taken into account in
measurements. Tunnel events to the unphysical phase are suppressed 
on large lattices and weak coupling $g$, i.e.\ in the continuum limit,
similarly as for Wess-Zumino models.
Nevertheless, these observations suggest that similar problems
may arise in simulations of all lattice models with exact supersymmetry 
that are constructed from a $\cQ$-exact action, 
e.g.\ two dimensional $\mathcal{N}=2$ Super Yang-Mills \cite{Sugino:2003yb}.\\
But why did this instability not show up in the results of
\cite{Catterall:2006sj}? The answer may be that in the rational 
HMC algorithm used in the simulations spectral
bounds must be chosen to cover the spectrum of $M^\dagger M$. 
Typically these are obtained by test runs with  rather pessimistic
bounds and \emph{small statistics}, such that only the physical phase
with $\Delta S \approx 0$ is present. But for the simulation that is shown
in Fig.~\ref{fig:sigma:catterallImprovement} the lowest eigenvalue of $M^\dagger
M$ decreases by a factor of $10^{-5}$ when entering the unphysical
phase.\footnote{The largest eigenvalue of $M^\dagger M$ is kept at the same
order of magnitude in the unphysical phase.} For that reason the rational hybrid
Monte-Carlo algorithm with spectral bounds that are not applicable to 
the whole simulation will be an \emph{inexact} algorithm and will give 
an arbitrarily small acceptance rate for the unphysical configurations 
that dominate the path integral.
Furthermore $\sign \det M$ is not positive definite and a \emph{deflated}
rational hybrid Monte-Carlo algorithm is necessary to get reliable expectation
values.\\ 
These results imply that for a reliable measurement of $\vev{\bar n_3}$
large $g^{-2}$ must be used in order to suppress the unphysical phase.
Since the continuum limit is reached for $g\to 0$, measurements will be affected by finite
size corrections. The corrections of the observable $\bar n_3$ are assumed to vanish
exponentially with growing volume,
\begin{equation}
\label{eq:sigma:nzExtrapolation}
\vev{\bar n_3}(\Ns) = \vev{\bar n_3}(\infty) + A\ee^{-B \Ns},
\end{equation}
such that a fit to this functional form for $\Ns \in \{10,11,12,13,14,16\}$
and corresponding lattice volumes $N=\Ns^2$ reveals the infinite volume
value for $g^{-2}\in\{3.5,4.0,4.5,5.0\}$, see Fig.~\ref{fig:sigma:cattermMeanZ}
(left panel). Although  $\vev{\bar n_3}$ decreases for fixed $\Ns$ and growing
$g^{-2}$, the infinite volume values tend to grow for larger
$g^{-2}$, see Fig.~\ref{fig:sigma:cattermMeanZ} (right panel). Therefore the
O$(3)$ symmetry will \emph{not} be restored in the infinite volume continuum
limit of the lattice model \eqref{eq:sigma:qExactModel}.
\FIGURE{
\hfill
\input{catterallMeanZ}
\hfill
\input{catterallMeanZinfiniteV}
\hfill
\phantom.
\caption{\label{fig:sigma:cattermMeanZ} Left panel: $\vev{\bar n_3}$ 
for different lattice volumes $N=\Ns^2$ at four couplings $g^{-2}$. The
infinite volume extrapolation according to Eq.~\eqref{eq:sigma:nzExtrapolation} is indicated
by the shaded areas. Right panel: Infinite volume values of 
$\vev{\bar n_3}$ for different $g^{-2}$. }
}
All these results have a crucial implication. Although the formulation
\eqref{eq:sigma:qExactModel} may restore the full $\mathcal{N} = (2,2)$
supersymmetry on the lattice, the resulting continuum model is \emph{not} the
supersymmetric O$(3)$ model, because the global
O$(3)$ symmetry, that cannot be broken spontaneously in the continuum
model \cite{Mermin:1966fe}, is \emph{not restored} in the continuum limit.

%% file: sigmaMeanZquenchedAM.tex
\begingroup
\footnotesize
  \makeatletter
  \providecommand\color[2][]{%
    \GenericError{(gnuplot) \space\space\space\@spaces}{%
      Package color not loaded in conjunction with
      terminal option `colourtext'%
    }{See the gnuplot documentation for explanation.%
    }{Either use 'blacktext' in gnuplot or load the package
      color.sty in LaTeX.}%
    \renewcommand\color[2][]{}%
  }%
  \providecommand\includegraphics[2][]{%
    \GenericError{(gnuplot) \space\space\space\@spaces}{%
      Package graphicx or graphics not loaded%
    }{See the gnuplot documentation for explanation.%
    }{The gnuplot epslatex terminal needs graphicx.sty or graphics.sty.}%
    \renewcommand\includegraphics[2][]{}%
  }%
  \providecommand\rotatebox[2]{#2}%
  \@ifundefined{ifGPcolor}{%
    \newif\ifGPcolor
    \GPcolortrue
  }{}%
  \@ifundefined{ifGPblacktext}{%
    \newif\ifGPblacktext
    \GPblacktexttrue
  }{}%
  \let\gplgaddtomacro\g@addto@macro
  \gdef\gplbacktext{}%
  \gdef\gplfronttext{}%
  \makeatother
  \ifGPblacktext
    \def\colorrgb#1{}%
    \def\colorgray#1{}%
  \else
    \ifGPcolor
      \def\colorrgb#1{\color[rgb]{#1}}%
      \def\colorgray#1{\color[gray]{#1}}%
      \expandafter\def\csname LTw\endcsname{\color{white}}%
      \expandafter\def\csname LTb\endcsname{\color{black}}%
      \expandafter\def\csname LTa\endcsname{\color{black}}%
      \expandafter\def\csname LT0\endcsname{\color[rgb]{1,0,0}}%
      \expandafter\def\csname LT1\endcsname{\color[rgb]{0,1,0}}%
      \expandafter\def\csname LT2\endcsname{\color[rgb]{0,0,1}}%
      \expandafter\def\csname LT3\endcsname{\color[rgb]{1,0,1}}%
      \expandafter\def\csname LT4\endcsname{\color[rgb]{0,1,1}}%
      \expandafter\def\csname LT5\endcsname{\color[rgb]{1,1,0}}%
      \expandafter\def\csname LT6\endcsname{\color[rgb]{0,0,0}}%
      \expandafter\def\csname LT7\endcsname{\color[rgb]{1,0.3,0}}%
      \expandafter\def\csname LT8\endcsname{\color[rgb]{0.5,0.5,0.5}}%
    \else
      \def\colorrgb#1{\color{black}}%
      \def\colorgray#1{\color[gray]{#1}}%
      \expandafter\def\csname LTw\endcsname{\color{white}}%
      \expandafter\def\csname LTb\endcsname{\color{black}}%
      \expandafter\def\csname LTa\endcsname{\color{black}}%
      \expandafter\def\csname LT0\endcsname{\color{black}}%
      \expandafter\def\csname LT1\endcsname{\color{black}}%
      \expandafter\def\csname LT2\endcsname{\color{black}}%
      \expandafter\def\csname LT3\endcsname{\color{black}}%
      \expandafter\def\csname LT4\endcsname{\color{black}}%
      \expandafter\def\csname LT5\endcsname{\color{black}}%
      \expandafter\def\csname LT6\endcsname{\color{black}}%
      \expandafter\def\csname LT7\endcsname{\color{black}}%
      \expandafter\def\csname LT8\endcsname{\color{black}}%
    \fi
  \fi
  \setlength{\unitlength}{0.0500bp}%
  \begin{picture}(3400.00,3400.00)%
    \gplgaddtomacro\gplbacktext{%
      \csname LTb\endcsname%
      \put(394,480){\makebox(0,0)[r]{\strut{}$-1.0$}}%
      \put(394,1198){\makebox(0,0)[r]{\strut{}$-0.5$}}%
      \put(394,1916){\makebox(0,0)[r]{\strut{}$0.0$}}%
      \put(394,2633){\makebox(0,0)[r]{\strut{}$0.5$}}%
      \put(394,3351){\makebox(0,0)[r]{\strut{}$1.0$}}%
      \put(490,320){\makebox(0,0){\strut{}$-1.0$}}%
      \put(1208,320){\makebox(0,0){\strut{}$-0.5$}}%
      \put(1926,320){\makebox(0,0){\strut{}$0.0$}}%
      \put(2643,320){\makebox(0,0){\strut{}$0.5$}}%
      \put(3361,320){\makebox(0,0){\strut{}$1.0$}}%
      \put(-118,1915){\makebox(0,0){\strut{}$\bar n_3$}}%
      \put(1925,80){\makebox(0,0){\strut{}$\bar n_1$}}%
    }%
    \gplgaddtomacro\gplfronttext{%
    }%
    \gplbacktext
    \put(0,0){\includegraphics{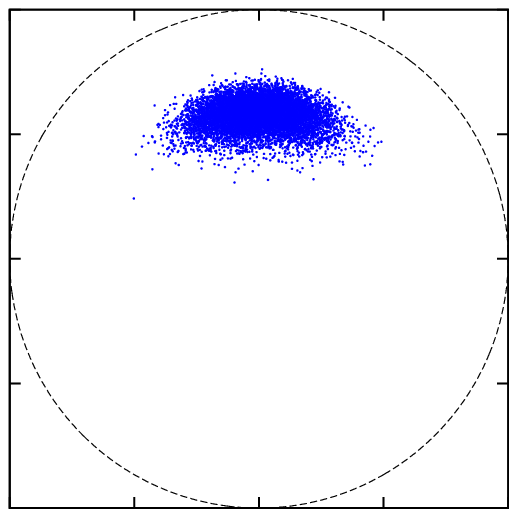}}%
    \gplfronttext
  \end{picture}%
\endgroup

%% file: sigmaMeanZquenchedGM.tex
\begingroup
\footnotesize
  \makeatletter
  \providecommand\color[2][]{%
    \GenericError{(gnuplot) \space\space\space\@spaces}{%
      Package color not loaded in conjunction with
      terminal option `colourtext'%
    }{See the gnuplot documentation for explanation.%
    }{Either use 'blacktext' in gnuplot or load the package
      color.sty in LaTeX.}%
    \renewcommand\color[2][]{}%
  }%
  \providecommand\includegraphics[2][]{%
    \GenericError{(gnuplot) \space\space\space\@spaces}{%
      Package graphicx or graphics not loaded%
    }{See the gnuplot documentation for explanation.%
    }{The gnuplot epslatex terminal needs graphicx.sty or graphics.sty.}%
    \renewcommand\includegraphics[2][]{}%
  }%
  \providecommand\rotatebox[2]{#2}%
  \@ifundefined{ifGPcolor}{%
    \newif\ifGPcolor
    \GPcolortrue
  }{}%
  \@ifundefined{ifGPblacktext}{%
    \newif\ifGPblacktext
    \GPblacktexttrue
  }{}%
  \let\gplgaddtomacro\g@addto@macro
  \gdef\gplbacktext{}%
  \gdef\gplfronttext{}%
  \makeatother
  \ifGPblacktext
    \def\colorrgb#1{}%
    \def\colorgray#1{}%
  \else
    \ifGPcolor
      \def\colorrgb#1{\color[rgb]{#1}}%
      \def\colorgray#1{\color[gray]{#1}}%
      \expandafter\def\csname LTw\endcsname{\color{white}}%
      \expandafter\def\csname LTb\endcsname{\color{black}}%
      \expandafter\def\csname LTa\endcsname{\color{black}}%
      \expandafter\def\csname LT0\endcsname{\color[rgb]{1,0,0}}%
      \expandafter\def\csname LT1\endcsname{\color[rgb]{0,1,0}}%
      \expandafter\def\csname LT2\endcsname{\color[rgb]{0,0,1}}%
      \expandafter\def\csname LT3\endcsname{\color[rgb]{1,0,1}}%
      \expandafter\def\csname LT4\endcsname{\color[rgb]{0,1,1}}%
      \expandafter\def\csname LT5\endcsname{\color[rgb]{1,1,0}}%
      \expandafter\def\csname LT6\endcsname{\color[rgb]{0,0,0}}%
      \expandafter\def\csname LT7\endcsname{\color[rgb]{1,0.3,0}}%
      \expandafter\def\csname LT8\endcsname{\color[rgb]{0.5,0.5,0.5}}%
    \else
      \def\colorrgb#1{\color{black}}%
      \def\colorgray#1{\color[gray]{#1}}%
      \expandafter\def\csname LTw\endcsname{\color{white}}%
      \expandafter\def\csname LTb\endcsname{\color{black}}%
      \expandafter\def\csname LTa\endcsname{\color{black}}%
      \expandafter\def\csname LT0\endcsname{\color{black}}%
      \expandafter\def\csname LT1\endcsname{\color{black}}%
      \expandafter\def\csname LT2\endcsname{\color{black}}%
      \expandafter\def\csname LT3\endcsname{\color{black}}%
      \expandafter\def\csname LT4\endcsname{\color{black}}%
      \expandafter\def\csname LT5\endcsname{\color{black}}%
      \expandafter\def\csname LT6\endcsname{\color{black}}%
      \expandafter\def\csname LT7\endcsname{\color{black}}%
      \expandafter\def\csname LT8\endcsname{\color{black}}%
    \fi
  \fi
  \setlength{\unitlength}{0.0500bp}%
  \begin{picture}(3400.00,3400.00)%
    \gplgaddtomacro\gplbacktext{%
      \csname LTb\endcsname%
      \put(394,480){\makebox(0,0)[r]{\strut{}$-1.0$}}%
      \put(394,1198){\makebox(0,0)[r]{\strut{}$-0.5$}}%
      \put(394,1916){\makebox(0,0)[r]{\strut{}$0.0$}}%
      \put(394,2633){\makebox(0,0)[r]{\strut{}$0.5$}}%
      \put(394,3351){\makebox(0,0)[r]{\strut{}$1.0$}}%
      \put(490,320){\makebox(0,0){\strut{}$-1.0$}}%
      \put(1208,320){\makebox(0,0){\strut{}$-0.5$}}%
      \put(1926,320){\makebox(0,0){\strut{}$0.0$}}%
      \put(2643,320){\makebox(0,0){\strut{}$0.5$}}%
      \put(3361,320){\makebox(0,0){\strut{}$1.0$}}%
      \put(-118,1915){\makebox(0,0){\strut{}$\bar n_3$}}%
      \put(1925,80){\makebox(0,0){\strut{}$\bar n_1$}}%
    }%
    \gplgaddtomacro\gplfronttext{%
    }%
    \gplbacktext
    \put(0,0){\includegraphics{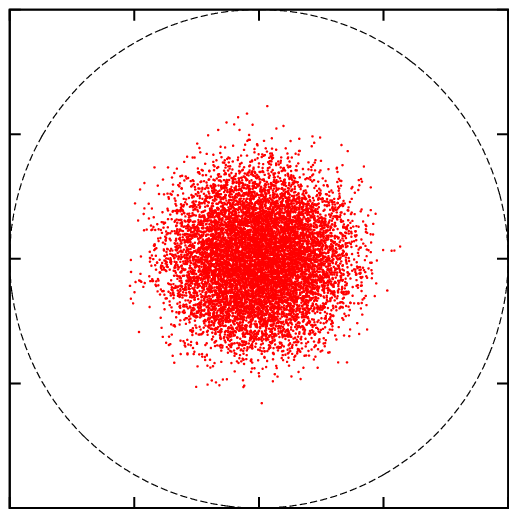}}%
    \gplfronttext
  \end{picture}%
\endgroup

%% file: sigmaArithmeticMeanLimitsBeta.tex
\begingroup
\footnotesize
  \makeatletter
  \providecommand\color[2][]{%
    \GenericError{(gnuplot) \space\space\space\@spaces}{%
      Package color not loaded in conjunction with
      terminal option `colourtext'%
    }{See the gnuplot documentation for explanation.%
    }{Either use 'blacktext' in gnuplot or load the package
      color.sty in LaTeX.}%
    \renewcommand\color[2][]{}%
  }%
  \providecommand\includegraphics[2][]{%
    \GenericError{(gnuplot) \space\space\space\@spaces}{%
      Package graphicx or graphics not loaded%
    }{See the gnuplot documentation for explanation.%
    }{The gnuplot epslatex terminal needs graphicx.sty or graphics.sty.}%
    \renewcommand\includegraphics[2][]{}%
  }%
  \providecommand\rotatebox[2]{#2}%
  \@ifundefined{ifGPcolor}{%
    \newif\ifGPcolor
    \GPcolortrue
  }{}%
  \@ifundefined{ifGPblacktext}{%
    \newif\ifGPblacktext
    \GPblacktexttrue
  }{}%
  \let\gplgaddtomacro\g@addto@macro
  \gdef\gplbacktext{}%
  \gdef\gplfronttext{}%
  \makeatother
  \ifGPblacktext
    \def\colorrgb#1{}%
    \def\colorgray#1{}%
  \else
    \ifGPcolor
      \def\colorrgb#1{\color[rgb]{#1}}%
      \def\colorgray#1{\color[gray]{#1}}%
      \expandafter\def\csname LTw\endcsname{\color{white}}%
      \expandafter\def\csname LTb\endcsname{\color{black}}%
      \expandafter\def\csname LTa\endcsname{\color{black}}%
      \expandafter\def\csname LT0\endcsname{\color[rgb]{1,0,0}}%
      \expandafter\def\csname LT1\endcsname{\color[rgb]{0,1,0}}%
      \expandafter\def\csname LT2\endcsname{\color[rgb]{0,0,1}}%
      \expandafter\def\csname LT3\endcsname{\color[rgb]{1,0,1}}%
      \expandafter\def\csname LT4\endcsname{\color[rgb]{0,1,1}}%
      \expandafter\def\csname LT5\endcsname{\color[rgb]{1,1,0}}%
      \expandafter\def\csname LT6\endcsname{\color[rgb]{0,0,0}}%
      \expandafter\def\csname LT7\endcsname{\color[rgb]{1,0.3,0}}%
      \expandafter\def\csname LT8\endcsname{\color[rgb]{0.5,0.5,0.5}}%
    \else
      \def\colorrgb#1{\color{black}}%
      \def\colorgray#1{\color[gray]{#1}}%
      \expandafter\def\csname LTw\endcsname{\color{white}}%
      \expandafter\def\csname LTb\endcsname{\color{black}}%
      \expandafter\def\csname LTa\endcsname{\color{black}}%
      \expandafter\def\csname LT0\endcsname{\color{black}}%
      \expandafter\def\csname LT1\endcsname{\color{black}}%
      \expandafter\def\csname LT2\endcsname{\color{black}}%
      \expandafter\def\csname LT3\endcsname{\color{black}}%
      \expandafter\def\csname LT4\endcsname{\color{black}}%
      \expandafter\def\csname LT5\endcsname{\color{black}}%
      \expandafter\def\csname LT6\endcsname{\color{black}}%
      \expandafter\def\csname LT7\endcsname{\color{black}}%
      \expandafter\def\csname LT8\endcsname{\color{black}}%
    \fi
  \fi
  \setlength{\unitlength}{0.0500bp}%
  \begin{picture}(3968.00,2976.00)%
    \gplgaddtomacro\gplbacktext{%
      \csname LTb\endcsname%
      \put(768,480){\makebox(0,0)[r]{\strut{}$0.65$}}%
      \put(768,969){\makebox(0,0)[r]{\strut{}$0.70$}}%
      \put(768,1459){\makebox(0,0)[r]{\strut{}$0.75$}}%
      \put(768,1948){\makebox(0,0)[r]{\strut{}$0.80$}}%
      \put(768,2438){\makebox(0,0)[r]{\strut{}$0.85$}}%
      \put(768,2927){\makebox(0,0)[r]{\strut{}$0.90$}}%
      \put(864,320){\makebox(0,0){\strut{}$1.5$}}%
      \put(1206,320){\makebox(0,0){\strut{}$2.0$}}%
      \put(1547,320){\makebox(0,0){\strut{}$2.5$}}%
      \put(1889,320){\makebox(0,0){\strut{}$3.0$}}%
      \put(2231,320){\makebox(0,0){\strut{}$3.5$}}%
      \put(2572,320){\makebox(0,0){\strut{}$4.0$}}%
      \put(2914,320){\makebox(0,0){\strut{}$4.5$}}%
      \put(3256,320){\makebox(0,0){\strut{}$5.0$}}%
      \put(3597,320){\makebox(0,0){\strut{}$5.5$}}%
      \put(3939,320){\makebox(0,0){\strut{}$6.0$}}%
      \put(208,1703){\makebox(0,0){\strut{}$\vev{\bar n_3}$}}%
      \put(2401,80){\makebox(0,0){\strut{}$g^{-2}$}}%
    }%
    \gplgaddtomacro\gplfronttext{%
      \csname LTb\endcsname%
      \put(3204,943){\makebox(0,0)[r]{\strut{}$N=32^2$}}%
      \csname LTb\endcsname%
      \put(3204,783){\makebox(0,0)[r]{\strut{}$N=48^2$}}%
      \csname LTb\endcsname%
      \put(3204,623){\makebox(0,0)[r]{\strut{}$N=64^2$}}%
    }%
    \gplbacktext
    \put(0,0){\includegraphics{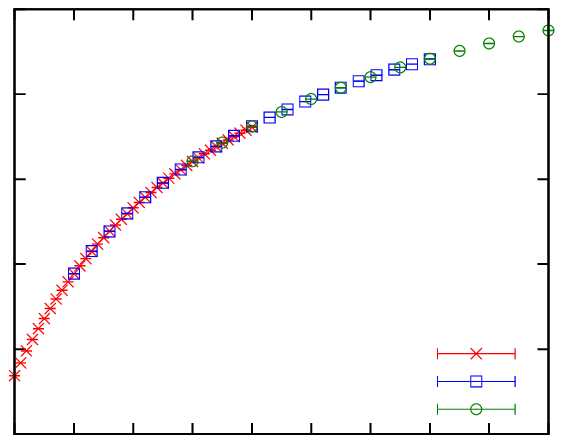}}%
    \gplfronttext
  \end{picture}%
\endgroup

%% file: sigmaArithmeticMeanLimitsML.tex
\begingroup
\footnotesize
  \makeatletter
  \providecommand\color[2][]{%
    \GenericError{(gnuplot) \space\space\space\@spaces}{%
      Package color not loaded in conjunction with
      terminal option `colourtext'%
    }{See the gnuplot documentation for explanation.%
    }{Either use 'blacktext' in gnuplot or load the package
      color.sty in LaTeX.}%
    \renewcommand\color[2][]{}%
  }%
  \providecommand\includegraphics[2][]{%
    \GenericError{(gnuplot) \space\space\space\@spaces}{%
      Package graphicx or graphics not loaded%
    }{See the gnuplot documentation for explanation.%
    }{The gnuplot epslatex terminal needs graphicx.sty or graphics.sty.}%
    \renewcommand\includegraphics[2][]{}%
  }%
  \providecommand\rotatebox[2]{#2}%
  \@ifundefined{ifGPcolor}{%
    \newif\ifGPcolor
    \GPcolortrue
  }{}%
  \@ifundefined{ifGPblacktext}{%
    \newif\ifGPblacktext
    \GPblacktexttrue
  }{}%
  \let\gplgaddtomacro\g@addto@macro
  \gdef\gplbacktext{}%
  \gdef\gplfronttext{}%
  \makeatother
  \ifGPblacktext
    \def\colorrgb#1{}%
    \def\colorgray#1{}%
  \else
    \ifGPcolor
      \def\colorrgb#1{\color[rgb]{#1}}%
      \def\colorgray#1{\color[gray]{#1}}%
      \expandafter\def\csname LTw\endcsname{\color{white}}%
      \expandafter\def\csname LTb\endcsname{\color{black}}%
      \expandafter\def\csname LTa\endcsname{\color{black}}%
      \expandafter\def\csname LT0\endcsname{\color[rgb]{1,0,0}}%
      \expandafter\def\csname LT1\endcsname{\color[rgb]{0,1,0}}%
      \expandafter\def\csname LT2\endcsname{\color[rgb]{0,0,1}}%
      \expandafter\def\csname LT3\endcsname{\color[rgb]{1,0,1}}%
      \expandafter\def\csname LT4\endcsname{\color[rgb]{0,1,1}}%
      \expandafter\def\csname LT5\endcsname{\color[rgb]{1,1,0}}%
      \expandafter\def\csname LT6\endcsname{\color[rgb]{0,0,0}}%
      \expandafter\def\csname LT7\endcsname{\color[rgb]{1,0.3,0}}%
      \expandafter\def\csname LT8\endcsname{\color[rgb]{0.5,0.5,0.5}}%
    \else
      \def\colorrgb#1{\color{black}}%
      \def\colorgray#1{\color[gray]{#1}}%
      \expandafter\def\csname LTw\endcsname{\color{white}}%
      \expandafter\def\csname LTb\endcsname{\color{black}}%
      \expandafter\def\csname LTa\endcsname{\color{black}}%
      \expandafter\def\csname LT0\endcsname{\color{black}}%
      \expandafter\def\csname LT1\endcsname{\color{black}}%
      \expandafter\def\csname LT2\endcsname{\color{black}}%
      \expandafter\def\csname LT3\endcsname{\color{black}}%
      \expandafter\def\csname LT4\endcsname{\color{black}}%
      \expandafter\def\csname LT5\endcsname{\color{black}}%
      \expandafter\def\csname LT6\endcsname{\color{black}}%
      \expandafter\def\csname LT7\endcsname{\color{black}}%
      \expandafter\def\csname LT8\endcsname{\color{black}}%
    \fi
  \fi
  \setlength{\unitlength}{0.0500bp}%
  \begin{picture}(3968.00,2976.00)%
    \gplgaddtomacro\gplbacktext{%
      \csname LTb\endcsname%
      \put(768,480){\makebox(0,0)[r]{\strut{}$0.65$}}%
      \put(768,969){\makebox(0,0)[r]{\strut{}$0.70$}}%
      \put(768,1459){\makebox(0,0)[r]{\strut{}$0.75$}}%
      \put(768,1948){\makebox(0,0)[r]{\strut{}$0.80$}}%
      \put(768,2438){\makebox(0,0)[r]{\strut{}$0.85$}}%
      \put(768,2927){\makebox(0,0)[r]{\strut{}$0.90$}}%
      \put(864,320){\makebox(0,0){\strut{}$4$}}%
      \put(1377,320){\makebox(0,0){\strut{}$6$}}%
      \put(1889,320){\makebox(0,0){\strut{}$8$}}%
      \put(2402,320){\makebox(0,0){\strut{}$10$}}%
      \put(2914,320){\makebox(0,0){\strut{}$12$}}%
      \put(3427,320){\makebox(0,0){\strut{}$14$}}%
      \put(3939,320){\makebox(0,0){\strut{}$16$}}%
      \put(208,1703){\makebox(0,0){\strut{}$\vev{\bar n_3}$}}%
      \put(2401,80){\makebox(0,0){\strut{}$mL$}}%
    }%
    \gplgaddtomacro\gplfronttext{%
      \csname LTb\endcsname%
      \put(1728,943){\makebox(0,0)[r]{\strut{}$N=32^2$}}%
      \csname LTb\endcsname%
      \put(1728,783){\makebox(0,0)[r]{\strut{}$N=48^2$}}%
      \csname LTb\endcsname%
      \put(1728,623){\makebox(0,0)[r]{\strut{}$N=64^2$}}%
    }%
    \gplbacktext
    \put(0,0){\includegraphics{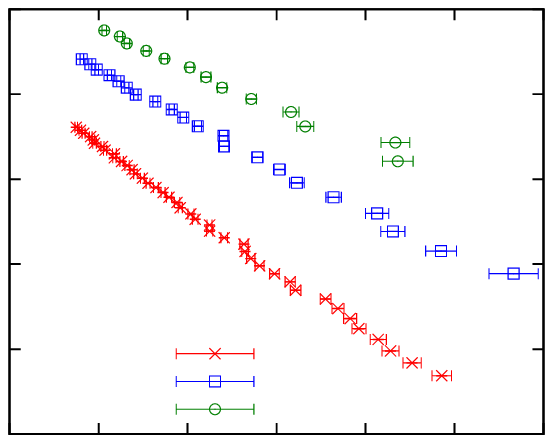}}%
    \gplfronttext
  \end{picture}%
\endgroup

%% file: catterallImprovementNZ.tex
\begingroup
\footnotesize
  \makeatletter
  \providecommand\color[2][]{%
    \GenericError{(gnuplot) \space\space\space\@spaces}{%
      Package color not loaded in conjunction with
      terminal option `colourtext'%
    }{See the gnuplot documentation for explanation.%
    }{Either use 'blacktext' in gnuplot or load the package
      color.sty in LaTeX.}%
    \renewcommand\color[2][]{}%
  }%
  \providecommand\includegraphics[2][]{%
    \GenericError{(gnuplot) \space\space\space\@spaces}{%
      Package graphicx or graphics not loaded%
    }{See the gnuplot documentation for explanation.%
    }{The gnuplot epslatex terminal needs graphicx.sty or graphics.sty.}%
    \renewcommand\includegraphics[2][]{}%
  }%
  \providecommand\rotatebox[2]{#2}%
  \@ifundefined{ifGPcolor}{%
    \newif\ifGPcolor
    \GPcolortrue
  }{}%
  \@ifundefined{ifGPblacktext}{%
    \newif\ifGPblacktext
    \GPblacktexttrue
  }{}%
  \let\gplgaddtomacro\g@addto@macro
  \gdef\gplbacktext{}%
  \gdef\gplfronttext{}%
  \makeatother
  \ifGPblacktext
    \def\colorrgb#1{}%
    \def\colorgray#1{}%
  \else
    \ifGPcolor
      \def\colorrgb#1{\color[rgb]{#1}}%
      \def\colorgray#1{\color[gray]{#1}}%
      \expandafter\def\csname LTw\endcsname{\color{white}}%
      \expandafter\def\csname LTb\endcsname{\color{black}}%
      \expandafter\def\csname LTa\endcsname{\color{black}}%
      \expandafter\def\csname LT0\endcsname{\color[rgb]{1,0,0}}%
      \expandafter\def\csname LT1\endcsname{\color[rgb]{0,1,0}}%
      \expandafter\def\csname LT2\endcsname{\color[rgb]{0,0,1}}%
      \expandafter\def\csname LT3\endcsname{\color[rgb]{1,0,1}}%
      \expandafter\def\csname LT4\endcsname{\color[rgb]{0,1,1}}%
      \expandafter\def\csname LT5\endcsname{\color[rgb]{1,1,0}}%
      \expandafter\def\csname LT6\endcsname{\color[rgb]{0,0,0}}%
      \expandafter\def\csname LT7\endcsname{\color[rgb]{1,0.3,0}}%
      \expandafter\def\csname LT8\endcsname{\color[rgb]{0.5,0.5,0.5}}%
    \else
      \def\colorrgb#1{\color{black}}%
      \def\colorgray#1{\color[gray]{#1}}%
      \expandafter\def\csname LTw\endcsname{\color{white}}%
      \expandafter\def\csname LTb\endcsname{\color{black}}%
      \expandafter\def\csname LTa\endcsname{\color{black}}%
      \expandafter\def\csname LT0\endcsname{\color{black}}%
      \expandafter\def\csname LT1\endcsname{\color{black}}%
      \expandafter\def\csname LT2\endcsname{\color{black}}%
      \expandafter\def\csname LT3\endcsname{\color{black}}%
      \expandafter\def\csname LT4\endcsname{\color{black}}%
      \expandafter\def\csname LT5\endcsname{\color{black}}%
      \expandafter\def\csname LT6\endcsname{\color{black}}%
      \expandafter\def\csname LT7\endcsname{\color{black}}%
      \expandafter\def\csname LT8\endcsname{\color{black}}%
    \fi
  \fi
  \setlength{\unitlength}{0.0500bp}%
  \begin{picture}(3968.00,2976.00)%
    \gplgaddtomacro\gplbacktext{%
      \csname LTb\endcsname%
      \put(768,480){\makebox(0,0)[r]{\strut{}$-1.0$}}%
      \put(768,1092){\makebox(0,0)[r]{\strut{}$-0.5$}}%
      \put(768,1704){\makebox(0,0)[r]{\strut{}$0.0$}}%
      \put(768,2315){\makebox(0,0)[r]{\strut{}$0.5$}}%
      \put(768,2927){\makebox(0,0)[r]{\strut{}$1.0$}}%
      \put(864,320){\makebox(0,0){\strut{}$0$}}%
      \put(1377,320){\makebox(0,0){\strut{}$1000$}}%
      \put(1889,320){\makebox(0,0){\strut{}$2000$}}%
      \put(2402,320){\makebox(0,0){\strut{}$3000$}}%
      \put(2914,320){\makebox(0,0){\strut{}$4000$}}%
      \put(3427,320){\makebox(0,0){\strut{}$5000$}}%
      \put(3939,320){\makebox(0,0){\strut{}$6000$}}%
      \put(400,2023){\makebox(0,0){\strut{}$\bar n_3$}}%
      \put(2401,80){\makebox(0,0){\strut{}config \#}}%
    }%
    \gplgaddtomacro\gplfronttext{%
    }%
    \gplbacktext
    \put(0,0){\includegraphics{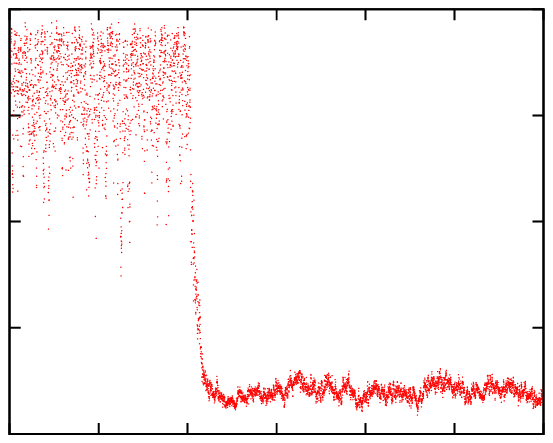}}%
    \gplfronttext
  \end{picture}%
\endgroup

%% file: catterallImprovementDS.tex
\begingroup
\footnotesize
  \makeatletter
  \providecommand\color[2][]{%
    \GenericError{(gnuplot) \space\space\space\@spaces}{%
      Package color not loaded in conjunction with
      terminal option `colourtext'%
    }{See the gnuplot documentation for explanation.%
    }{Either use 'blacktext' in gnuplot or load the package
      color.sty in LaTeX.}%
    \renewcommand\color[2][]{}%
  }%
  \providecommand\includegraphics[2][]{%
    \GenericError{(gnuplot) \space\space\space\@spaces}{%
      Package graphicx or graphics not loaded%
    }{See the gnuplot documentation for explanation.%
    }{The gnuplot epslatex terminal needs graphicx.sty or graphics.sty.}%
    \renewcommand\includegraphics[2][]{}%
  }%
  \providecommand\rotatebox[2]{#2}%
  \@ifundefined{ifGPcolor}{%
    \newif\ifGPcolor
    \GPcolortrue
  }{}%
  \@ifundefined{ifGPblacktext}{%
    \newif\ifGPblacktext
    \GPblacktexttrue
  }{}%
  \let\gplgaddtomacro\g@addto@macro
  \gdef\gplbacktext{}%
  \gdef\gplfronttext{}%
  \makeatother
  \ifGPblacktext
    \def\colorrgb#1{}%
    \def\colorgray#1{}%
  \else
    \ifGPcolor
      \def\colorrgb#1{\color[rgb]{#1}}%
      \def\colorgray#1{\color[gray]{#1}}%
      \expandafter\def\csname LTw\endcsname{\color{white}}%
      \expandafter\def\csname LTb\endcsname{\color{black}}%
      \expandafter\def\csname LTa\endcsname{\color{black}}%
      \expandafter\def\csname LT0\endcsname{\color[rgb]{1,0,0}}%
      \expandafter\def\csname LT1\endcsname{\color[rgb]{0,1,0}}%
      \expandafter\def\csname LT2\endcsname{\color[rgb]{0,0,1}}%
      \expandafter\def\csname LT3\endcsname{\color[rgb]{1,0,1}}%
      \expandafter\def\csname LT4\endcsname{\color[rgb]{0,1,1}}%
      \expandafter\def\csname LT5\endcsname{\color[rgb]{1,1,0}}%
      \expandafter\def\csname LT6\endcsname{\color[rgb]{0,0,0}}%
      \expandafter\def\csname LT7\endcsname{\color[rgb]{1,0.3,0}}%
      \expandafter\def\csname LT8\endcsname{\color[rgb]{0.5,0.5,0.5}}%
    \else
      \def\colorrgb#1{\color{black}}%
      \def\colorgray#1{\color[gray]{#1}}%
      \expandafter\def\csname LTw\endcsname{\color{white}}%
      \expandafter\def\csname LTb\endcsname{\color{black}}%
      \expandafter\def\csname LTa\endcsname{\color{black}}%
      \expandafter\def\csname LT0\endcsname{\color{black}}%
      \expandafter\def\csname LT1\endcsname{\color{black}}%
      \expandafter\def\csname LT2\endcsname{\color{black}}%
      \expandafter\def\csname LT3\endcsname{\color{black}}%
      \expandafter\def\csname LT4\endcsname{\color{black}}%
      \expandafter\def\csname LT5\endcsname{\color{black}}%
      \expandafter\def\csname LT6\endcsname{\color{black}}%
      \expandafter\def\csname LT7\endcsname{\color{black}}%
      \expandafter\def\csname LT8\endcsname{\color{black}}%
    \fi
  \fi
  \setlength{\unitlength}{0.0500bp}%
  \begin{picture}(3968.00,2976.00)%
    \gplgaddtomacro\gplbacktext{%
      \csname LTb\endcsname%
      \put(768,480){\makebox(0,0)[r]{\strut{}$-400$}}%
      \put(768,969){\makebox(0,0)[r]{\strut{}$-300$}}%
      \put(768,1459){\makebox(0,0)[r]{\strut{}$-200$}}%
      \put(768,1948){\makebox(0,0)[r]{\strut{}$-100$}}%
      \put(768,2438){\makebox(0,0)[r]{\strut{}$0$}}%
      \put(768,2927){\makebox(0,0)[r]{\strut{}$100$}}%
      \put(864,320){\makebox(0,0){\strut{}$0$}}%
      \put(1377,320){\makebox(0,0){\strut{}$1000$}}%
      \put(1889,320){\makebox(0,0){\strut{}$2000$}}%
      \put(2402,320){\makebox(0,0){\strut{}$3000$}}%
      \put(2914,320){\makebox(0,0){\strut{}$4000$}}%
      \put(3427,320){\makebox(0,0){\strut{}$5000$}}%
      \put(3939,320){\makebox(0,0){\strut{}$6000$}}%
      \put(304,1703){\makebox(0,0){\strut{}$\Delta S/N$}}%
      \put(2401,80){\makebox(0,0){\strut{}config \#}}%
    }%
    \gplgaddtomacro\gplfronttext{%
    }%
    \gplbacktext
    \put(0,0){\includegraphics{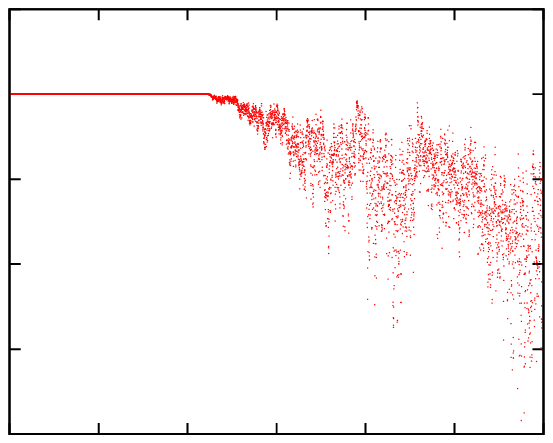}}%
    \gplfronttext
  \end{picture}%
\endgroup

%% file: catterallMeanZ.tex
\begingroup
\footnotesize
  \makeatletter
  \providecommand\color[2][]{%
    \GenericError{(gnuplot) \space\space\space\@spaces}{%
      Package color not loaded in conjunction with
      terminal option `colourtext'%
    }{See the gnuplot documentation for explanation.%
    }{Either use 'blacktext' in gnuplot or load the package
      color.sty in LaTeX.}%
    \renewcommand\color[2][]{}%
  }%
  \providecommand\includegraphics[2][]{%
    \GenericError{(gnuplot) \space\space\space\@spaces}{%
      Package graphicx or graphics not loaded%
    }{See the gnuplot documentation for explanation.%
    }{The gnuplot epslatex terminal needs graphicx.sty or graphics.sty.}%
    \renewcommand\includegraphics[2][]{}%
  }%
  \providecommand\rotatebox[2]{#2}%
  \@ifundefined{ifGPcolor}{%
    \newif\ifGPcolor
    \GPcolortrue
  }{}%
  \@ifundefined{ifGPblacktext}{%
    \newif\ifGPblacktext
    \GPblacktexttrue
  }{}%
  \let\gplgaddtomacro\g@addto@macro
  \gdef\gplbacktext{}%
  \gdef\gplfronttext{}%
  \makeatother
  \ifGPblacktext
    \def\colorrgb#1{}%
    \def\colorgray#1{}%
  \else
    \ifGPcolor
      \def\colorrgb#1{\color[rgb]{#1}}%
      \def\colorgray#1{\color[gray]{#1}}%
      \expandafter\def\csname LTw\endcsname{\color{white}}%
      \expandafter\def\csname LTb\endcsname{\color{black}}%
      \expandafter\def\csname LTa\endcsname{\color{black}}%
      \expandafter\def\csname LT0\endcsname{\color[rgb]{1,0,0}}%
      \expandafter\def\csname LT1\endcsname{\color[rgb]{0,1,0}}%
      \expandafter\def\csname LT2\endcsname{\color[rgb]{0,0,1}}%
      \expandafter\def\csname LT3\endcsname{\color[rgb]{1,0,1}}%
      \expandafter\def\csname LT4\endcsname{\color[rgb]{0,1,1}}%
      \expandafter\def\csname LT5\endcsname{\color[rgb]{1,1,0}}%
      \expandafter\def\csname LT6\endcsname{\color[rgb]{0,0,0}}%
      \expandafter\def\csname LT7\endcsname{\color[rgb]{1,0.3,0}}%
      \expandafter\def\csname LT8\endcsname{\color[rgb]{0.5,0.5,0.5}}%
    \else
      \def\colorrgb#1{\color{black}}%
      \def\colorgray#1{\color[gray]{#1}}%
      \expandafter\def\csname LTw\endcsname{\color{white}}%
      \expandafter\def\csname LTb\endcsname{\color{black}}%
      \expandafter\def\csname LTa\endcsname{\color{black}}%
      \expandafter\def\csname LT0\endcsname{\color{black}}%
      \expandafter\def\csname LT1\endcsname{\color{black}}%
      \expandafter\def\csname LT2\endcsname{\color{black}}%
      \expandafter\def\csname LT3\endcsname{\color{black}}%
      \expandafter\def\csname LT4\endcsname{\color{black}}%
      \expandafter\def\csname LT5\endcsname{\color{black}}%
      \expandafter\def\csname LT6\endcsname{\color{black}}%
      \expandafter\def\csname LT7\endcsname{\color{black}}%
      \expandafter\def\csname LT8\endcsname{\color{black}}%
    \fi
  \fi
  \setlength{\unitlength}{0.0500bp}%
  \begin{picture}(3968.00,2976.00)%
    \gplgaddtomacro\gplbacktext{%
      \csname LTb\endcsname%
      \put(768,480){\makebox(0,0)[r]{\strut{}$0.2$}}%
      \put(768,888){\makebox(0,0)[r]{\strut{}$0.3$}}%
      \put(768,1296){\makebox(0,0)[r]{\strut{}$0.4$}}%
      \put(768,1703){\makebox(0,0)[r]{\strut{}$0.5$}}%
      \put(768,2111){\makebox(0,0)[r]{\strut{}$0.6$}}%
      \put(768,2519){\makebox(0,0)[r]{\strut{}$0.7$}}%
      \put(768,2927){\makebox(0,0)[r]{\strut{}$0.8$}}%
      \put(864,320){\makebox(0,0){\strut{}$8$}}%
      \put(1479,320){\makebox(0,0){\strut{}$10$}}%
      \put(2094,320){\makebox(0,0){\strut{}$12$}}%
      \put(2709,320){\makebox(0,0){\strut{}$14$}}%
      \put(3324,320){\makebox(0,0){\strut{}$16$}}%
      \put(3939,320){\makebox(0,0){\strut{}$18$}}%
      \put(400,1895){\makebox(0,0){\strut{}$\vev{\tilde n_3}$}}%
      \put(2401,80){\makebox(0,0){\strut{}$\Ns$}}%
    }%
    \gplgaddtomacro\gplfronttext{%
      \csname LTb\endcsname%
      \put(3204,1215){\makebox(0,0)[r]{\strut{}$g^{-2}=3.5$}}%
      \csname LTb\endcsname%
      \put(3204,1023){\makebox(0,0)[r]{\strut{}$g^{-2}=4.0$}}%
      \csname LTb\endcsname%
      \put(3204,831){\makebox(0,0)[r]{\strut{}$g^{-2}=4.5$}}%
      \csname LTb\endcsname%
      \put(3204,639){\makebox(0,0)[r]{\strut{}$g^{-2}=5.0$}}%
    }%
    \gplbacktext
    \put(0,0){\includegraphics{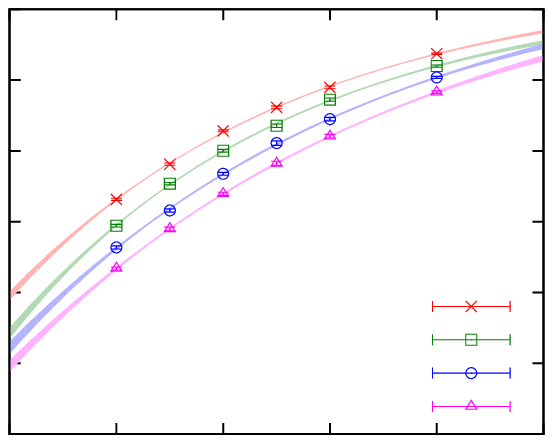}}%
    \gplfronttext
  \end{picture}%
\endgroup

%% file: catterallMeanZinfiniteV.tex
\begingroup
\footnotesize
  \makeatletter
  \providecommand\color[2][]{%
    \GenericError{(gnuplot) \space\space\space\@spaces}{%
      Package color not loaded in conjunction with
      terminal option `colourtext'%
    }{See the gnuplot documentation for explanation.%
    }{Either use 'blacktext' in gnuplot or load the package
      color.sty in LaTeX.}%
    \renewcommand\color[2][]{}%
  }%
  \providecommand\includegraphics[2][]{%
    \GenericError{(gnuplot) \space\space\space\@spaces}{%
      Package graphicx or graphics not loaded%
    }{See the gnuplot documentation for explanation.%
    }{The gnuplot epslatex terminal needs graphicx.sty or graphics.sty.}%
    \renewcommand\includegraphics[2][]{}%
  }%
  \providecommand\rotatebox[2]{#2}%
  \@ifundefined{ifGPcolor}{%
    \newif\ifGPcolor
    \GPcolortrue
  }{}%
  \@ifundefined{ifGPblacktext}{%
    \newif\ifGPblacktext
    \GPblacktexttrue
  }{}%
  \let\gplgaddtomacro\g@addto@macro
  \gdef\gplbacktext{}%
  \gdef\gplfronttext{}%
  \makeatother
  \ifGPblacktext
    \def\colorrgb#1{}%
    \def\colorgray#1{}%
  \else
    \ifGPcolor
      \def\colorrgb#1{\color[rgb]{#1}}%
      \def\colorgray#1{\color[gray]{#1}}%
      \expandafter\def\csname LTw\endcsname{\color{white}}%
      \expandafter\def\csname LTb\endcsname{\color{black}}%
      \expandafter\def\csname LTa\endcsname{\color{black}}%
      \expandafter\def\csname LT0\endcsname{\color[rgb]{1,0,0}}%
      \expandafter\def\csname LT1\endcsname{\color[rgb]{0,1,0}}%
      \expandafter\def\csname LT2\endcsname{\color[rgb]{0,0,1}}%
      \expandafter\def\csname LT3\endcsname{\color[rgb]{1,0,1}}%
      \expandafter\def\csname LT4\endcsname{\color[rgb]{0,1,1}}%
      \expandafter\def\csname LT5\endcsname{\color[rgb]{1,1,0}}%
      \expandafter\def\csname LT6\endcsname{\color[rgb]{0,0,0}}%
      \expandafter\def\csname LT7\endcsname{\color[rgb]{1,0.3,0}}%
      \expandafter\def\csname LT8\endcsname{\color[rgb]{0.5,0.5,0.5}}%
    \else
      \def\colorrgb#1{\color{black}}%
      \def\colorgray#1{\color[gray]{#1}}%
      \expandafter\def\csname LTw\endcsname{\color{white}}%
      \expandafter\def\csname LTb\endcsname{\color{black}}%
      \expandafter\def\csname LTa\endcsname{\color{black}}%
      \expandafter\def\csname LT0\endcsname{\color{black}}%
      \expandafter\def\csname LT1\endcsname{\color{black}}%
      \expandafter\def\csname LT2\endcsname{\color{black}}%
      \expandafter\def\csname LT3\endcsname{\color{black}}%
      \expandafter\def\csname LT4\endcsname{\color{black}}%
      \expandafter\def\csname LT5\endcsname{\color{black}}%
      \expandafter\def\csname LT6\endcsname{\color{black}}%
      \expandafter\def\csname LT7\endcsname{\color{black}}%
      \expandafter\def\csname LT8\endcsname{\color{black}}%
    \fi
  \fi
  \setlength{\unitlength}{0.0500bp}%
  \begin{picture}(3968.00,2976.00)%
    \gplgaddtomacro\gplbacktext{%
      \csname LTb\endcsname%
      \put(768,480){\makebox(0,0)[r]{\strut{}$0.75$}}%
      \put(768,969){\makebox(0,0)[r]{\strut{}$0.80$}}%
      \put(768,1459){\makebox(0,0)[r]{\strut{}$0.85$}}%
      \put(768,1948){\makebox(0,0)[r]{\strut{}$0.90$}}%
      \put(768,2438){\makebox(0,0)[r]{\strut{}$0.95$}}%
      \put(768,2927){\makebox(0,0)[r]{\strut{}$1.00$}}%
      \put(864,320){\makebox(0,0){\strut{}$3.0$}}%
      \put(1479,320){\makebox(0,0){\strut{}$3.5$}}%
      \put(2094,320){\makebox(0,0){\strut{}$4.0$}}%
      \put(2709,320){\makebox(0,0){\strut{}$4.5$}}%
      \put(3324,320){\makebox(0,0){\strut{}$5.0$}}%
      \put(3939,320){\makebox(0,0){\strut{}$5.5$}}%
      \put(304,1703){\makebox(0,0){\strut{}$\vev{\tilde n_3}$}}%
      \put(2401,80){\makebox(0,0){\strut{}$g^{-2}$}}%
    }%
    \gplgaddtomacro\gplfronttext{%
    }%
    \gplbacktext
    \put(0,0){\includegraphics{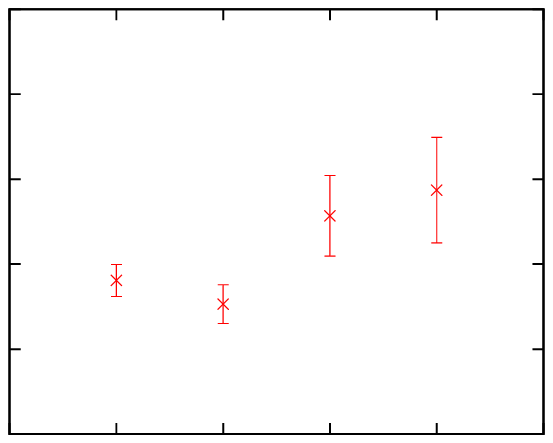}}%
    \gplfronttext
  \end{picture}%
\endgroup

%% file: o3symmetry.tex
\subsection{O$(3)$ symmetric formulations}
The previous studies have stressed the importance of an O$(3)$ invariant 
formulation of the theory. The discretizations given in section \ref{ch:discretization} 
respect the global flavor symmetry and in order to check whether the 
corresponding algorithms respect it as well we record the expectation 
value of $\bn$. For the coset construction with group valued field $R$ the
expectation value vanishes and the autocorrelation times
are small. For the stereographically projected fields $\bu$
we generate O$(3)$ symmetric configurations but observe 
large autocorrelation times for observables which depend 
on the field component corresponding to the projection axis. 
This stems from the interplay between stereographic projection and 
molecular dynamics steps which introduce two pseudo-momenta and a
pseudo-Hamiltonian for $\bu$ to generate test configurations. 
Computing the pseudo-momenta for the constrained variables, 
we see that the momentum corresponding to the projection axis used in the 
HMC algorithm takes values roughly one half of
the other momenta. In the language of unconstrained variables, evolution 
of configurations is glued to a hyperplane of constant $\bu^2$, 
which can be seen from figure \ref{fig:sigma:stereo_dist2} (left panel), 
where $\bu^2$, $\bv^2$ and $\bw^2$ correspond to the three possible (canonical) 
projection axes, while the actual projection axis of the HMC algorithm is fixed.
\FIGURE{\hfill
\input{stereo_dist2_timeline_6x6_1_bad}\hfill
\input{stereo_dist2_timeline_6x6_1_imp}\hfill\hfill\phantom.
\caption{\label{fig:sigma:stereo_dist2}MC-histories for the lattice
averages of $\bu^2,\,\bv^2$ and $\bw^2$ using the three
different projection axes in the usual HMC algorithm (left
panel) and the improved algorithm (right panel). Note that fluctuations for
$\bu^2$ which corresponds to the projection axis chosen in the HMC
algorithm are severely suppressed in the left panel.}}
To circumvent this problem, the projection axis in the molecular dynamics step
is changed repeatedly between successive trial and acceptance steps. While
it is in principle possible to choose random projection axes for every update,
we rotate through the three canonical axes, saving only every third
configuration which corresponds to a fixed projection axis. A
Metropolis acceptance step is yet needed after each change of the projection axis. 
The right panel of figure \ref{fig:sigma:stereo_dist2} illustrates how this 
improved update scheme restores the balance between the three projection axes. 
The configuration space is traversed quickly and the expectation values 
$\langle n_i \rangle$ decrease with ongoing MC time. 


%% file: stereo_dist2_timeline_6x6_1_bad.tex
\begingroup
\footnotesize
  \makeatletter
  \providecommand\color[2][]{%
    \GenericError{(gnuplot) \space\space\space\@spaces}{%
      Package color not loaded in conjunction with
      terminal option `colourtext'%
    }{See the gnuplot documentation for explanation.%
    }{Either use 'blacktext' in gnuplot or load the package
      color.sty in LaTeX.}%
    \renewcommand\color[2][]{}%
  }%
  \providecommand\includegraphics[2][]{%
    \GenericError{(gnuplot) \space\space\space\@spaces}{%
      Package graphicx or graphics not loaded%
    }{See the gnuplot documentation for explanation.%
    }{The gnuplot epslatex terminal needs graphicx.sty or graphics.sty.}%
    \renewcommand\includegraphics[2][]{}%
  }%
  \providecommand\rotatebox[2]{#2}%
  \@ifundefined{ifGPcolor}{%
    \newif\ifGPcolor
    \GPcolortrue
  }{}%
  \@ifundefined{ifGPblacktext}{%
    \newif\ifGPblacktext
    \GPblacktexttrue
  }{}%
  \let\gplgaddtomacro\g@addto@macro
  \gdef\gplbacktext{}%
  \gdef\gplfronttext{}%
  \makeatother
  \ifGPblacktext
    \def\colorrgb#1{}%
    \def\colorgray#1{}%
  \else
    \ifGPcolor
      \def\colorrgb#1{\color[rgb]{#1}}%
      \def\colorgray#1{\color[gray]{#1}}%
      \expandafter\def\csname LTw\endcsname{\color{white}}%
      \expandafter\def\csname LTb\endcsname{\color{black}}%
      \expandafter\def\csname LTa\endcsname{\color{black}}%
      \expandafter\def\csname LT0\endcsname{\color[rgb]{1,0,0}}%
      \expandafter\def\csname LT1\endcsname{\color[rgb]{0,1,0}}%
      \expandafter\def\csname LT2\endcsname{\color[rgb]{0,0,1}}%
      \expandafter\def\csname LT3\endcsname{\color[rgb]{1,0,1}}%
      \expandafter\def\csname LT4\endcsname{\color[rgb]{0,1,1}}%
      \expandafter\def\csname LT5\endcsname{\color[rgb]{1,1,0}}%
      \expandafter\def\csname LT6\endcsname{\color[rgb]{0,0,0}}%
      \expandafter\def\csname LT7\endcsname{\color[rgb]{1,0.3,0}}%
      \expandafter\def\csname LT8\endcsname{\color[rgb]{0.5,0.5,0.5}}%
    \else
      \def\colorrgb#1{\color{black}}%
      \def\colorgray#1{\color[gray]{#1}}%
      \expandafter\def\csname LTw\endcsname{\color{white}}%
      \expandafter\def\csname LTb\endcsname{\color{black}}%
      \expandafter\def\csname LTa\endcsname{\color{black}}%
      \expandafter\def\csname LT0\endcsname{\color{black}}%
      \expandafter\def\csname LT1\endcsname{\color{black}}%
      \expandafter\def\csname LT2\endcsname{\color{black}}%
      \expandafter\def\csname LT3\endcsname{\color{black}}%
      \expandafter\def\csname LT4\endcsname{\color{black}}%
      \expandafter\def\csname LT5\endcsname{\color{black}}%
      \expandafter\def\csname LT6\endcsname{\color{black}}%
      \expandafter\def\csname LT7\endcsname{\color{black}}%
      \expandafter\def\csname LT8\endcsname{\color{black}}%
    \fi
  \fi
  \setlength{\unitlength}{0.0500bp}%
  \begin{picture}(3968.00,2976.00)%
    \gplgaddtomacro\gplbacktext{%
      \csname LTb\endcsname%
      \put(768,480){\makebox(0,0)[r]{\strut{} 0}}%
      \put(768,969){\makebox(0,0)[r]{\strut{} 20}}%
      \put(768,1459){\makebox(0,0)[r]{\strut{} 40}}%
      \put(768,1948){\makebox(0,0)[r]{\strut{} 60}}%
      \put(768,2438){\makebox(0,0)[r]{\strut{} 80}}%
      \put(768,2927){\makebox(0,0)[r]{\strut{} 100}}%
      \put(864,320){\makebox(0,0){\strut{} 0}}%
      \put(1633,320){\makebox(0,0){\strut{} 50}}%
      \put(2402,320){\makebox(0,0){\strut{} 100}}%
      \put(3170,320){\makebox(0,0){\strut{} 150}}%
      \put(3939,320){\makebox(0,0){\strut{} 200}}%
      \put(208,1703){\rotatebox{90}{\makebox(0,0){\strut{}lattice averaged squared field}}}%
      \put(2401,80){\makebox(0,0){\strut{}Monte Carlo time parameter}}%
    }%
    \gplgaddtomacro\gplfronttext{%
      \csname LTb\endcsname%
      \put(3204,2784){\makebox(0,0)[r]{\strut{}$\mathbf{u}^2$}}%
      \csname LTb\endcsname%
      \put(3204,2624){\makebox(0,0)[r]{\strut{}$\mathbf{v}^2$}}%
      \csname LTb\endcsname%
      \put(3204,2464){\makebox(0,0)[r]{\strut{}$\mathbf{w}^2$}}%
    }%
    \gplbacktext
    \put(0,0){\includegraphics{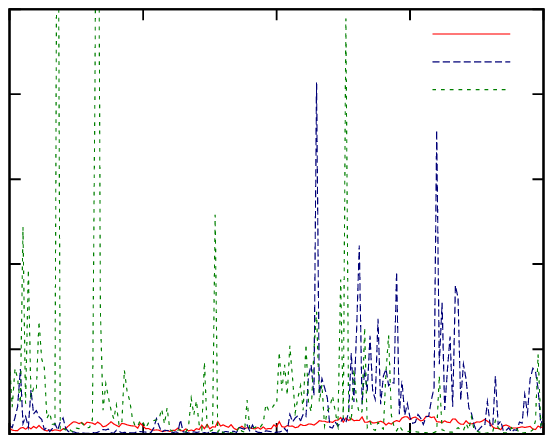}}%
    \gplfronttext
  \end{picture}%
\endgroup

%% file: stereo_dist2_timeline_6x6_1_imp.tex
\begingroup
\footnotesize
  \makeatletter
  \providecommand\color[2][]{%
    \GenericError{(gnuplot) \space\space\space\@spaces}{%
      Package color not loaded in conjunction with
      terminal option `colourtext'%
    }{See the gnuplot documentation for explanation.%
    }{Either use 'blacktext' in gnuplot or load the package
      color.sty in LaTeX.}%
    \renewcommand\color[2][]{}%
  }%
  \providecommand\includegraphics[2][]{%
    \GenericError{(gnuplot) \space\space\space\@spaces}{%
      Package graphicx or graphics not loaded%
    }{See the gnuplot documentation for explanation.%
    }{The gnuplot epslatex terminal needs graphicx.sty or graphics.sty.}%
    \renewcommand\includegraphics[2][]{}%
  }%
  \providecommand\rotatebox[2]{#2}%
  \@ifundefined{ifGPcolor}{%
    \newif\ifGPcolor
    \GPcolortrue
  }{}%
  \@ifundefined{ifGPblacktext}{%
    \newif\ifGPblacktext
    \GPblacktexttrue
  }{}%
  \let\gplgaddtomacro\g@addto@macro
  \gdef\gplbacktext{}%
  \gdef\gplfronttext{}%
  \makeatother
  \ifGPblacktext
    \def\colorrgb#1{}%
    \def\colorgray#1{}%
  \else
    \ifGPcolor
      \def\colorrgb#1{\color[rgb]{#1}}%
      \def\colorgray#1{\color[gray]{#1}}%
      \expandafter\def\csname LTw\endcsname{\color{white}}%
      \expandafter\def\csname LTb\endcsname{\color{black}}%
      \expandafter\def\csname LTa\endcsname{\color{black}}%
      \expandafter\def\csname LT0\endcsname{\color[rgb]{1,0,0}}%
      \expandafter\def\csname LT1\endcsname{\color[rgb]{0,1,0}}%
      \expandafter\def\csname LT2\endcsname{\color[rgb]{0,0,1}}%
      \expandafter\def\csname LT3\endcsname{\color[rgb]{1,0,1}}%
      \expandafter\def\csname LT4\endcsname{\color[rgb]{0,1,1}}%
      \expandafter\def\csname LT5\endcsname{\color[rgb]{1,1,0}}%
      \expandafter\def\csname LT6\endcsname{\color[rgb]{0,0,0}}%
      \expandafter\def\csname LT7\endcsname{\color[rgb]{1,0.3,0}}%
      \expandafter\def\csname LT8\endcsname{\color[rgb]{0.5,0.5,0.5}}%
    \else
      \def\colorrgb#1{\color{black}}%
      \def\colorgray#1{\color[gray]{#1}}%
      \expandafter\def\csname LTw\endcsname{\color{white}}%
      \expandafter\def\csname LTb\endcsname{\color{black}}%
      \expandafter\def\csname LTa\endcsname{\color{black}}%
      \expandafter\def\csname LT0\endcsname{\color{black}}%
      \expandafter\def\csname LT1\endcsname{\color{black}}%
      \expandafter\def\csname LT2\endcsname{\color{black}}%
      \expandafter\def\csname LT3\endcsname{\color{black}}%
      \expandafter\def\csname LT4\endcsname{\color{black}}%
      \expandafter\def\csname LT5\endcsname{\color{black}}%
      \expandafter\def\csname LT6\endcsname{\color{black}}%
      \expandafter\def\csname LT7\endcsname{\color{black}}%
      \expandafter\def\csname LT8\endcsname{\color{black}}%
    \fi
  \fi
  \setlength{\unitlength}{0.0500bp}%
  \begin{picture}(3968.00,2976.00)%
    \gplgaddtomacro\gplbacktext{%
      \csname LTb\endcsname%
      \put(768,480){\makebox(0,0)[r]{\strut{} 0}}%
      \put(768,969){\makebox(0,0)[r]{\strut{} 20}}%
      \put(768,1459){\makebox(0,0)[r]{\strut{} 40}}%
      \put(768,1948){\makebox(0,0)[r]{\strut{} 60}}%
      \put(768,2438){\makebox(0,0)[r]{\strut{} 80}}%
      \put(768,2927){\makebox(0,0)[r]{\strut{} 100}}%
      \put(864,320){\makebox(0,0){\strut{} 0}}%
      \put(1633,320){\makebox(0,0){\strut{} 50}}%
      \put(2402,320){\makebox(0,0){\strut{} 100}}%
      \put(3170,320){\makebox(0,0){\strut{} 150}}%
      \put(3939,320){\makebox(0,0){\strut{} 200}}%
      \put(208,1703){\rotatebox{90}{\makebox(0,0){\strut{}lattice averaged squared field}}}%
      \put(2401,80){\makebox(0,0){\strut{}Monte Carlo time parameter}}%
    }%
    \gplgaddtomacro\gplfronttext{%
      \csname LTb\endcsname%
      \put(3204,2784){\makebox(0,0)[r]{\strut{}$\mathbf{u}^2$}}%
      \csname LTb\endcsname%
      \put(3204,2624){\makebox(0,0)[r]{\strut{}$\mathbf{v}^2$}}%
      \csname LTb\endcsname%
      \put(3204,2464){\makebox(0,0)[r]{\strut{}$\mathbf{w}^2$}}%
    }%
    \gplbacktext
    \put(0,0){\includegraphics{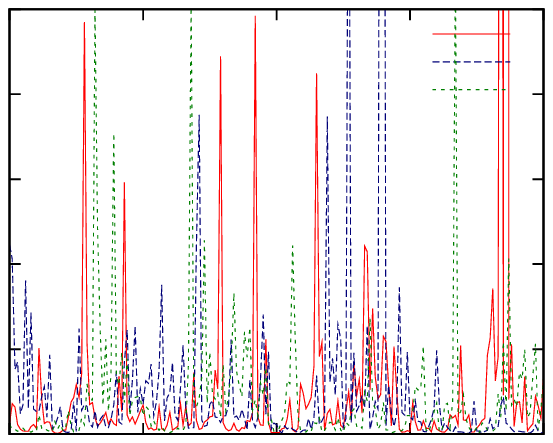}}%
    \gplfronttext
  \end{picture}%
\endgroup

%% file: Fmasses.tex
\subsection{Fermionic masses}

The classical theory has no intrinsic mass scale.
But there is a relation between mass gap and bare coupling by dimensional
transmutation. In the $\overline{\text{MS}}$ scheme it is possible to
compute the mass gap \cite{Evans:1994sv,Evans:1994sy} in relation to
$\Lambda_{\overline{\text{MS}}}$. Investigations of the $\mathcal{N}=1$ 
Wess-Zumino model have revealed that the fermionic mass is less affected by 
finite size effects than the bosonic mass. For that reason the fermionic mass is 
used to set the physical scale of the lattice, i.e. the lattice spacing and the 
physical box length.
The O$(3)$ symmetric fermionic correlator for group-valued fields is constructed
as
\begin{equation}
\begin{aligned}
\vev{\ii \bar\bpsi_x  \bpsi_y} &= \ZZ^{-1} \int\cD \bn \,\cD \sigma\,
\cD\bpsi\,\delta(\bn\bpsi) \delta (\bn^2-1)\, (\ii \bar\bpsi_x \bpsi_y)
\,\ee^{-S}\\ 
&= \ZZ^{-1} \int\cD R \,\cD \sigma\, \ee^{-\SB}\int \cD\bchi\,~ \ii\bar
\bchi_x \kappa^\trnsp R_x^\trnsp R_y \kappa
\bchi_y\, \ee^{-\SF}\\ 
&=g^2 \vev{\tr_\text{f,s} ( R_x \kappa CQ^{-1}_{xy} \kappa^\trnsp R_y^\trnsp)},\quad
\kappa^\trnsp =\begin{pmatrix}0&1&0\\0&0&1\end{pmatrix}\,,\label{fermion2pt}
\end{aligned}
\end{equation}
with $\SF$ given in (\ref{rotaction}) and `$\tr_\text{f,s}$' indicates
the trace over flavor and spinor indices. The corresponding timeslice correlator is 
given by $C_\text{F}(t) = \Ns^{-2}\sum_{xy} \vev{\ii \bar\bpsi_{(t,x)} \bpsi_{(0,y)}}$.
In order to measure the mass in \emph{one} of the ground states the configurations are
projected, without loss of generality, onto the sector with
$\Xi>0$ for $a \Xi = N^{-1} \sum_x \bar\bpsi_x
\bpsi_x$. This can be achieved by flipping the sign of $\sigma$ for configurations 
with $\Xi<0$. Using these
definitions the fermionic masses are obtained by a $\cosh$ fit to the
correlator over the range $t\in]0,\Nt[$. A similar procedure is utilized in the case of 
stereographically projected fermions, where we have to use the projected fermion
correlator instead of (\ref{fermion2pt}),
\begin{align}
\label{stereocond}
\big\langle \ii\bar\bpsi_{x,\perp}\bpsi_{y,\perp}\big\rangle =& \big\langle 4\rho_x\ii\bar{\blam}_x \blam_y \rho_y -
8\rho_x (\bu_x\ii\bar\blam_x) (\blam_y \bu_y) \rho_y^2 - 8\rho_x^2 (\bu_x\ii\bar\blam_x) (\blam_y \bu_y) \rho_y \nonumber\\ 
& +16 \rho_x^2 (\bu_x\ii\bar\blam_x) (\bu_x\bu_y) (\blam_y \bu_y) \rho_y^2\big\rangle,\nonumber\\
\big\langle \ii\bar{\psi}_{x,1}\psi_{y,1}\big\rangle =& \big\langle
16\rho_x^2 (\bu_x \ii\bar\blam_x)(\blam_y \bu_y) \rho_y^2 \big\rangle .
\end{align}
In this case non-diagonal elements contribute to the
full correlator, which relate unconstrained fields of different flavor.
A comparison of the fermionic masses for different lattice sizes using the SLAC derivative
(see Figure \ref{fig:sigma:fermionMasses}) reveals that 
finite size effects on $m_\text{F} a$ are within statistical error bars if
$m_\text{F} L\gtrsim 5$.
\FIGURE{\hfill
\input{sigmaFermionMassesML}\hfill
\input{sigmaFermionMassesMa}\hfill\hfill\phantom.
\caption{\label{fig:sigma:fermionMasses}Fermionic masses from SLAC ensembles in units of the box
length (left panel) and lattice spacing (right panel) for different lattice
volumes and bare couplings $g^{-2}$ computed using up to $2\cdot 10^7$
configurations.}}

%% file: sigmaFermionMassesML.tex
\begingroup
\footnotesize
  \makeatletter
  \providecommand\color[2][]{%
    \GenericError{(gnuplot) \space\space\space\@spaces}{%
      Package color not loaded in conjunction with
      terminal option `colourtext'%
    }{See the gnuplot documentation for explanation.%
    }{Either use 'blacktext' in gnuplot or load the package
      color.sty in LaTeX.}%
    \renewcommand\color[2][]{}%
  }%
  \providecommand\includegraphics[2][]{%
    \GenericError{(gnuplot) \space\space\space\@spaces}{%
      Package graphicx or graphics not loaded%
    }{See the gnuplot documentation for explanation.%
    }{The gnuplot epslatex terminal needs graphicx.sty or graphics.sty.}%
    \renewcommand\includegraphics[2][]{}%
  }%
  \providecommand\rotatebox[2]{#2}%
  \@ifundefined{ifGPcolor}{%
    \newif\ifGPcolor
    \GPcolortrue
  }{}%
  \@ifundefined{ifGPblacktext}{%
    \newif\ifGPblacktext
    \GPblacktexttrue
  }{}%
  \let\gplgaddtomacro\g@addto@macro
  \gdef\gplbacktext{}%
  \gdef\gplfronttext{}%
  \makeatother
  \ifGPblacktext
    \def\colorrgb#1{}%
    \def\colorgray#1{}%
  \else
    \ifGPcolor
      \def\colorrgb#1{\color[rgb]{#1}}%
      \def\colorgray#1{\color[gray]{#1}}%
      \expandafter\def\csname LTw\endcsname{\color{white}}%
      \expandafter\def\csname LTb\endcsname{\color{black}}%
      \expandafter\def\csname LTa\endcsname{\color{black}}%
      \expandafter\def\csname LT0\endcsname{\color[rgb]{1,0,0}}%
      \expandafter\def\csname LT1\endcsname{\color[rgb]{0,1,0}}%
      \expandafter\def\csname LT2\endcsname{\color[rgb]{0,0,1}}%
      \expandafter\def\csname LT3\endcsname{\color[rgb]{1,0,1}}%
      \expandafter\def\csname LT4\endcsname{\color[rgb]{0,1,1}}%
      \expandafter\def\csname LT5\endcsname{\color[rgb]{1,1,0}}%
      \expandafter\def\csname LT6\endcsname{\color[rgb]{0,0,0}}%
      \expandafter\def\csname LT7\endcsname{\color[rgb]{1,0.3,0}}%
      \expandafter\def\csname LT8\endcsname{\color[rgb]{0.5,0.5,0.5}}%
    \else
      \def\colorrgb#1{\color{black}}%
      \def\colorgray#1{\color[gray]{#1}}%
      \expandafter\def\csname LTw\endcsname{\color{white}}%
      \expandafter\def\csname LTb\endcsname{\color{black}}%
      \expandafter\def\csname LTa\endcsname{\color{black}}%
      \expandafter\def\csname LT0\endcsname{\color{black}}%
      \expandafter\def\csname LT1\endcsname{\color{black}}%
      \expandafter\def\csname LT2\endcsname{\color{black}}%
      \expandafter\def\csname LT3\endcsname{\color{black}}%
      \expandafter\def\csname LT4\endcsname{\color{black}}%
      \expandafter\def\csname LT5\endcsname{\color{black}}%
      \expandafter\def\csname LT6\endcsname{\color{black}}%
      \expandafter\def\csname LT7\endcsname{\color{black}}%
      \expandafter\def\csname LT8\endcsname{\color{black}}%
    \fi
  \fi
  \setlength{\unitlength}{0.0500bp}%
  \begin{picture}(3968.00,2976.00)%
    \gplgaddtomacro\gplbacktext{%
      \csname LTb\endcsname%
      \put(768,480){\makebox(0,0)[r]{\strut{}$0$}}%
      \put(768,830){\makebox(0,0)[r]{\strut{}$2$}}%
      \put(768,1179){\makebox(0,0)[r]{\strut{}$4$}}%
      \put(768,1529){\makebox(0,0)[r]{\strut{}$6$}}%
      \put(768,1878){\makebox(0,0)[r]{\strut{}$8$}}%
      \put(768,2228){\makebox(0,0)[r]{\strut{}$10$}}%
      \put(768,2577){\makebox(0,0)[r]{\strut{}$12$}}%
      \put(768,2927){\makebox(0,0)[r]{\strut{}$14$}}%
      \put(864,320){\makebox(0,0){\strut{}$0.3$}}%
      \put(1206,320){\makebox(0,0){\strut{}$0.4$}}%
      \put(1547,320){\makebox(0,0){\strut{}$0.5$}}%
      \put(1889,320){\makebox(0,0){\strut{}$0.6$}}%
      \put(2231,320){\makebox(0,0){\strut{}$0.7$}}%
      \put(2572,320){\makebox(0,0){\strut{}$0.8$}}%
      \put(2914,320){\makebox(0,0){\strut{}$0.9$}}%
      \put(3256,320){\makebox(0,0){\strut{}$1.0$}}%
      \put(3597,320){\makebox(0,0){\strut{}$1.1$}}%
      \put(3939,320){\makebox(0,0){\strut{}$1.2$}}%
      \put(352,1703){\makebox(0,0){\strut{}$m_\text{F}L$}}%
      \put(2401,80){\makebox(0,0){\strut{}$g^{-2}$}}%
    }%
    \gplgaddtomacro\gplfronttext{%
      \csname LTb\endcsname%
      \put(3204,2784){\makebox(0,0)[r]{\strut{}$N=5^2$}}%
      \csname LTb\endcsname%
      \put(3204,2624){\makebox(0,0)[r]{\strut{}$N=7^2$}}%
      \csname LTb\endcsname%
      \put(3204,2464){\makebox(0,0)[r]{\strut{}$N=9^2$}}%
    }%
    \gplbacktext
    \put(0,0){\includegraphics{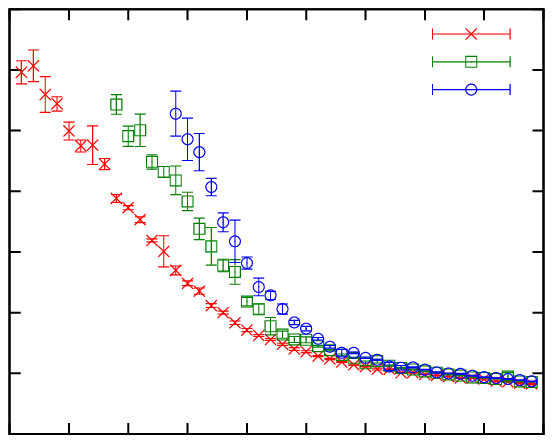}}%
    \gplfronttext
  \end{picture}%
\endgroup

%% file: sigmaFermionMassesMa.tex
\begingroup
\footnotesize
  \makeatletter
  \providecommand\color[2][]{%
    \GenericError{(gnuplot) \space\space\space\@spaces}{%
      Package color not loaded in conjunction with
      terminal option `colourtext'%
    }{See the gnuplot documentation for explanation.%
    }{Either use 'blacktext' in gnuplot or load the package
      color.sty in LaTeX.}%
    \renewcommand\color[2][]{}%
  }%
  \providecommand\includegraphics[2][]{%
    \GenericError{(gnuplot) \space\space\space\@spaces}{%
      Package graphicx or graphics not loaded%
    }{See the gnuplot documentation for explanation.%
    }{The gnuplot epslatex terminal needs graphicx.sty or graphics.sty.}%
    \renewcommand\includegraphics[2][]{}%
  }%
  \providecommand\rotatebox[2]{#2}%
  \@ifundefined{ifGPcolor}{%
    \newif\ifGPcolor
    \GPcolortrue
  }{}%
  \@ifundefined{ifGPblacktext}{%
    \newif\ifGPblacktext
    \GPblacktexttrue
  }{}%
  \let\gplgaddtomacro\g@addto@macro
  \gdef\gplbacktext{}%
  \gdef\gplfronttext{}%
  \makeatother
  \ifGPblacktext
    \def\colorrgb#1{}%
    \def\colorgray#1{}%
  \else
    \ifGPcolor
      \def\colorrgb#1{\color[rgb]{#1}}%
      \def\colorgray#1{\color[gray]{#1}}%
      \expandafter\def\csname LTw\endcsname{\color{white}}%
      \expandafter\def\csname LTb\endcsname{\color{black}}%
      \expandafter\def\csname LTa\endcsname{\color{black}}%
      \expandafter\def\csname LT0\endcsname{\color[rgb]{1,0,0}}%
      \expandafter\def\csname LT1\endcsname{\color[rgb]{0,1,0}}%
      \expandafter\def\csname LT2\endcsname{\color[rgb]{0,0,1}}%
      \expandafter\def\csname LT3\endcsname{\color[rgb]{1,0,1}}%
      \expandafter\def\csname LT4\endcsname{\color[rgb]{0,1,1}}%
      \expandafter\def\csname LT5\endcsname{\color[rgb]{1,1,0}}%
      \expandafter\def\csname LT6\endcsname{\color[rgb]{0,0,0}}%
      \expandafter\def\csname LT7\endcsname{\color[rgb]{1,0.3,0}}%
      \expandafter\def\csname LT8\endcsname{\color[rgb]{0.5,0.5,0.5}}%
    \else
      \def\colorrgb#1{\color{black}}%
      \def\colorgray#1{\color[gray]{#1}}%
      \expandafter\def\csname LTw\endcsname{\color{white}}%
      \expandafter\def\csname LTb\endcsname{\color{black}}%
      \expandafter\def\csname LTa\endcsname{\color{black}}%
      \expandafter\def\csname LT0\endcsname{\color{black}}%
      \expandafter\def\csname LT1\endcsname{\color{black}}%
      \expandafter\def\csname LT2\endcsname{\color{black}}%
      \expandafter\def\csname LT3\endcsname{\color{black}}%
      \expandafter\def\csname LT4\endcsname{\color{black}}%
      \expandafter\def\csname LT5\endcsname{\color{black}}%
      \expandafter\def\csname LT6\endcsname{\color{black}}%
      \expandafter\def\csname LT7\endcsname{\color{black}}%
      \expandafter\def\csname LT8\endcsname{\color{black}}%
    \fi
  \fi
  \setlength{\unitlength}{0.0500bp}%
  \begin{picture}(3968.00,2976.00)%
    \gplgaddtomacro\gplbacktext{%
      \csname LTb\endcsname%
      \put(768,480){\makebox(0,0)[r]{\strut{}$0.0$}}%
      \put(768,888){\makebox(0,0)[r]{\strut{}$0.5$}}%
      \put(768,1296){\makebox(0,0)[r]{\strut{}$1.0$}}%
      \put(768,1704){\makebox(0,0)[r]{\strut{}$1.5$}}%
      \put(768,2111){\makebox(0,0)[r]{\strut{}$2.0$}}%
      \put(768,2519){\makebox(0,0)[r]{\strut{}$2.5$}}%
      \put(768,2927){\makebox(0,0)[r]{\strut{}$3.0$}}%
      \put(864,320){\makebox(0,0){\strut{}$0.3$}}%
      \put(1206,320){\makebox(0,0){\strut{}$0.4$}}%
      \put(1547,320){\makebox(0,0){\strut{}$0.5$}}%
      \put(1889,320){\makebox(0,0){\strut{}$0.6$}}%
      \put(2231,320){\makebox(0,0){\strut{}$0.7$}}%
      \put(2572,320){\makebox(0,0){\strut{}$0.8$}}%
      \put(2914,320){\makebox(0,0){\strut{}$0.9$}}%
      \put(3256,320){\makebox(0,0){\strut{}$1.0$}}%
      \put(3597,320){\makebox(0,0){\strut{}$1.1$}}%
      \put(3939,320){\makebox(0,0){\strut{}$1.2$}}%
      \put(256,1863){\makebox(0,0){\strut{}$m_\text{F}a$}}%
      \put(2401,80){\makebox(0,0){\strut{}$g^{-2}$}}%
    }%
    \gplgaddtomacro\gplfronttext{%
      \csname LTb\endcsname%
      \put(3204,2784){\makebox(0,0)[r]{\strut{}$N=5^2$}}%
      \csname LTb\endcsname%
      \put(3204,2624){\makebox(0,0)[r]{\strut{}$N=7^2$}}%
      \csname LTb\endcsname%
      \put(3204,2464){\makebox(0,0)[r]{\strut{}$N=9^2$}}%
    }%
    \gplbacktext
    \put(0,0){\includegraphics{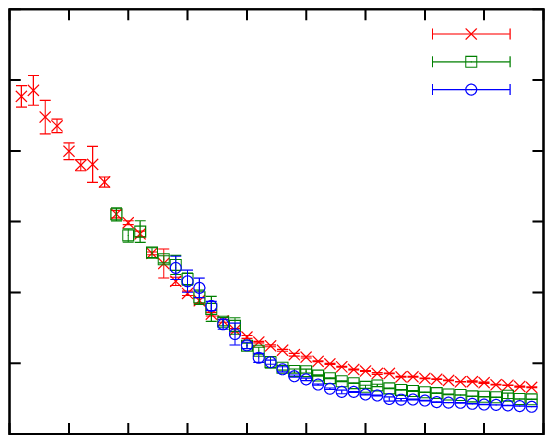}}%
    \gplfronttext
  \end{picture}%
\endgroup

%% file: chiral.tex
\subsection{Chiral symmetry breaking}
Due to the orthogonality of $R_x$ the chiral condensate in the coset formulation simplifies to 
\begin{equation}
\vev{\ii\bar\bpsi_x\bpsi_x} 
=\ZZ^{-1}\!\! \int\!\cD R \,\cD \sigma\,\ee^{-\SB}\int \cD\bchi\, 
\ii\bar\bchi_x\bchi_x \,\ee^{-\SF} =
g^2\vev{\tr_\text{f,s}(CQ^{-1}_{xx})}.
\end{equation} 
Using stereographic projection, however, we have to compute the projected condensate
given in terms of unconstrained fields. This is obtained by using the trace of the
correlator given in equation (\ref{stereocond}). 
If we rewrite the action \eqref{stereoaction} as $S = S_B[u,\sigma] + \blam^{\trnsp} P \blam$, 
we can replace the quadratic fermion operator by 
\begin{align}
\big\langle ~...~\bar{\lambda}_{x,i}\lambda_{y,j} ~...~ \big\rangle =
\big\langle ~...~ C \big( P^{-1} \big)_{xy,ij} ~...~ \big\rangle.
\end{align}
\FIGURE{\hfill \input{sigmaChiralHistoReweight} \hfill
\input{sigmaChiralEffectivePot}\hfill\hfill\phantom.
\caption{\label{fig:sigma:chiralCondensate}Left panel: Probability density
of the volume averaged chiral condensate for lattice size $9\times 9$ at a bare coupling
$g^{-2}=1$ in the sign quenched and reweighted ensemble using the SLAC
derivative. Right panel: Constraint effective potential
(normalised to $\min_{(a\Xi)} \hat U(a\Xi) = 0$) of the chiral condensate for
different lattice volumes at $g^{-2}=0.64$ computed using up to $3\cdot 10^7$
configurations and the SLAC derivative.}}

\noindent The continuum model is invariant under a discrete chiral symmetry $\bpsi \to i \gamma_* \bpsi$. 
It is spontaneously broken in the infinite volume limit and the
supersymmetric ground states correspond to the two ground states of this broken
symmetry \cite{Witten:1982df}. A discretization based on the SLAC derivative maintains chiral
symmetry on the lattice by the cost of having a non-local derivative. For every finite lattice volume 
the expectation value $\vev{\ii\bar\bpsi \bpsi}$ will vanish and is hence not the appropriate measure to trace
a broken symmetry. One should instead analyze the histograms of
the volume average $a\Xi = N^{-1} \sum_x \bar\bpsi_x
\bpsi_x$. Fig.~\ref{fig:sigma:chiralCondensate} (left panel) clearly shows a double peak
structure of the corresponding distribution $\rho(a\Xi)$, coinciding with
the two ground states. The reweighting process reveals that a cancellation
between positive and negative Pfaffians happens mostly for $a\Xi\approx 0$.
In the analysis of the constrained effective potential $\hat U(a\Xi) =
-\ln(\rho(a\Xi))/N$ for several lattice volumes at fixed coupling (see
Fig.~\ref{fig:sigma:chiralCondensate}, right panel) no running of the two minima
is visible, such that there will be a spontaneous chiral symmetry breaking in
the infinite volume limit of the lattice model.\\

%% file: sigmaChiralHistoReweight.tex
\begingroup
\footnotesize
  \makeatletter
  \providecommand\color[2][]{%
    \GenericError{(gnuplot) \space\space\space\@spaces}{%
      Package color not loaded in conjunction with
      terminal option `colourtext'%
    }{See the gnuplot documentation for explanation.%
    }{Either use 'blacktext' in gnuplot or load the package
      color.sty in LaTeX.}%
    \renewcommand\color[2][]{}%
  }%
  \providecommand\includegraphics[2][]{%
    \GenericError{(gnuplot) \space\space\space\@spaces}{%
      Package graphicx or graphics not loaded%
    }{See the gnuplot documentation for explanation.%
    }{The gnuplot epslatex terminal needs graphicx.sty or graphics.sty.}%
    \renewcommand\includegraphics[2][]{}%
  }%
  \providecommand\rotatebox[2]{#2}%
  \@ifundefined{ifGPcolor}{%
    \newif\ifGPcolor
    \GPcolortrue
  }{}%
  \@ifundefined{ifGPblacktext}{%
    \newif\ifGPblacktext
    \GPblacktexttrue
  }{}%
  \let\gplgaddtomacro\g@addto@macro
  \gdef\gplbacktext{}%
  \gdef\gplfronttext{}%
  \makeatother
  \ifGPblacktext
    \def\colorrgb#1{}%
    \def\colorgray#1{}%
  \else
    \ifGPcolor
      \def\colorrgb#1{\color[rgb]{#1}}%
      \def\colorgray#1{\color[gray]{#1}}%
      \expandafter\def\csname LTw\endcsname{\color{white}}%
      \expandafter\def\csname LTb\endcsname{\color{black}}%
      \expandafter\def\csname LTa\endcsname{\color{black}}%
      \expandafter\def\csname LT0\endcsname{\color[rgb]{1,0,0}}%
      \expandafter\def\csname LT1\endcsname{\color[rgb]{0,1,0}}%
      \expandafter\def\csname LT2\endcsname{\color[rgb]{0,0,1}}%
      \expandafter\def\csname LT3\endcsname{\color[rgb]{1,0,1}}%
      \expandafter\def\csname LT4\endcsname{\color[rgb]{0,1,1}}%
      \expandafter\def\csname LT5\endcsname{\color[rgb]{1,1,0}}%
      \expandafter\def\csname LT6\endcsname{\color[rgb]{0,0,0}}%
      \expandafter\def\csname LT7\endcsname{\color[rgb]{1,0.3,0}}%
      \expandafter\def\csname LT8\endcsname{\color[rgb]{0.5,0.5,0.5}}%
    \else
      \def\colorrgb#1{\color{black}}%
      \def\colorgray#1{\color[gray]{#1}}%
      \expandafter\def\csname LTw\endcsname{\color{white}}%
      \expandafter\def\csname LTb\endcsname{\color{black}}%
      \expandafter\def\csname LTa\endcsname{\color{black}}%
      \expandafter\def\csname LT0\endcsname{\color{black}}%
      \expandafter\def\csname LT1\endcsname{\color{black}}%
      \expandafter\def\csname LT2\endcsname{\color{black}}%
      \expandafter\def\csname LT3\endcsname{\color{black}}%
      \expandafter\def\csname LT4\endcsname{\color{black}}%
      \expandafter\def\csname LT5\endcsname{\color{black}}%
      \expandafter\def\csname LT6\endcsname{\color{black}}%
      \expandafter\def\csname LT7\endcsname{\color{black}}%
      \expandafter\def\csname LT8\endcsname{\color{black}}%
    \fi
  \fi
  \setlength{\unitlength}{0.0500bp}%
  \begin{picture}(3968.00,2976.00)%
    \gplgaddtomacro\gplbacktext{%
      \csname LTb\endcsname%
      \put(768,480){\makebox(0,0)[r]{\strut{}$0.0$}}%
      \put(768,1092){\makebox(0,0)[r]{\strut{}$0.5$}}%
      \put(768,1704){\makebox(0,0)[r]{\strut{}$1.0$}}%
      \put(768,2315){\makebox(0,0)[r]{\strut{}$1.5$}}%
      \put(768,2927){\makebox(0,0)[r]{\strut{}$2.0$}}%
      \put(1120,320){\makebox(0,0){\strut{}$-1.0$}}%
      \put(1761,320){\makebox(0,0){\strut{}$-0.5$}}%
      \put(2402,320){\makebox(0,0){\strut{}$0.0$}}%
      \put(3042,320){\makebox(0,0){\strut{}$0.5$}}%
      \put(3683,320){\makebox(0,0){\strut{}$1.0$}}%
      \put(304,2023){\makebox(0,0){\strut{}$\rho(\Xi a)$}}%
      \put(2401,80){\makebox(0,0){\strut{}$\Xi a$}}%
    }%
    \gplgaddtomacro\gplfronttext{%
      \csname LTb\endcsname%
      \put(3204,2784){\makebox(0,0)[r]{\strut{}sign quenched}}%
      \csname LTb\endcsname%
      \put(3204,2624){\makebox(0,0)[r]{\strut{}reweighted}}%
    }%
    \gplbacktext
    \put(0,0){\includegraphics{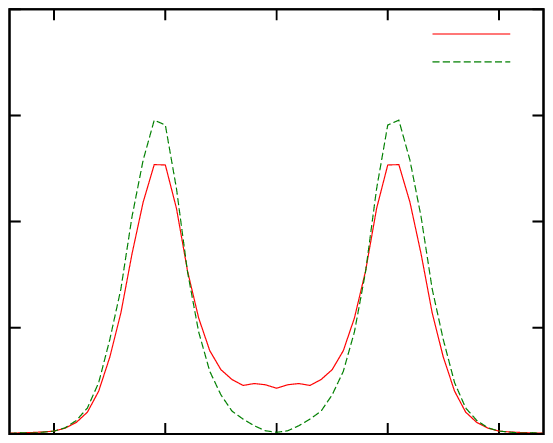}}%
    \gplfronttext
  \end{picture}%
\endgroup

%% file: sigmaChiralEffectivePot.tex
\begingroup
\footnotesize
  \makeatletter
  \providecommand\color[2][]{%
    \GenericError{(gnuplot) \space\space\space\@spaces}{%
      Package color not loaded in conjunction with
      terminal option `colourtext'%
    }{See the gnuplot documentation for explanation.%
    }{Either use 'blacktext' in gnuplot or load the package
      color.sty in LaTeX.}%
    \renewcommand\color[2][]{}%
  }%
  \providecommand\includegraphics[2][]{%
    \GenericError{(gnuplot) \space\space\space\@spaces}{%
      Package graphicx or graphics not loaded%
    }{See the gnuplot documentation for explanation.%
    }{The gnuplot epslatex terminal needs graphicx.sty or graphics.sty.}%
    \renewcommand\includegraphics[2][]{}%
  }%
  \providecommand\rotatebox[2]{#2}%
  \@ifundefined{ifGPcolor}{%
    \newif\ifGPcolor
    \GPcolortrue
  }{}%
  \@ifundefined{ifGPblacktext}{%
    \newif\ifGPblacktext
    \GPblacktexttrue
  }{}%
  \let\gplgaddtomacro\g@addto@macro
  \gdef\gplbacktext{}%
  \gdef\gplfronttext{}%
  \makeatother
  \ifGPblacktext
    \def\colorrgb#1{}%
    \def\colorgray#1{}%
  \else
    \ifGPcolor
      \def\colorrgb#1{\color[rgb]{#1}}%
      \def\colorgray#1{\color[gray]{#1}}%
      \expandafter\def\csname LTw\endcsname{\color{white}}%
      \expandafter\def\csname LTb\endcsname{\color{black}}%
      \expandafter\def\csname LTa\endcsname{\color{black}}%
      \expandafter\def\csname LT0\endcsname{\color[rgb]{1,0,0}}%
      \expandafter\def\csname LT1\endcsname{\color[rgb]{0,1,0}}%
      \expandafter\def\csname LT2\endcsname{\color[rgb]{0,0,1}}%
      \expandafter\def\csname LT3\endcsname{\color[rgb]{1,0,1}}%
      \expandafter\def\csname LT4\endcsname{\color[rgb]{0,1,1}}%
      \expandafter\def\csname LT5\endcsname{\color[rgb]{1,1,0}}%
      \expandafter\def\csname LT6\endcsname{\color[rgb]{0,0,0}}%
      \expandafter\def\csname LT7\endcsname{\color[rgb]{1,0.3,0}}%
      \expandafter\def\csname LT8\endcsname{\color[rgb]{0.5,0.5,0.5}}%
    \else
      \def\colorrgb#1{\color{black}}%
      \def\colorgray#1{\color[gray]{#1}}%
      \expandafter\def\csname LTw\endcsname{\color{white}}%
      \expandafter\def\csname LTb\endcsname{\color{black}}%
      \expandafter\def\csname LTa\endcsname{\color{black}}%
      \expandafter\def\csname LT0\endcsname{\color{black}}%
      \expandafter\def\csname LT1\endcsname{\color{black}}%
      \expandafter\def\csname LT2\endcsname{\color{black}}%
      \expandafter\def\csname LT3\endcsname{\color{black}}%
      \expandafter\def\csname LT4\endcsname{\color{black}}%
      \expandafter\def\csname LT5\endcsname{\color{black}}%
      \expandafter\def\csname LT6\endcsname{\color{black}}%
      \expandafter\def\csname LT7\endcsname{\color{black}}%
      \expandafter\def\csname LT8\endcsname{\color{black}}%
    \fi
  \fi
  \setlength{\unitlength}{0.0500bp}%
  \begin{picture}(3968.00,2976.00)%
    \gplgaddtomacro\gplbacktext{%
      \csname LTb\endcsname%
      \put(768,480){\makebox(0,0)[r]{\strut{}$0.00$}}%
      \put(768,1092){\makebox(0,0)[r]{\strut{}$0.05$}}%
      \put(768,1704){\makebox(0,0)[r]{\strut{}$0.10$}}%
      \put(768,2315){\makebox(0,0)[r]{\strut{}$0.15$}}%
      \put(768,2927){\makebox(0,0)[r]{\strut{}$0.20$}}%
      \put(1172,320){\makebox(0,0){\strut{}$-2.0$}}%
      \put(1787,320){\makebox(0,0){\strut{}$-1.0$}}%
      \put(2402,320){\makebox(0,0){\strut{}$0.0$}}%
      \put(3017,320){\makebox(0,0){\strut{}$1.0$}}%
      \put(3632,320){\makebox(0,0){\strut{}$2.0$}}%
      \put(208,2023){\makebox(0,0){\strut{}$\hat U(\Xi a)$}}%
      \put(2401,80){\makebox(0,0){\strut{}$\Xi a$}}%
    }%
    \gplgaddtomacro\gplfronttext{%
      \csname LTb\endcsname%
      \put(3204,2784){\makebox(0,0)[r]{\strut{}$N=5^2$}}%
      \csname LTb\endcsname%
      \put(3204,2624){\makebox(0,0)[r]{\strut{}$N=7^2$}}%
      \csname LTb\endcsname%
      \put(3204,2464){\makebox(0,0)[r]{\strut{}$N=9^2$}}%
    }%
    \gplbacktext
    \put(0,0){\includegraphics{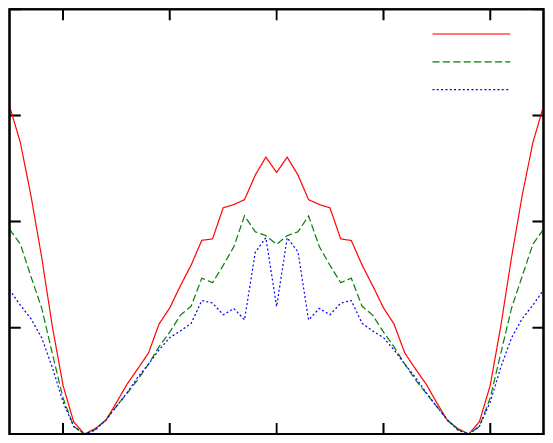}}%
    \gplfronttext
  \end{picture}%
\endgroup

%% file: Bmasses.tex
\subsection{Bosonic masses}
In order to test whether the chosen discretization corresponds to a supersymmetric theory, one should 
check the degeneracy of the bosonic and fermionic masses. The bosonic
masses $m_\text{B}L$ are extracted from the O$(3)$ invariant correlator
\eqref{eq:sigma:bosonCorrelator} via a $\cosh$ fit over the range
$t\in]0,\Nt[$. The bosonic correlator is unaffected by a change in
$\sigma$, such that no projection on one of the two ground states is necessary. 
Having computed the boson masses, we can compare them with the fermionic ones for different 
couplings and lattice sizes.

\paragraph{Simulations using SLAC fermions}
Calculations have been performed on lattice sizes $N\in\{5^2,7^2,9^2\}$ over a
coupling range $g^{-2}\in[0.4,1.2]$. The direct comparison is shown in
Fig.~\ref{fig:sigma:massComparison} (left panel) and the results seem to be
disappointing at first sight. The bosonic masses lie considerably below the
fermionic partners and this deviation becomes even more pronounced for larger 
lattices. However, this does not necessarily imply that supersymmetry will be broken in the continuum limit.
Already for the simple lattice $\Neqtwo$ Wess-Zumino model with spontaneously broken $\bbZ_2$ 
symmetry and one exact supersymmetry the masses split in the
strong coupling regime at finite physical box sizes \cite{Kastner:2008zc}. E.g.\ for
couplings for which the one-loop perturbation theory fails and for a box size
$m_\text{F}L \approx 10$, a $20\%$ splitting with a smaller bosonic mass is
observed. From that point of view the supersymmetric O$(3)$ nonlinear sigma model could be 
similar to a strongly coupled $\mathcal{N}=2$ Wess-Zumino model.
The finite size effects may be even larger and a mass splitting of much more
than $20\%$ would not be surprising for $m_\text{F}L<10$. Only an analysis of the mass
ratio $m_\text{B}/m_\text{F}$ in the large volume limit can uncover a 
restoration of degenerate masses. This is exemplarily shown for the results
on the $5\times 5$ lattice in Fig.~\ref{fig:sigma:massComparison} (right panel).
Despite the fact that lattice artefacts are sizeable the basic mechanism
becomes clear. In the limit of large volumes a relation
\begin{equation}
\label{eq:sigma:massRatioRelation}
m_\text{B}L = m_\text{F}L-\Delta M\quad \Rightarrow
\quad \frac{m_\text{B}}{m_\text{F}} = 1 -\frac{\Delta M}{m_\text{F}L}
\end{equation}
with constant $\Delta M$ is found, such that the ratio tends to $1$ and the
masses will be degenerate in the infinite volume limit.\footnote{For the
$5\times 5$ lattice a fit to Eq.~\eqref{eq:sigma:massRatioRelation} for
$m_\text{F}L>6$ gives $\Delta M=2.56(10)$.}

\FIGURE{\hfill
\input{sigmaMassComparisonDirect}\hfill
\input{sigmaMassComparisonRatio}\hfill\hfill\phantom.
\caption{\label{fig:sigma:massComparison} Left panel: Direct comparison of
bosonic and fermionic mass from the SLAC ensemble in units of the box size for three different lattice
sizes. The dotted line denotes the case $m_\text{F}=m_\text{B}$. Right panel:
Difference $m_\text{B}L-m_\text{F}L$ for varying box size $m_\text{F}L$ on a
$5\times 5$ lattice. The shaded area denotes a fit according to
Eq.~\eqref{eq:sigma:massRatioRelation} for $m_\text{F}L>6$.}}
\noindent But the accessible physical volumes at larger lattices still do not allow for a
reliable extrapolation of the corresponding $\Delta M$ and no continuum limit
can be taken at the moment. 
It is hence an open question if
eq.~\eqref{eq:sigma:massRatioRelation} also holds true in the continuum limit and
if supersymmetry will be restored. These questions should be resolved by computations on 
larger lattices. Such computations, however, become unfeasible because of the sign problem,
which worsens with increasing lattice size in case of the SLAC derivative, see 
appendix \ref{ch:eval}. Nevertheless, further information about the supersymmetric
features of the implementation can be obtained by studying a corresponding Ward identity,
cf. section \ref{ch:ward}.

\paragraph{Simulations using Wilson fermions}
While the applicability of the SLAC 
derivative is confined to small lattice volumes due to the strong sign problem, 
Wilson fermions offer the possibility to explore larger volumes by utilizing 
efficiently preconditioned pseudofermion algorithms (compare to appendix \ref{app:algo}), 
however at the 
cost of larger lattice artifacts. Bosonic masses are extracted as explained 
beforehand by calculating the stereographically projected correlators
\begin{align}
\big< \bn_{x,\perp} \bn_{y,\perp} \big> &= 4\big< \rho_{x} \bu_x \bu_y \rho_{y} \big>, \nonumber\\
\big< n_{x,1}n_{y,1} \big> &= \big< \rho_{x}\big( 1-\bu_{x}^2 \big)\big(
1-\bu_{y}^2 \big)\rho_{y}\big> .
\end{align}
\rFIGURE{
\input{stereo_mfLmbL}
\caption{\label{fig:sigma:mfLmbL} Comparison of
bosonic and fermionic mass in units of the box size for three different lattice
sizes using Wilson fermions. The dotted line denotes the case $m_\text{F}=m_\text{B}$. }}
For lattice volumes up to $32^2$ we obtain a discrepancy between bosonic and fermionic masses
with lighter bosons, see Figure \ref{fig:sigma:mfLmbL}. This gap increases for larger
lattices and there is no indication of a degeneracy in the continuum limit, so that supersymmetry 
is seemingly not restored. This is not too surprising, though, since Wilson fermions may 
break supersymmetry in a way that is not dissolved in the continuum limit.\\
In section \ref{ch:exactsusy} we showed that it is not possible to construct a discretized action 
which simultaneously respects at least one exact supersymmetry as well as the O$(3)$ symmetry. 
An improved action thus relies on adding fine tuning terms as a compensation for renormalized 
couplings that arise from symmetry breaking terms on the lattice.

%% file: sigmaMassComparisonDirect.tex
\begingroup
\footnotesize
  \makeatletter
  \providecommand\color[2][]{%
    \GenericError{(gnuplot) \space\space\space\@spaces}{%
      Package color not loaded in conjunction with
      terminal option `colourtext'%
    }{See the gnuplot documentation for explanation.%
    }{Either use 'blacktext' in gnuplot or load the package
      color.sty in LaTeX.}%
    \renewcommand\color[2][]{}%
  }%
  \providecommand\includegraphics[2][]{%
    \GenericError{(gnuplot) \space\space\space\@spaces}{%
      Package graphicx or graphics not loaded%
    }{See the gnuplot documentation for explanation.%
    }{The gnuplot epslatex terminal needs graphicx.sty or graphics.sty.}%
    \renewcommand\includegraphics[2][]{}%
  }%
  \providecommand\rotatebox[2]{#2}%
  \@ifundefined{ifGPcolor}{%
    \newif\ifGPcolor
    \GPcolortrue
  }{}%
  \@ifundefined{ifGPblacktext}{%
    \newif\ifGPblacktext
    \GPblacktexttrue
  }{}%
  \let\gplgaddtomacro\g@addto@macro
  \gdef\gplbacktext{}%
  \gdef\gplfronttext{}%
  \makeatother
  \ifGPblacktext
    \def\colorrgb#1{}%
    \def\colorgray#1{}%
  \else
    \ifGPcolor
      \def\colorrgb#1{\color[rgb]{#1}}%
      \def\colorgray#1{\color[gray]{#1}}%
      \expandafter\def\csname LTw\endcsname{\color{white}}%
      \expandafter\def\csname LTb\endcsname{\color{black}}%
      \expandafter\def\csname LTa\endcsname{\color{black}}%
      \expandafter\def\csname LT0\endcsname{\color[rgb]{1,0,0}}%
      \expandafter\def\csname LT1\endcsname{\color[rgb]{0,1,0}}%
      \expandafter\def\csname LT2\endcsname{\color[rgb]{0,0,1}}%
      \expandafter\def\csname LT3\endcsname{\color[rgb]{1,0,1}}%
      \expandafter\def\csname LT4\endcsname{\color[rgb]{0,1,1}}%
      \expandafter\def\csname LT5\endcsname{\color[rgb]{1,1,0}}%
      \expandafter\def\csname LT6\endcsname{\color[rgb]{0,0,0}}%
      \expandafter\def\csname LT7\endcsname{\color[rgb]{1,0.3,0}}%
      \expandafter\def\csname LT8\endcsname{\color[rgb]{0.5,0.5,0.5}}%
    \else
      \def\colorrgb#1{\color{black}}%
      \def\colorgray#1{\color[gray]{#1}}%
      \expandafter\def\csname LTw\endcsname{\color{white}}%
      \expandafter\def\csname LTb\endcsname{\color{black}}%
      \expandafter\def\csname LTa\endcsname{\color{black}}%
      \expandafter\def\csname LT0\endcsname{\color{black}}%
      \expandafter\def\csname LT1\endcsname{\color{black}}%
      \expandafter\def\csname LT2\endcsname{\color{black}}%
      \expandafter\def\csname LT3\endcsname{\color{black}}%
      \expandafter\def\csname LT4\endcsname{\color{black}}%
      \expandafter\def\csname LT5\endcsname{\color{black}}%
      \expandafter\def\csname LT6\endcsname{\color{black}}%
      \expandafter\def\csname LT7\endcsname{\color{black}}%
      \expandafter\def\csname LT8\endcsname{\color{black}}%
    \fi
  \fi
  \setlength{\unitlength}{0.0500bp}%
  \begin{picture}(3968.00,2976.00)%
    \gplgaddtomacro\gplbacktext{%
      \csname LTb\endcsname%
      \put(768,480){\makebox(0,0)[r]{\strut{}$0$}}%
      \put(768,830){\makebox(0,0)[r]{\strut{}$2$}}%
      \put(768,1179){\makebox(0,0)[r]{\strut{}$4$}}%
      \put(768,1529){\makebox(0,0)[r]{\strut{}$6$}}%
      \put(768,1878){\makebox(0,0)[r]{\strut{}$8$}}%
      \put(768,2228){\makebox(0,0)[r]{\strut{}$10$}}%
      \put(768,2577){\makebox(0,0)[r]{\strut{}$12$}}%
      \put(768,2927){\makebox(0,0)[r]{\strut{}$14$}}%
      \put(864,320){\makebox(0,0){\strut{}$0$}}%
      \put(1303,320){\makebox(0,0){\strut{}$2$}}%
      \put(1743,320){\makebox(0,0){\strut{}$4$}}%
      \put(2182,320){\makebox(0,0){\strut{}$6$}}%
      \put(2621,320){\makebox(0,0){\strut{}$8$}}%
      \put(3060,320){\makebox(0,0){\strut{}$10$}}%
      \put(3500,320){\makebox(0,0){\strut{}$12$}}%
      \put(3939,320){\makebox(0,0){\strut{}$14$}}%
      \put(352,1703){\makebox(0,0){\strut{}$m_\text{B}L$}}%
      \put(2401,80){\makebox(0,0){\strut{}$m_\text{F}L$}}%
    }%
    \gplgaddtomacro\gplfronttext{%
      \csname LTb\endcsname%
      \put(1632,2784){\makebox(0,0)[r]{\strut{}$N=5^2$}}%
      \csname LTb\endcsname%
      \put(1632,2624){\makebox(0,0)[r]{\strut{}$N=7^2$}}%
      \csname LTb\endcsname%
      \put(1632,2464){\makebox(0,0)[r]{\strut{}$N=9^2$}}%
    }%
    \gplbacktext
    \put(0,0){\includegraphics{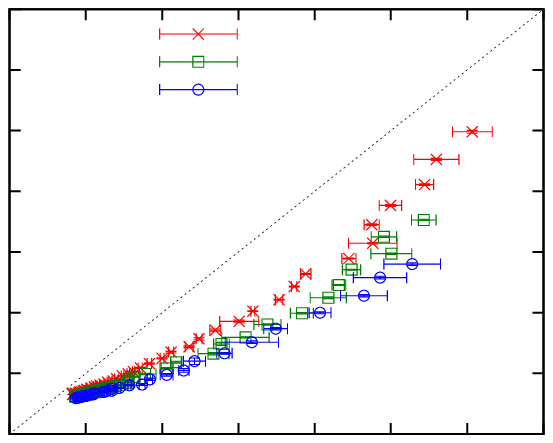}}%
    \gplfronttext
  \end{picture}%
\endgroup

%% file: sigmaMassComparisonRatio.tex
\begingroup
\footnotesize
  \makeatletter
  \providecommand\color[2][]{%
    \GenericError{(gnuplot) \space\space\space\@spaces}{%
      Package color not loaded in conjunction with
      terminal option `colourtext'%
    }{See the gnuplot documentation for explanation.%
    }{Either use 'blacktext' in gnuplot or load the package
      color.sty in LaTeX.}%
    \renewcommand\color[2][]{}%
  }%
  \providecommand\includegraphics[2][]{%
    \GenericError{(gnuplot) \space\space\space\@spaces}{%
      Package graphicx or graphics not loaded%
    }{See the gnuplot documentation for explanation.%
    }{The gnuplot epslatex terminal needs graphicx.sty or graphics.sty.}%
    \renewcommand\includegraphics[2][]{}%
  }%
  \providecommand\rotatebox[2]{#2}%
  \@ifundefined{ifGPcolor}{%
    \newif\ifGPcolor
    \GPcolortrue
  }{}%
  \@ifundefined{ifGPblacktext}{%
    \newif\ifGPblacktext
    \GPblacktexttrue
  }{}%
  \let\gplgaddtomacro\g@addto@macro
  \gdef\gplbacktext{}%
  \gdef\gplfronttext{}%
  \makeatother
  \ifGPblacktext
    \def\colorrgb#1{}%
    \def\colorgray#1{}%
  \else
    \ifGPcolor
      \def\colorrgb#1{\color[rgb]{#1}}%
      \def\colorgray#1{\color[gray]{#1}}%
      \expandafter\def\csname LTw\endcsname{\color{white}}%
      \expandafter\def\csname LTb\endcsname{\color{black}}%
      \expandafter\def\csname LTa\endcsname{\color{black}}%
      \expandafter\def\csname LT0\endcsname{\color[rgb]{1,0,0}}%
      \expandafter\def\csname LT1\endcsname{\color[rgb]{0,1,0}}%
      \expandafter\def\csname LT2\endcsname{\color[rgb]{0,0,1}}%
      \expandafter\def\csname LT3\endcsname{\color[rgb]{1,0,1}}%
      \expandafter\def\csname LT4\endcsname{\color[rgb]{0,1,1}}%
      \expandafter\def\csname LT5\endcsname{\color[rgb]{1,1,0}}%
      \expandafter\def\csname LT6\endcsname{\color[rgb]{0,0,0}}%
      \expandafter\def\csname LT7\endcsname{\color[rgb]{1,0.3,0}}%
      \expandafter\def\csname LT8\endcsname{\color[rgb]{0.5,0.5,0.5}}%
    \else
      \def\colorrgb#1{\color{black}}%
      \def\colorgray#1{\color[gray]{#1}}%
      \expandafter\def\csname LTw\endcsname{\color{white}}%
      \expandafter\def\csname LTb\endcsname{\color{black}}%
      \expandafter\def\csname LTa\endcsname{\color{black}}%
      \expandafter\def\csname LT0\endcsname{\color{black}}%
      \expandafter\def\csname LT1\endcsname{\color{black}}%
      \expandafter\def\csname LT2\endcsname{\color{black}}%
      \expandafter\def\csname LT3\endcsname{\color{black}}%
      \expandafter\def\csname LT4\endcsname{\color{black}}%
      \expandafter\def\csname LT5\endcsname{\color{black}}%
      \expandafter\def\csname LT6\endcsname{\color{black}}%
      \expandafter\def\csname LT7\endcsname{\color{black}}%
      \expandafter\def\csname LT8\endcsname{\color{black}}%
    \fi
  \fi
  \setlength{\unitlength}{0.0500bp}%
  \begin{picture}(3968.00,2976.00)%
    \gplgaddtomacro\gplbacktext{%
      \csname LTb\endcsname%
      \put(768,480){\makebox(0,0)[r]{\strut{}$-4.0$}}%
      \put(768,786){\makebox(0,0)[r]{\strut{}$-3.5$}}%
      \put(768,1092){\makebox(0,0)[r]{\strut{}$-3.0$}}%
      \put(768,1398){\makebox(0,0)[r]{\strut{}$-2.5$}}%
      \put(768,1704){\makebox(0,0)[r]{\strut{}$-2.0$}}%
      \put(768,2009){\makebox(0,0)[r]{\strut{}$-1.5$}}%
      \put(768,2315){\makebox(0,0)[r]{\strut{}$-1.0$}}%
      \put(768,2621){\makebox(0,0)[r]{\strut{}$-0.5$}}%
      \put(768,2927){\makebox(0,0)[r]{\strut{}$0.0$}}%
      \put(864,320){\makebox(0,0){\strut{}$4$}}%
      \put(1479,320){\makebox(0,0){\strut{}$6$}}%
      \put(2094,320){\makebox(0,0){\strut{}$8$}}%
      \put(2709,320){\makebox(0,0){\strut{}$10$}}%
      \put(3324,320){\makebox(0,0){\strut{}$12$}}%
      \put(3939,320){\makebox(0,0){\strut{}$14$}}%
      \put(160,1703){\rotatebox{90}{\makebox(0,0){\strut{}$m_\text{B}L-m_\text{F}L$}}}%
      \put(2401,80){\makebox(0,0){\strut{}$m_\text{F}L$}}%
    }%
    \gplgaddtomacro\gplfronttext{%
    }%
    \gplbacktext
    \put(0,0){\includegraphics{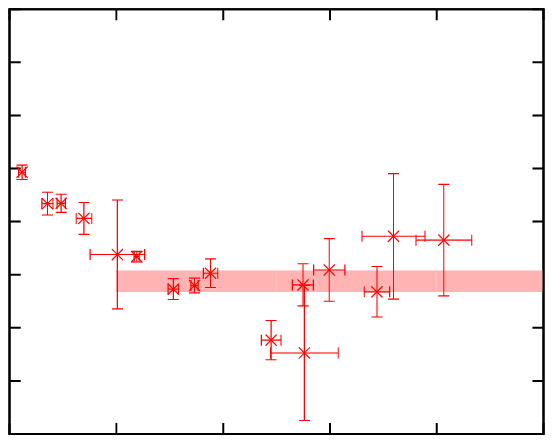}}%
    \gplfronttext
  \end{picture}%
\endgroup

%% file: stereo_mfLmbL.tex
\begingroup
\footnotesize
  \makeatletter
  \providecommand\color[2][]{%
    \GenericError{(gnuplot) \space\space\space\@spaces}{%
      Package color not loaded in conjunction with
      terminal option `colourtext'%
    }{See the gnuplot documentation for explanation.%
    }{Either use 'blacktext' in gnuplot or load the package
      color.sty in LaTeX.}%
    \renewcommand\color[2][]{}%
  }%
  \providecommand\includegraphics[2][]{%
    \GenericError{(gnuplot) \space\space\space\@spaces}{%
      Package graphicx or graphics not loaded%
    }{See the gnuplot documentation for explanation.%
    }{The gnuplot epslatex terminal needs graphicx.sty or graphics.sty.}%
    \renewcommand\includegraphics[2][]{}%
  }%
  \providecommand\rotatebox[2]{#2}%
  \@ifundefined{ifGPcolor}{%
    \newif\ifGPcolor
    \GPcolortrue
  }{}%
  \@ifundefined{ifGPblacktext}{%
    \newif\ifGPblacktext
    \GPblacktexttrue
  }{}%
  \let\gplgaddtomacro\g@addto@macro
  \gdef\gplbacktext{}%
  \gdef\gplfronttext{}%
  \makeatother
  \ifGPblacktext
    \def\colorrgb#1{}%
    \def\colorgray#1{}%
  \else
    \ifGPcolor
      \def\colorrgb#1{\color[rgb]{#1}}%
      \def\colorgray#1{\color[gray]{#1}}%
      \expandafter\def\csname LTw\endcsname{\color{white}}%
      \expandafter\def\csname LTb\endcsname{\color{black}}%
      \expandafter\def\csname LTa\endcsname{\color{black}}%
      \expandafter\def\csname LT0\endcsname{\color[rgb]{1,0,0}}%
      \expandafter\def\csname LT1\endcsname{\color[rgb]{0,1,0}}%
      \expandafter\def\csname LT2\endcsname{\color[rgb]{0,0,1}}%
      \expandafter\def\csname LT3\endcsname{\color[rgb]{1,0,1}}%
      \expandafter\def\csname LT4\endcsname{\color[rgb]{0,1,1}}%
      \expandafter\def\csname LT5\endcsname{\color[rgb]{1,1,0}}%
      \expandafter\def\csname LT6\endcsname{\color[rgb]{0,0,0}}%
      \expandafter\def\csname LT7\endcsname{\color[rgb]{1,0.3,0}}%
      \expandafter\def\csname LT8\endcsname{\color[rgb]{0.5,0.5,0.5}}%
    \else
      \def\colorrgb#1{\color{black}}%
      \def\colorgray#1{\color[gray]{#1}}%
      \expandafter\def\csname LTw\endcsname{\color{white}}%
      \expandafter\def\csname LTb\endcsname{\color{black}}%
      \expandafter\def\csname LTa\endcsname{\color{black}}%
      \expandafter\def\csname LT0\endcsname{\color{black}}%
      \expandafter\def\csname LT1\endcsname{\color{black}}%
      \expandafter\def\csname LT2\endcsname{\color{black}}%
      \expandafter\def\csname LT3\endcsname{\color{black}}%
      \expandafter\def\csname LT4\endcsname{\color{black}}%
      \expandafter\def\csname LT5\endcsname{\color{black}}%
      \expandafter\def\csname LT6\endcsname{\color{black}}%
      \expandafter\def\csname LT7\endcsname{\color{black}}%
      \expandafter\def\csname LT8\endcsname{\color{black}}%
    \fi
  \fi
  \setlength{\unitlength}{0.0500bp}%
  \begin{picture}(3968.00,2976.00)%
    \gplgaddtomacro\gplbacktext{%
      \csname LTb\endcsname%
      \put(768,480){\makebox(0,0)[r]{\strut{}$0$}}%
      \put(768,1092){\makebox(0,0)[r]{\strut{}$5$}}%
      \put(768,1704){\makebox(0,0)[r]{\strut{}$10$}}%
      \put(768,2315){\makebox(0,0)[r]{\strut{}$15$}}%
      \put(768,2927){\makebox(0,0)[r]{\strut{}$20$}}%
      \put(864,320){\makebox(0,0){\strut{}$0$}}%
      \put(1377,320){\makebox(0,0){\strut{}$5$}}%
      \put(1889,320){\makebox(0,0){\strut{}$10$}}%
      \put(2402,320){\makebox(0,0){\strut{}$15$}}%
      \put(2914,320){\makebox(0,0){\strut{}$20$}}%
      \put(3427,320){\makebox(0,0){\strut{}$25$}}%
      \put(3939,320){\makebox(0,0){\strut{}$30$}}%
      \put(352,1703){\makebox(0,0){\strut{}$m_\text{B}L$}}%
      \put(2401,80){\makebox(0,0){\strut{}$m_\text{F}L$}}%
    }%
    \gplgaddtomacro\gplfronttext{%
      \csname LTb\endcsname%
      \put(1728,2784){\makebox(0,0)[r]{\strut{}$N=8^2$}}%
      \csname LTb\endcsname%
      \put(1728,2624){\makebox(0,0)[r]{\strut{}$N=16^2$}}%
      \csname LTb\endcsname%
      \put(1728,2464){\makebox(0,0)[r]{\strut{}$N=32^2$}}%
    }%
    \gplbacktext
    \put(0,0){\includegraphics{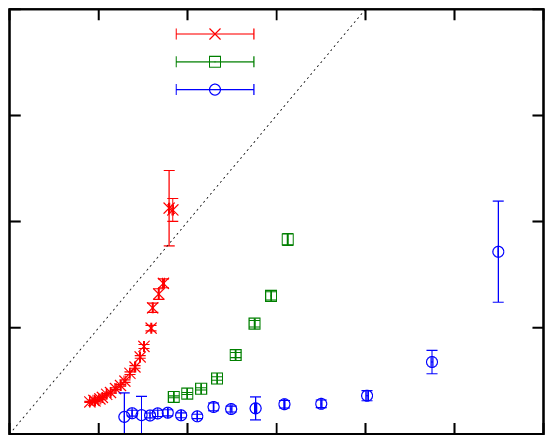}}%
    \gplfronttext
  \end{picture}%
\endgroup

%% file: finetuning.tex
\subsection{Fine tuning of the Wilson derivative}

From $\Neqone$ Super Yang-Mills theories it is known that the correct continuum limit may be
achieved by a fine tuning term that resembles an explicit fermionic mass
such that the renormalized gluino mass is zero \cite{Donini:1997}.
A similar procedure will be performed in the present case by
deforming the fermionic derivative using a fine tuning mass $m$,
\begin{equation}
M_{xy}^{\alpha\beta}=\gamma_\mu^{\alpha\beta}\big(\partial_{\mu}^{sym}\big)_{xy} 
+\delta^{\alpha\beta}\frac{ra}{2}\Delta_{xy}
+\delta^{\alpha\beta}m\delta_{xy},
\end{equation}
which enters the Hopping parameter $\kappa=(4+2m)^{-1}$.
Additional degrees of freedom in the fine tuning procedure increase the numerical
effort considerably, such that a careful choice of tuning parameters is necessary in 
order to keep the RHMC algorithm exact and efficient. In particular, for different $\kappa$,
we have monitored the sign of the Pfaffian determinant as well as the spectrum 
of the Dirac operator, which is approximated by rational functions \cite{Clark:2006fx}.
To guide our efforts, we measure the chiral 
condensate, bosonic and fermionic masses as well as the bosonic action for several values
of the fine tuning parameter.\\
The chiral condensate is extracted from the trace of the projected correlator, as described 
in \eqref{stereocond}, while the sign of the Pfaffian is taken into account by a reweighting procedure,
\begin{equation}
	\big< \bar{\psi}\psi \big> = \frac{\big< \sgn \Pf \bar{\psi}\psi\big>_{q}}{\big< \sgn \Pf \big>_{q}}.
\end{equation}
Here, $\langle ... \rangle_{q}$ denotes the sign-quenched ensemble. For the Wilson derivative we 
expect that one of the ground state energies is raised due
to the explicit breaking of chiral symmetry, so that we get fluctuations
around one of the minima in Figure \ref{fig:sigma:chiralCondensate} only. Switching on 
the fine tuning mass, the condensate is driven to larger values, showing a jump
at some distinct point (see Figure \ref{fig:sigma:cc_tuned}, left panel). 
\FIGURE{\hfill
\input{stereo_chiralCondensate_finetuned}\hfill
\input{stereo_chiralCondensate_histogram}\hfill\hfill\phantom.
\caption{\label{fig:sigma:cc_tuned} Left Panel: Expectation value of the chiral condensate $a\Xi$ for lattice volume $16^2$ and
different couplings $g^{-2}$. A jump corresponding to a first order transition is visible. Right Panel: Histograms of the 
chiral condensate $a\Xi$ for different values of the Hopping parameter ($N=24^2$, $g^{-2}=2$). }}
\noindent To gain a better understanding of this behaviour, 
we utilize histograms of the chiral condensate to measure the distribution function $\rho(a\Xi)$,
which is formally obtained by introducing a delta function into the partition sum,
\begin{equation*}
	\rho(a\Xi) = \frac{1}{Z}\int \cD\bn\cD\bpsi\;\delta\!\left(a\Xi - \ii\bar\bpsi\bpsi\right)\;e^{-S} 
\longrightarrow \frac{1}{M} 
	\sum_{i=1}^{M} \delta\!\left(a\Xi - \ii(\bar\bpsi\bpsi)_{i}\right).
\end{equation*}
It is henceforth possible to express the expectation value for the chiral condensate using this quantity as 
$\langle \ii\bar\bpsi\bpsi \rangle = \sum_{i=1}^M  a\Xi_i\rho( a\Xi_i) / \sum_{i=1}^M \rho(a\Xi_i)$. Simulating
the sign-quenched ensemble, we need to use the reweighted distribution function,
\begin{equation}
\rho(a\Xi) = \frac{1}{M} \sum_{i=1}^M \delta\left(a\Xi - \ii(\bar\bpsi\bpsi)_{i}\right)
\frac{\sgn \Pf_i}{\langle \sgn \Pf \rangle_q},
\end{equation}
which may no longer be interpreted as a probability distribution,
since $\sgn\Pf$ can be negative. 
\rFIGURE{\hfill
\input{stereo_critical_kappa}\hfill
\caption{\label{fig:sigma:kappa}Critical value of the Hopping parameter 
for several lattice sizes.}}
In contrast to the case of SLAC fermions, where no problem arises due to
intact $\Z_{2}$-symmetry, the Wilson prescription leads to an additive 
renormalization, such that configurations
with negative sign yield entries corresponding to a sign-flipped 
renormalization constant. To avoid this behaviour, we will omit the sign 
information when considering histograms, seeing 
that the Wilson ensemble shows frequent sign fluctuations only in the 
direct vicinity of the critical Hopping parameter and therefore 
allows to extrapolate from a region where omitting the sign is safe. 
Expectation values are however \emph{always} determined using the 
reweighting procedure and are hence not affected by this approximation. 
By an appropriate choice of the tuning parameter 
we obtain the spontaneously broken signature that is 
expected in the continuum limit (see Figure \ref{fig:sigma:cc_tuned}, 
right panel), 
modified by the additive renormalization. The point of steep increase is 
identified as a first order 
phase transition in the chiral condensate, which is analogous to 
the case of Super-Yang-Mills theories. Using this signature, we have 
determined the critical value 
of the fine tuning parameter $\kappa$
for lattice sizes $8^2$, $16^2$ and $24^2$ and coupling 
$g^{-2}=1\ldots 2$, see Figure \ref{fig:sigma:kappa}.

\noindent Regarding the masses of the elementary excitations, we see that the bosonic correlator is not affected by 
the fine tuning procedure and the bosonic mass takes a constant value within error bars. For the fermionic mass, 
we observe linear scaling
behaviour for $\kappa < \kappa_c$ and $\kappa > \kappa_c$, however in the vicinity of $\kappa\approx\kappa_c$, the 
scaling breaks down (Figure \ref{fig:sigma:finetuning}, left panel). Therefore, a reliable extrapolation of the 
fermionic mass based on small values of $\kappa$ is not possible, at least for the lattice volumes considered. We will
hence utilize the bosonic mass to fix the physical box size $m_{B}L$ for our investigations involving fine tuning. In
contrast to e.g. $\Neqone$ Super Yang-Mills \cite{Donini:1997}, simulations at the critical point are feasible 
since the theory inhibits
a finite mass gap even in the continuum.
\FIGURE{\hfill
\input{stereo_finetuning_linear_16x16}\hfill
\input{stereo_finetuned_mfLmbL}\hfill\hfill\phantom.
\caption{\label{fig:sigma:finetuning} Left Panel: Scaling behaviour of the bosonic and fermionic mass 
for $N=16^{2}$ and $g^{-2}=1.4$ with $\kappa_{c}=0.382(1)$, marked by the black dot. The dashed line denotes the lattice cutoff of $1/16$. 
Right Panel: Comparison of bosonic and fermionic masses in units of the box size for three different lattice
sizes at $\kappa=\kappa_{c}$. }}
In the vicinity of the phase transition, mixing between the
two ground states occurs and the $\ii\bar{\psi}\psi$ correlator must be projected onto one of the ground 
states, analogously to the SLAC case (see chapter \ref{ch:fmasses}). However, since we may only determine the 
additive renormalization constant
from the chiral condensate distribution function up to finite precision, fermion mass extraction is 
affected by a systematic error which 
comes from configurations with close to vanishing renormalized chiral condensate that are 
erroneously weighted as belonging to the false ground state. 
This error is negligible for large lattices and $g^{-2}$ though, since these configurations 
are suppressed by the finite tunneling probability between both ground states. We observe 
a great improvement regarding the 
degeneracy of the masses (Figure \ref{fig:sigma:finetuning}, right panel), which is expected in the continuum limit from 
supersymmetry restoration. In particular,
the fermionic masses no longer "run away" if the volume is increased, 
which hints at a proper cancellation of 
the divergent operator causing these problems (compare to Figure 
\ref{fig:sigma:mfLmbL}). Nevertheless, a true proof
of this conjecture may only be provided by a study of all divergent 
operators based on lattice perturbation theory, 
which is not pursued here. For finite box sizes of $m_FL>2$ a thermal
mass-splitting similar to the SLAC case seems to emerge, however showing
a slightly smaller gap. To explore this region, further large volume 
simulations would be needed in order to suppress lattice artifacts.


%% file: stereo_chiralCondensate_finetuned.tex
\begingroup
\footnotesize
  \makeatletter
  \providecommand\color[2][]{%
    \GenericError{(gnuplot) \space\space\space\@spaces}{%
      Package color not loaded in conjunction with
      terminal option `colourtext'%
    }{See the gnuplot documentation for explanation.%
    }{Either use 'blacktext' in gnuplot or load the package
      color.sty in LaTeX.}%
    \renewcommand\color[2][]{}%
  }%
  \providecommand\includegraphics[2][]{%
    \GenericError{(gnuplot) \space\space\space\@spaces}{%
      Package graphicx or graphics not loaded%
    }{See the gnuplot documentation for explanation.%
    }{The gnuplot epslatex terminal needs graphicx.sty or graphics.sty.}%
    \renewcommand\includegraphics[2][]{}%
  }%
  \providecommand\rotatebox[2]{#2}%
  \@ifundefined{ifGPcolor}{%
    \newif\ifGPcolor
    \GPcolortrue
  }{}%
  \@ifundefined{ifGPblacktext}{%
    \newif\ifGPblacktext
    \GPblacktexttrue
  }{}%
  \let\gplgaddtomacro\g@addto@macro
  \gdef\gplbacktext{}%
  \gdef\gplfronttext{}%
  \makeatother
  \ifGPblacktext
    \def\colorrgb#1{}%
    \def\colorgray#1{}%
  \else
    \ifGPcolor
      \def\colorrgb#1{\color[rgb]{#1}}%
      \def\colorgray#1{\color[gray]{#1}}%
      \expandafter\def\csname LTw\endcsname{\color{white}}%
      \expandafter\def\csname LTb\endcsname{\color{black}}%
      \expandafter\def\csname LTa\endcsname{\color{black}}%
      \expandafter\def\csname LT0\endcsname{\color[rgb]{1,0,0}}%
      \expandafter\def\csname LT1\endcsname{\color[rgb]{0,1,0}}%
      \expandafter\def\csname LT2\endcsname{\color[rgb]{0,0,1}}%
      \expandafter\def\csname LT3\endcsname{\color[rgb]{1,0,1}}%
      \expandafter\def\csname LT4\endcsname{\color[rgb]{0,1,1}}%
      \expandafter\def\csname LT5\endcsname{\color[rgb]{1,1,0}}%
      \expandafter\def\csname LT6\endcsname{\color[rgb]{0,0,0}}%
      \expandafter\def\csname LT7\endcsname{\color[rgb]{1,0.3,0}}%
      \expandafter\def\csname LT8\endcsname{\color[rgb]{0.5,0.5,0.5}}%
    \else
      \def\colorrgb#1{\color{black}}%
      \def\colorgray#1{\color[gray]{#1}}%
      \expandafter\def\csname LTw\endcsname{\color{white}}%
      \expandafter\def\csname LTb\endcsname{\color{black}}%
      \expandafter\def\csname LTa\endcsname{\color{black}}%
      \expandafter\def\csname LT0\endcsname{\color{black}}%
      \expandafter\def\csname LT1\endcsname{\color{black}}%
      \expandafter\def\csname LT2\endcsname{\color{black}}%
      \expandafter\def\csname LT3\endcsname{\color{black}}%
      \expandafter\def\csname LT4\endcsname{\color{black}}%
      \expandafter\def\csname LT5\endcsname{\color{black}}%
      \expandafter\def\csname LT6\endcsname{\color{black}}%
      \expandafter\def\csname LT7\endcsname{\color{black}}%
      \expandafter\def\csname LT8\endcsname{\color{black}}%
    \fi
  \fi
  \setlength{\unitlength}{0.0500bp}%
  \begin{picture}(3968.00,2976.00)%
    \gplgaddtomacro\gplbacktext{%
      \csname LTb\endcsname%
      \put(768,480){\makebox(0,0)[r]{\strut{}$-0.6$}}%
      \put(768,969){\makebox(0,0)[r]{\strut{}$-0.5$}}%
      \put(768,1459){\makebox(0,0)[r]{\strut{}$-0.4$}}%
      \put(768,1948){\makebox(0,0)[r]{\strut{}$-0.3$}}%
      \put(768,2438){\makebox(0,0)[r]{\strut{}$-0.2$}}%
      \put(768,2927){\makebox(0,0)[r]{\strut{}$-0.1$}}%
      \put(864,320){\makebox(0,0){\strut{}$0.25$}}%
      \put(1479,320){\makebox(0,0){\strut{}$0.30$}}%
      \put(2094,320){\makebox(0,0){\strut{}$0.35$}}%
      \put(2709,320){\makebox(0,0){\strut{}$0.40$}}%
      \put(3324,320){\makebox(0,0){\strut{}$0.45$}}%
      \put(3939,320){\makebox(0,0){\strut{}$0.50$}}%
      \put(160,1703){\makebox(0,0){\strut{}$a\Xi$}}%
      \put(2401,80){\makebox(0,0){\strut{}$\kappa$}}%
    }%
    \gplgaddtomacro\gplfronttext{%
      \csname LTb\endcsname%
      \put(3204,2704){\makebox(0,0)[r]{\strut{}$g^{-2}=1.2$}}%
      \csname LTb\endcsname%
      \put(3204,2544){\makebox(0,0)[r]{\strut{}$g^{-2}=1.3$}}%
      \csname LTb\endcsname%
      \put(3204,2384){\makebox(0,0)[r]{\strut{}$g^{-2}=1.4$}}%
      \csname LTb\endcsname%
      \put(3204,2224){\makebox(0,0)[r]{\strut{}$g^{-2}=2.0$}}%
    }%
    \gplbacktext
    \put(0,0){\includegraphics{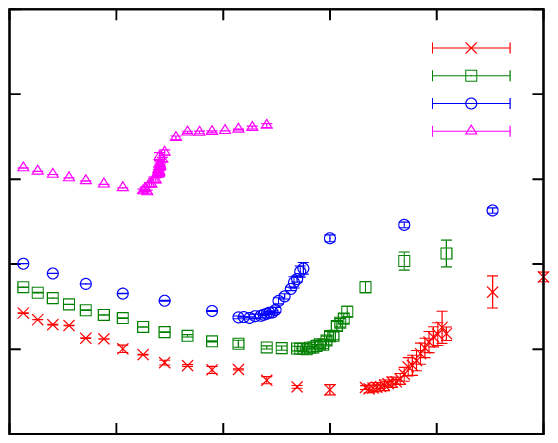}}%
    \gplfronttext
  \end{picture}%
\endgroup

%% file: stereo_chiralCondensate_histogram.tex
\begingroup
\footnotesize
  \makeatletter
  \providecommand\color[2][]{%
    \GenericError{(gnuplot) \space\space\space\@spaces}{%
      Package color not loaded in conjunction with
      terminal option `colourtext'%
    }{See the gnuplot documentation for explanation.%
    }{Either use 'blacktext' in gnuplot or load the package
      color.sty in LaTeX.}%
    \renewcommand\color[2][]{}%
  }%
  \providecommand\includegraphics[2][]{%
    \GenericError{(gnuplot) \space\space\space\@spaces}{%
      Package graphicx or graphics not loaded%
    }{See the gnuplot documentation for explanation.%
    }{The gnuplot epslatex terminal needs graphicx.sty or graphics.sty.}%
    \renewcommand\includegraphics[2][]{}%
  }%
  \providecommand\rotatebox[2]{#2}%
  \@ifundefined{ifGPcolor}{%
    \newif\ifGPcolor
    \GPcolortrue
  }{}%
  \@ifundefined{ifGPblacktext}{%
    \newif\ifGPblacktext
    \GPblacktexttrue
  }{}%
  \let\gplgaddtomacro\g@addto@macro
  \gdef\gplbacktext{}%
  \gdef\gplfronttext{}%
  \makeatother
  \ifGPblacktext
    \def\colorrgb#1{}%
    \def\colorgray#1{}%
  \else
    \ifGPcolor
      \def\colorrgb#1{\color[rgb]{#1}}%
      \def\colorgray#1{\color[gray]{#1}}%
      \expandafter\def\csname LTw\endcsname{\color{white}}%
      \expandafter\def\csname LTb\endcsname{\color{black}}%
      \expandafter\def\csname LTa\endcsname{\color{black}}%
      \expandafter\def\csname LT0\endcsname{\color[rgb]{1,0,0}}%
      \expandafter\def\csname LT1\endcsname{\color[rgb]{0,1,0}}%
      \expandafter\def\csname LT2\endcsname{\color[rgb]{0,0,1}}%
      \expandafter\def\csname LT3\endcsname{\color[rgb]{1,0,1}}%
      \expandafter\def\csname LT4\endcsname{\color[rgb]{0,1,1}}%
      \expandafter\def\csname LT5\endcsname{\color[rgb]{1,1,0}}%
      \expandafter\def\csname LT6\endcsname{\color[rgb]{0,0,0}}%
      \expandafter\def\csname LT7\endcsname{\color[rgb]{1,0.3,0}}%
      \expandafter\def\csname LT8\endcsname{\color[rgb]{0.5,0.5,0.5}}%
    \else
      \def\colorrgb#1{\color{black}}%
      \def\colorgray#1{\color[gray]{#1}}%
      \expandafter\def\csname LTw\endcsname{\color{white}}%
      \expandafter\def\csname LTb\endcsname{\color{black}}%
      \expandafter\def\csname LTa\endcsname{\color{black}}%
      \expandafter\def\csname LT0\endcsname{\color{black}}%
      \expandafter\def\csname LT1\endcsname{\color{black}}%
      \expandafter\def\csname LT2\endcsname{\color{black}}%
      \expandafter\def\csname LT3\endcsname{\color{black}}%
      \expandafter\def\csname LT4\endcsname{\color{black}}%
      \expandafter\def\csname LT5\endcsname{\color{black}}%
      \expandafter\def\csname LT6\endcsname{\color{black}}%
      \expandafter\def\csname LT7\endcsname{\color{black}}%
      \expandafter\def\csname LT8\endcsname{\color{black}}%
    \fi
  \fi
  \setlength{\unitlength}{0.0500bp}%
  \begin{picture}(3968.00,2976.00)%
    \gplgaddtomacro\gplbacktext{%
      \csname LTb\endcsname%
      \put(768,480){\makebox(0,0)[r]{\strut{}$0.00$}}%
      \put(768,830){\makebox(0,0)[r]{\strut{}$0.02$}}%
      \put(768,1179){\makebox(0,0)[r]{\strut{}$0.04$}}%
      \put(768,1529){\makebox(0,0)[r]{\strut{}$0.06$}}%
      \put(768,1878){\makebox(0,0)[r]{\strut{}$0.08$}}%
      \put(768,2228){\makebox(0,0)[r]{\strut{}$0.10$}}%
      \put(768,2577){\makebox(0,0)[r]{\strut{}$0.12$}}%
      \put(768,2927){\makebox(0,0)[r]{\strut{}$0.14$}}%
      \put(1120,320){\makebox(0,0){\strut{}$-0.34$}}%
      \put(1633,320){\makebox(0,0){\strut{}$-0.32$}}%
      \put(2145,320){\makebox(0,0){\strut{}$-0.30$}}%
      \put(2658,320){\makebox(0,0){\strut{}$-0.28$}}%
      \put(3170,320){\makebox(0,0){\strut{}$-0.26$}}%
      \put(3683,320){\makebox(0,0){\strut{}$-0.24$}}%
      \put(160,1703){\makebox(0,0){\strut{}$\rho(a\Xi)$}}%
      \put(2401,80){\makebox(0,0){\strut{}$a\Xi$}}%
    }%
    \gplgaddtomacro\gplfronttext{%
      \csname LTb\endcsname%
      \put(1920,2704){\makebox(0,0)[r]{\strut{}$\kappa=0.3205$}}%
      \csname LTb\endcsname%
      \put(1920,2544){\makebox(0,0)[r]{\strut{}$\kappa=0.3215$}}%
      \csname LTb\endcsname%
      \put(1920,2384){\makebox(0,0)[r]{\strut{}$\kappa=0.3225$}}%
    }%
    \gplbacktext
    \put(0,0){\includegraphics{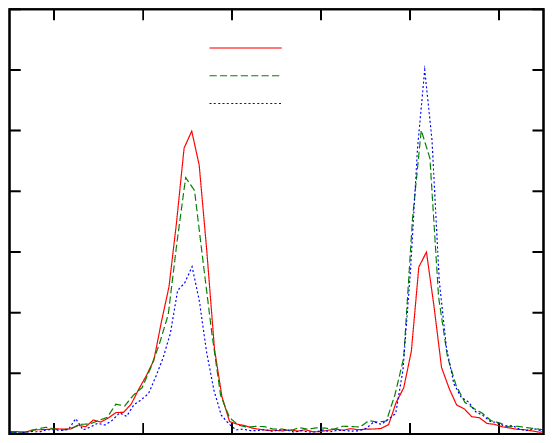}}%
    \gplfronttext
  \end{picture}%
\endgroup

%% file: stereo_critical_kappa.tex
\begingroup
\footnotesize
  \makeatletter
  \providecommand\color[2][]{%
    \GenericError{(gnuplot) \space\space\space\@spaces}{%
      Package color not loaded in conjunction with
      terminal option `colourtext'%
    }{See the gnuplot documentation for explanation.%
    }{Either use 'blacktext' in gnuplot or load the package
      color.sty in LaTeX.}%
    \renewcommand\color[2][]{}%
  }%
  \providecommand\includegraphics[2][]{%
    \GenericError{(gnuplot) \space\space\space\@spaces}{%
      Package graphicx or graphics not loaded%
    }{See the gnuplot documentation for explanation.%
    }{The gnuplot epslatex terminal needs graphicx.sty or graphics.sty.}%
    \renewcommand\includegraphics[2][]{}%
  }%
  \providecommand\rotatebox[2]{#2}%
  \@ifundefined{ifGPcolor}{%
    \newif\ifGPcolor
    \GPcolortrue
  }{}%
  \@ifundefined{ifGPblacktext}{%
    \newif\ifGPblacktext
    \GPblacktexttrue
  }{}%
  \let\gplgaddtomacro\g@addto@macro
  \gdef\gplbacktext{}%
  \gdef\gplfronttext{}%
  \makeatother
  \ifGPblacktext
    \def\colorrgb#1{}%
    \def\colorgray#1{}%
  \else
    \ifGPcolor
      \def\colorrgb#1{\color[rgb]{#1}}%
      \def\colorgray#1{\color[gray]{#1}}%
      \expandafter\def\csname LTw\endcsname{\color{white}}%
      \expandafter\def\csname LTb\endcsname{\color{black}}%
      \expandafter\def\csname LTa\endcsname{\color{black}}%
      \expandafter\def\csname LT0\endcsname{\color[rgb]{1,0,0}}%
      \expandafter\def\csname LT1\endcsname{\color[rgb]{0,1,0}}%
      \expandafter\def\csname LT2\endcsname{\color[rgb]{0,0,1}}%
      \expandafter\def\csname LT3\endcsname{\color[rgb]{1,0,1}}%
      \expandafter\def\csname LT4\endcsname{\color[rgb]{0,1,1}}%
      \expandafter\def\csname LT5\endcsname{\color[rgb]{1,1,0}}%
      \expandafter\def\csname LT6\endcsname{\color[rgb]{0,0,0}}%
      \expandafter\def\csname LT7\endcsname{\color[rgb]{1,0.3,0}}%
      \expandafter\def\csname LT8\endcsname{\color[rgb]{0.5,0.5,0.5}}%
    \else
      \def\colorrgb#1{\color{black}}%
      \def\colorgray#1{\color[gray]{#1}}%
      \expandafter\def\csname LTw\endcsname{\color{white}}%
      \expandafter\def\csname LTb\endcsname{\color{black}}%
      \expandafter\def\csname LTa\endcsname{\color{black}}%
      \expandafter\def\csname LT0\endcsname{\color{black}}%
      \expandafter\def\csname LT1\endcsname{\color{black}}%
      \expandafter\def\csname LT2\endcsname{\color{black}}%
      \expandafter\def\csname LT3\endcsname{\color{black}}%
      \expandafter\def\csname LT4\endcsname{\color{black}}%
      \expandafter\def\csname LT5\endcsname{\color{black}}%
      \expandafter\def\csname LT6\endcsname{\color{black}}%
      \expandafter\def\csname LT7\endcsname{\color{black}}%
      \expandafter\def\csname LT8\endcsname{\color{black}}%
    \fi
  \fi
  \setlength{\unitlength}{0.0500bp}%
  \begin{picture}(3968.00,2976.00)%
    \gplgaddtomacro\gplbacktext{%
      \csname LTb\endcsname%
      \put(768,480){\makebox(0,0)[r]{\strut{}$0.25$}}%
      \put(768,830){\makebox(0,0)[r]{\strut{}$0.30$}}%
      \put(768,1179){\makebox(0,0)[r]{\strut{}$0.35$}}%
      \put(768,1529){\makebox(0,0)[r]{\strut{}$0.40$}}%
      \put(768,1878){\makebox(0,0)[r]{\strut{}$0.45$}}%
      \put(768,2228){\makebox(0,0)[r]{\strut{}$0.50$}}%
      \put(768,2577){\makebox(0,0)[r]{\strut{}$0.55$}}%
      \put(768,2927){\makebox(0,0)[r]{\strut{}$0.60$}}%
      \put(1120,320){\makebox(0,0){\strut{}$1.00$}}%
      \put(1633,320){\makebox(0,0){\strut{}$1.20$}}%
      \put(2145,320){\makebox(0,0){\strut{}$1.40$}}%
      \put(2658,320){\makebox(0,0){\strut{}$1.60$}}%
      \put(3170,320){\makebox(0,0){\strut{}$1.80$}}%
      \put(3683,320){\makebox(0,0){\strut{}$2.00$}}%
      \put(160,1703){\makebox(0,0){\strut{}$\kappa_c$}}%
      \put(2401,80){\makebox(0,0){\strut{}$g^{-2}$}}%
    }%
    \gplgaddtomacro\gplfronttext{%
      \csname LTb\endcsname%
      \put(3204,2704){\makebox(0,0)[r]{\strut{}$N=8^2$}}%
      \csname LTb\endcsname%
      \put(3204,2544){\makebox(0,0)[r]{\strut{}$N=16^2$}}%
      \csname LTb\endcsname%
      \put(3204,2384){\makebox(0,0)[r]{\strut{}$N=24^2$}}%
    }%
    \gplbacktext
    \put(0,0){\includegraphics{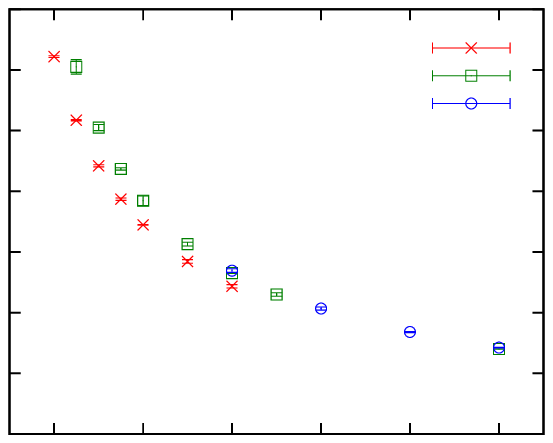}}%
    \gplfronttext
  \end{picture}%
\endgroup

%% file: stereo_finetuning_linear_16x16.tex
\begingroup
\footnotesize
  \makeatletter
  \providecommand\color[2][]{%
    \GenericError{(gnuplot) \space\space\space\@spaces}{%
      Package color not loaded in conjunction with
      terminal option `colourtext'%
    }{See the gnuplot documentation for explanation.%
    }{Either use 'blacktext' in gnuplot or load the package
      color.sty in LaTeX.}%
    \renewcommand\color[2][]{}%
  }%
  \providecommand\includegraphics[2][]{%
    \GenericError{(gnuplot) \space\space\space\@spaces}{%
      Package graphicx or graphics not loaded%
    }{See the gnuplot documentation for explanation.%
    }{The gnuplot epslatex terminal needs graphicx.sty or graphics.sty.}%
    \renewcommand\includegraphics[2][]{}%
  }%
  \providecommand\rotatebox[2]{#2}%
  \@ifundefined{ifGPcolor}{%
    \newif\ifGPcolor
    \GPcolortrue
  }{}%
  \@ifundefined{ifGPblacktext}{%
    \newif\ifGPblacktext
    \GPblacktexttrue
  }{}%
  \let\gplgaddtomacro\g@addto@macro
  \gdef\gplbacktext{}%
  \gdef\gplfronttext{}%
  \makeatother
  \ifGPblacktext
    \def\colorrgb#1{}%
    \def\colorgray#1{}%
  \else
    \ifGPcolor
      \def\colorrgb#1{\color[rgb]{#1}}%
      \def\colorgray#1{\color[gray]{#1}}%
      \expandafter\def\csname LTw\endcsname{\color{white}}%
      \expandafter\def\csname LTb\endcsname{\color{black}}%
      \expandafter\def\csname LTa\endcsname{\color{black}}%
      \expandafter\def\csname LT0\endcsname{\color[rgb]{1,0,0}}%
      \expandafter\def\csname LT1\endcsname{\color[rgb]{0,1,0}}%
      \expandafter\def\csname LT2\endcsname{\color[rgb]{0,0,1}}%
      \expandafter\def\csname LT3\endcsname{\color[rgb]{1,0,1}}%
      \expandafter\def\csname LT4\endcsname{\color[rgb]{0,1,1}}%
      \expandafter\def\csname LT5\endcsname{\color[rgb]{1,1,0}}%
      \expandafter\def\csname LT6\endcsname{\color[rgb]{0,0,0}}%
      \expandafter\def\csname LT7\endcsname{\color[rgb]{1,0.3,0}}%
      \expandafter\def\csname LT8\endcsname{\color[rgb]{0.5,0.5,0.5}}%
    \else
      \def\colorrgb#1{\color{black}}%
      \def\colorgray#1{\color[gray]{#1}}%
      \expandafter\def\csname LTw\endcsname{\color{white}}%
      \expandafter\def\csname LTb\endcsname{\color{black}}%
      \expandafter\def\csname LTa\endcsname{\color{black}}%
      \expandafter\def\csname LT0\endcsname{\color{black}}%
      \expandafter\def\csname LT1\endcsname{\color{black}}%
      \expandafter\def\csname LT2\endcsname{\color{black}}%
      \expandafter\def\csname LT3\endcsname{\color{black}}%
      \expandafter\def\csname LT4\endcsname{\color{black}}%
      \expandafter\def\csname LT5\endcsname{\color{black}}%
      \expandafter\def\csname LT6\endcsname{\color{black}}%
      \expandafter\def\csname LT7\endcsname{\color{black}}%
      \expandafter\def\csname LT8\endcsname{\color{black}}%
    \fi
  \fi
  \setlength{\unitlength}{0.0500bp}%
  \begin{picture}(3968.00,2976.00)%
    \gplgaddtomacro\gplbacktext{%
      \csname LTb\endcsname%
      \put(768,480){\makebox(0,0)[r]{\strut{}$0.0$}}%
      \put(768,752){\makebox(0,0)[r]{\strut{}$0.1$}}%
      \put(768,1024){\makebox(0,0)[r]{\strut{}$0.2$}}%
      \put(768,1296){\makebox(0,0)[r]{\strut{}$0.3$}}%
      \put(768,1568){\makebox(0,0)[r]{\strut{}$0.4$}}%
      \put(768,1839){\makebox(0,0)[r]{\strut{}$0.5$}}%
      \put(768,2111){\makebox(0,0)[r]{\strut{}$0.6$}}%
      \put(768,2383){\makebox(0,0)[r]{\strut{}$0.7$}}%
      \put(768,2655){\makebox(0,0)[r]{\strut{}$0.8$}}%
      \put(768,2927){\makebox(0,0)[r]{\strut{}$0.9$}}%
      \put(864,320){\makebox(0,0){\strut{}$0.25$}}%
      \put(1633,320){\makebox(0,0){\strut{}$0.30$}}%
      \put(2401,320){\makebox(0,0){\strut{}$0.35$}}%
      \put(3170,320){\makebox(0,0){\strut{}$0.40$}}%
      \put(3939,320){\makebox(0,0){\strut{}$0.45$}}%
      \put(2401,80){\makebox(0,0){\strut{}$\kappa$}}%
    }%
    \gplgaddtomacro\gplfronttext{%
      \csname LTb\endcsname%
      \put(1455,2725){\makebox(0,0)[r]{\strut{}$a m_\text{F}$}}%
      \csname LTb\endcsname%
      \put(1455,2565){\makebox(0,0)[r]{\strut{}$a m_\text{B}$}}%
    }%
    \gplgaddtomacro\gplbacktext{%
      \csname LTb\endcsname%
      \put(2553,1819){\makebox(0,0)[r]{\strut{}$0.08$}}%
      \put(2553,2093){\makebox(0,0)[r]{\strut{}$0.10$}}%
      \put(2553,2367){\makebox(0,0)[r]{\strut{}$0.12$}}%
      \put(2553,2641){\makebox(0,0)[r]{\strut{}$0.14$}}%
      \put(2649,1659){\makebox(0,0){\strut{}$0.37$}}%
      \put(3195,1659){\makebox(0,0){\strut{}$0.38$}}%
      \put(3740,1659){\makebox(0,0){\strut{}$0.39$}}%
    }%
    \gplgaddtomacro\gplfronttext{%
    }%
    \gplbacktext
    \put(0,0){\includegraphics{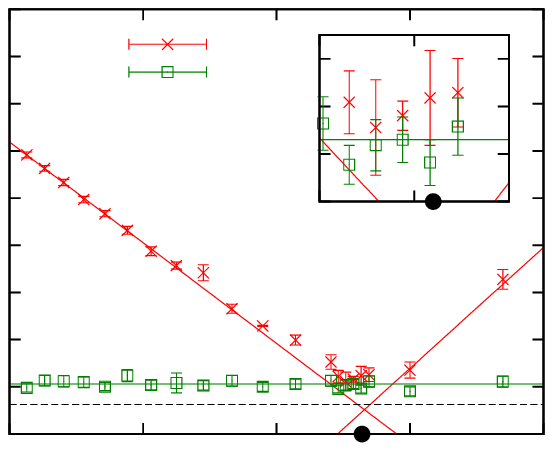}}%
    \gplfronttext
  \end{picture}%
\endgroup

%% file: stereo_finetuned_mfLmbL.tex
\begingroup
\footnotesize
  \makeatletter
  \providecommand\color[2][]{%
    \GenericError{(gnuplot) \space\space\space\@spaces}{%
      Package color not loaded in conjunction with
      terminal option `colourtext'%
    }{See the gnuplot documentation for explanation.%
    }{Either use 'blacktext' in gnuplot or load the package
      color.sty in LaTeX.}%
    \renewcommand\color[2][]{}%
  }%
  \providecommand\includegraphics[2][]{%
    \GenericError{(gnuplot) \space\space\space\@spaces}{%
      Package graphicx or graphics not loaded%
    }{See the gnuplot documentation for explanation.%
    }{The gnuplot epslatex terminal needs graphicx.sty or graphics.sty.}%
    \renewcommand\includegraphics[2][]{}%
  }%
  \providecommand\rotatebox[2]{#2}%
  \@ifundefined{ifGPcolor}{%
    \newif\ifGPcolor
    \GPcolortrue
  }{}%
  \@ifundefined{ifGPblacktext}{%
    \newif\ifGPblacktext
    \GPblacktexttrue
  }{}%
  \let\gplgaddtomacro\g@addto@macro
  \gdef\gplbacktext{}%
  \gdef\gplfronttext{}%
  \makeatother
  \ifGPblacktext
    \def\colorrgb#1{}%
    \def\colorgray#1{}%
  \else
    \ifGPcolor
      \def\colorrgb#1{\color[rgb]{#1}}%
      \def\colorgray#1{\color[gray]{#1}}%
      \expandafter\def\csname LTw\endcsname{\color{white}}%
      \expandafter\def\csname LTb\endcsname{\color{black}}%
      \expandafter\def\csname LTa\endcsname{\color{black}}%
      \expandafter\def\csname LT0\endcsname{\color[rgb]{1,0,0}}%
      \expandafter\def\csname LT1\endcsname{\color[rgb]{0,1,0}}%
      \expandafter\def\csname LT2\endcsname{\color[rgb]{0,0,1}}%
      \expandafter\def\csname LT3\endcsname{\color[rgb]{1,0,1}}%
      \expandafter\def\csname LT4\endcsname{\color[rgb]{0,1,1}}%
      \expandafter\def\csname LT5\endcsname{\color[rgb]{1,1,0}}%
      \expandafter\def\csname LT6\endcsname{\color[rgb]{0,0,0}}%
      \expandafter\def\csname LT7\endcsname{\color[rgb]{1,0.3,0}}%
      \expandafter\def\csname LT8\endcsname{\color[rgb]{0.5,0.5,0.5}}%
    \else
      \def\colorrgb#1{\color{black}}%
      \def\colorgray#1{\color[gray]{#1}}%
      \expandafter\def\csname LTw\endcsname{\color{white}}%
      \expandafter\def\csname LTb\endcsname{\color{black}}%
      \expandafter\def\csname LTa\endcsname{\color{black}}%
      \expandafter\def\csname LT0\endcsname{\color{black}}%
      \expandafter\def\csname LT1\endcsname{\color{black}}%
      \expandafter\def\csname LT2\endcsname{\color{black}}%
      \expandafter\def\csname LT3\endcsname{\color{black}}%
      \expandafter\def\csname LT4\endcsname{\color{black}}%
      \expandafter\def\csname LT5\endcsname{\color{black}}%
      \expandafter\def\csname LT6\endcsname{\color{black}}%
      \expandafter\def\csname LT7\endcsname{\color{black}}%
      \expandafter\def\csname LT8\endcsname{\color{black}}%
    \fi
  \fi
  \setlength{\unitlength}{0.0500bp}%
  \begin{picture}(3968.00,2976.00)%
    \gplgaddtomacro\gplbacktext{%
      \csname LTb\endcsname%
      \put(768,480){\makebox(0,0)[r]{\strut{}$0$}}%
      \put(768,969){\makebox(0,0)[r]{\strut{}$1$}}%
      \put(768,1459){\makebox(0,0)[r]{\strut{}$2$}}%
      \put(768,1948){\makebox(0,0)[r]{\strut{}$3$}}%
      \put(768,2438){\makebox(0,0)[r]{\strut{}$4$}}%
      \put(768,2927){\makebox(0,0)[r]{\strut{}$5$}}%
      \put(864,320){\makebox(0,0){\strut{}$0$}}%
      \put(1479,320){\makebox(0,0){\strut{}$1$}}%
      \put(2094,320){\makebox(0,0){\strut{}$2$}}%
      \put(2709,320){\makebox(0,0){\strut{}$3$}}%
      \put(3324,320){\makebox(0,0){\strut{}$4$}}%
      \put(3939,320){\makebox(0,0){\strut{}$5$}}%
      \put(448,1703){\makebox(0,0){\strut{}$m_\text{B}L$}}%
      \put(2401,80){\makebox(0,0){\strut{}$m_\text{F}L$}}%
    }%
    \gplgaddtomacro\gplfronttext{%
      \csname LTb\endcsname%
      \put(1728,2784){\makebox(0,0)[r]{\strut{}$N=8^2$}}%
      \csname LTb\endcsname%
      \put(1728,2624){\makebox(0,0)[r]{\strut{}$N=16^2$}}%
      \csname LTb\endcsname%
      \put(1728,2464){\makebox(0,0)[r]{\strut{}$N=24^2$}}%
    }%
    \gplbacktext
    \put(0,0){\includegraphics{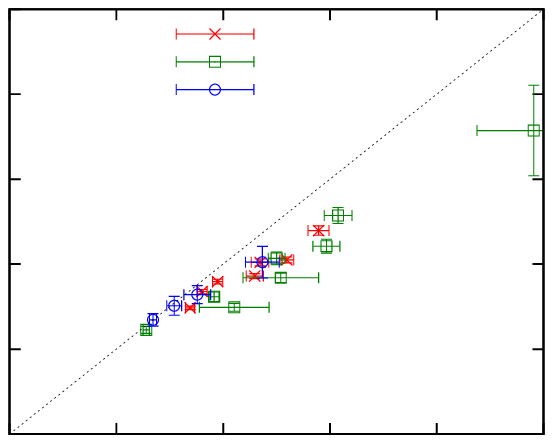}}%
    \gplfronttext
  \end{picture}%
\endgroup

%% file: ward.tex
\subsection{Path integral based Ward identity}
Similar to the Ward identity that is given in
\cite{Catterall:2006sj} an equivalent relation can be derived for
the present lattice model. Starting from the path integral in the continuum,
$Q$ exactness is given by the twisted supercharge
\cite{Catterall:2006sj} in the continuum, such that $S=\frac{1}{2 g^2} Q
\Lambda$ with $Q^2 = 0$. This implies the continuum Ward identity
\begin{equation}
\label{eq:sigma:wardContinuum}
\frac{\partial \ln \ZZ}{\partial(g^{-2})} = \vev{-\half Q \Lambda} = 0,
\end{equation}
since action and measure are invariant under the symmetry
transformation generated by $Q$. After putting the theory on the
lattice, integrating out the constrained auxiliary field $f$, and introducing
the unconstrained field $\sigma$ to get rid of the four-fermion interaction (in
that order) one is left with the path integral,
\vspace*{-0.5ex}
\begin{equation}
\ZZ = g^{N} \int \DD R\, \DD \sigma\, \DD\lambda \, \ee^{-S[R,\sigma,\lambda]}
\vspace*{-0.5ex}
\end{equation}
with $N$ as number of lattice sites and the action given in
eq.~\eqref{eq:sigma:pathIntegralInRvars}.\footnote{The factor $g^N$ in front of
the path integral stems from the Gaussian integrals that need to be carried out
for $f$ and be introduced for $\sigma$.} Here, the coupling
dependent part is important, so that only constant numerical factors may be dropped. 
The derivative of the Schwinger functional is
\begin{equation}
\label{eq:sigma:wardLattice}
\frac{\partial \ln \ZZ}{\partial (g^{-2})} = -\frac{Ng^2}{2} -
g^2\vev{S_\text{B}} + g^2 \frac{\dim Q_{F}}{2},
\end{equation}
where $Q_{F}$ denotes the Dirac operator in  \eqref{diracoperator}, 
and $\dim Q_{F}$ is its dimension in terms of flavor and spinor components
as well as lattices sites. 
In our case the Ward identity \eqref{eq:sigma:wardContinuum} reads
\begin{equation}
\label{eq:sigma:wardFull}
\vev{S_\text{B}} = \frac{3}{2}N,
\end{equation}
with $S_\text{B}$ defined in Eq.~\eqref{rotaction}. The same Ward identity can be 
derived for the formulation based on the stereographic projection.\\
In order to see a possible restoration of supersymmetry in the continuum limit, the bosonic
action has been calculated with the SLAC derivative for three lattice sizes in a coupling 
range where finite size effects should be negligible, i.e. for $m_\text{F}L>5$.
\FIGURE{\hfill
\input{sigmaWardIdentityFixedVolume}\hfill
\input{sigmaWardIdentityEpsilonRegime}\hfill\hfill\phantom.
\caption{\label{fig:sigma:wardIdentity} Normalized action $2\vev{\SB}/3N$, which
will take the value $1$ if supersymmetry is restored in the continuum limit as
required by the Ward identity \eqref{eq:sigma:wardFull}. Left panel:
Measurements for different lattice sizes with physical volumes
$5<m_\text{F}L<11$ in the SLAC ensemble. Right panel: Results for fixed 
lattice volume $N=7^2$ (SLAC) and $N=8^2$ (Wilson, $\kappa=0.25$ fixed)
and couplings $g^{-2}\in[0.4,100]$ that reach the regime of small physical
volumes (small $g$).}}
The results that are shown in
Fig.~\ref{fig:sigma:wardIdentity} (left panel) reveal that for the smallest
($5\times 5$) lattice Eq.~\eqref{eq:sigma:wardFull} is violated as much as
$10\%$ for $m_\text{F}L\approx 5$ and up to $20\%$ at $m_\text{F}L\approx 10$.
Therefore the Ward identity violation grows for coarser lattice spacings. However, in the
continuum limit at a fixed physical volume $m_\text{F}L$ the Ward identity
tends to be restored, cf. Fig~\ref{fig:sigma:wardIdentity} (left panel). Additionally one can explore the small volume regime of
this theory by sending $g\to 0$ at fixed lattice volume to reach
the continuum limit. This has been performed
on a $7\times 7$ lattice for a large range of couplings $g^{-2}\in[0.4,100]$,
see Fig.~\ref{fig:sigma:wardIdentity} (right panel). Here the Ward identity is
explicitly restored in the limit of large $g^{-2}$, although this result has to be 
taken with care since the physical box size becomes unreliably small.\\
The examined lattice sizes are rather 
small, but these observations at least imply that a supersymmetric continuum limit 
can be reached and that the non-degeneracy of bosonic and fermionic mass is a finite size 
effect. However, in order to obtain a definite answer simulations on larger lattices would 
be necessary. For instance, one ought to verify that $\vev{\SB}/N$ does not undershoot 
and drop below $1.5$ in the continuum limit at fixed $m_\text{F} L$.
Using the Wilson derivative, we measured the bosonic action in the finetuned ensemble at 
$\kappa=\kappa_{c}$ and observe similar behavior, although 
for the largest lattice considered we see a discrepancy of $7\%$ for fine lattice spacing ($g^{-2}=2$)
and up to $14\%$ for a coarser lattice spacing ($g^{-2}=1.4$), see figure \ref{fig:sigma:bosonic_action}.
However, unlike for SLAC fermions,
the Ward identity approaches its continuum value from above in the limit of $g^{-2}$ to infinity
and therefore is unlikely to undershoot (see Fig. \ref{fig:sigma:wardIdentity}, right panel). 
Hence a comparison of both derivatives is only valid
in the regime of large $g^{-2}$ where monotonic behaviour is encountered, which is however 
unavailable for the SLAC derivative due to either
fine-size effects for small lattices or the strong sign problem for large lattices.
Overall, the slope of the Ward identity 
clearly points to a restoration of supersymmetry in the continuum limit.
\FIGURE{\hfill
\input{stereo_bosonicAction}\hfill
\input{stereo_bosonicAction_mbL}\hfill\hfill\phantom.
\caption{\label{fig:sigma:bosonic_action} Expectation value of the bosonic action at $\kappa=\kappa_c$ for different $g^{-2}$ (left panel)
and different box sizes (right panel). $\frac{2\langle S_B\rangle}{3N}=1$ is expected in the supersymmetric continuum limit. }}

%% file: sigmaWardIdentityFixedVolume.tex
\begingroup
\footnotesize
  \makeatletter
  \providecommand\color[2][]{%
    \GenericError{(gnuplot) \space\space\space\@spaces}{%
      Package color not loaded in conjunction with
      terminal option `colourtext'%
    }{See the gnuplot documentation for explanation.%
    }{Either use 'blacktext' in gnuplot or load the package
      color.sty in LaTeX.}%
    \renewcommand\color[2][]{}%
  }%
  \providecommand\includegraphics[2][]{%
    \GenericError{(gnuplot) \space\space\space\@spaces}{%
      Package graphicx or graphics not loaded%
    }{See the gnuplot documentation for explanation.%
    }{The gnuplot epslatex terminal needs graphicx.sty or graphics.sty.}%
    \renewcommand\includegraphics[2][]{}%
  }%
  \providecommand\rotatebox[2]{#2}%
  \@ifundefined{ifGPcolor}{%
    \newif\ifGPcolor
    \GPcolortrue
  }{}%
  \@ifundefined{ifGPblacktext}{%
    \newif\ifGPblacktext
    \GPblacktexttrue
  }{}%
  \let\gplgaddtomacro\g@addto@macro
  \gdef\gplbacktext{}%
  \gdef\gplfronttext{}%
  \makeatother
  \ifGPblacktext
    \def\colorrgb#1{}%
    \def\colorgray#1{}%
  \else
    \ifGPcolor
      \def\colorrgb#1{\color[rgb]{#1}}%
      \def\colorgray#1{\color[gray]{#1}}%
      \expandafter\def\csname LTw\endcsname{\color{white}}%
      \expandafter\def\csname LTb\endcsname{\color{black}}%
      \expandafter\def\csname LTa\endcsname{\color{black}}%
      \expandafter\def\csname LT0\endcsname{\color[rgb]{1,0,0}}%
      \expandafter\def\csname LT1\endcsname{\color[rgb]{0,1,0}}%
      \expandafter\def\csname LT2\endcsname{\color[rgb]{0,0,1}}%
      \expandafter\def\csname LT3\endcsname{\color[rgb]{1,0,1}}%
      \expandafter\def\csname LT4\endcsname{\color[rgb]{0,1,1}}%
      \expandafter\def\csname LT5\endcsname{\color[rgb]{1,1,0}}%
      \expandafter\def\csname LT6\endcsname{\color[rgb]{0,0,0}}%
      \expandafter\def\csname LT7\endcsname{\color[rgb]{1,0.3,0}}%
      \expandafter\def\csname LT8\endcsname{\color[rgb]{0.5,0.5,0.5}}%
    \else
      \def\colorrgb#1{\color{black}}%
      \def\colorgray#1{\color[gray]{#1}}%
      \expandafter\def\csname LTw\endcsname{\color{white}}%
      \expandafter\def\csname LTb\endcsname{\color{black}}%
      \expandafter\def\csname LTa\endcsname{\color{black}}%
      \expandafter\def\csname LT0\endcsname{\color{black}}%
      \expandafter\def\csname LT1\endcsname{\color{black}}%
      \expandafter\def\csname LT2\endcsname{\color{black}}%
      \expandafter\def\csname LT3\endcsname{\color{black}}%
      \expandafter\def\csname LT4\endcsname{\color{black}}%
      \expandafter\def\csname LT5\endcsname{\color{black}}%
      \expandafter\def\csname LT6\endcsname{\color{black}}%
      \expandafter\def\csname LT7\endcsname{\color{black}}%
      \expandafter\def\csname LT8\endcsname{\color{black}}%
    \fi
  \fi
  \setlength{\unitlength}{0.0500bp}%
  \begin{picture}(3968.00,2976.00)%
    \gplgaddtomacro\gplbacktext{%
      \csname LTb\endcsname%
      \put(768,480){\makebox(0,0)[r]{\strut{}$0.90$}}%
      \put(768,786){\makebox(0,0)[r]{\strut{}$0.95$}}%
      \put(768,1092){\makebox(0,0)[r]{\strut{}$1.00$}}%
      \put(768,1398){\makebox(0,0)[r]{\strut{}$1.05$}}%
      \put(768,1704){\makebox(0,0)[r]{\strut{}$1.10$}}%
      \put(768,2009){\makebox(0,0)[r]{\strut{}$1.15$}}%
      \put(768,2315){\makebox(0,0)[r]{\strut{}$1.20$}}%
      \put(768,2621){\makebox(0,0)[r]{\strut{}$1.25$}}%
      \put(768,2927){\makebox(0,0)[r]{\strut{}$1.30$}}%
      \put(864,320){\makebox(0,0){\strut{}$5$}}%
      \put(1303,320){\makebox(0,0){\strut{}$6$}}%
      \put(1743,320){\makebox(0,0){\strut{}$7$}}%
      \put(2182,320){\makebox(0,0){\strut{}$8$}}%
      \put(2621,320){\makebox(0,0){\strut{}$9$}}%
      \put(3060,320){\makebox(0,0){\strut{}$10$}}%
      \put(3500,320){\makebox(0,0){\strut{}$11$}}%
      \put(3939,320){\makebox(0,0){\strut{}$12$}}%
      \put(112,1703){\makebox(0,0){\strut{}$\frac{2\vev{S_\text{B}}}{3N}$}}%
      \put(2401,80){\makebox(0,0){\strut{}$m_\text{F}L$}}%
    }%
    \gplgaddtomacro\gplfronttext{%
      \csname LTb\endcsname%
      \put(3204,943){\makebox(0,0)[r]{\strut{}$N=5^2$}}%
      \csname LTb\endcsname%
      \put(3204,783){\makebox(0,0)[r]{\strut{}$N=7^2$}}%
      \csname LTb\endcsname%
      \put(3204,623){\makebox(0,0)[r]{\strut{}$N=9^2$}}%
    }%
    \gplbacktext
    \put(0,0){\includegraphics{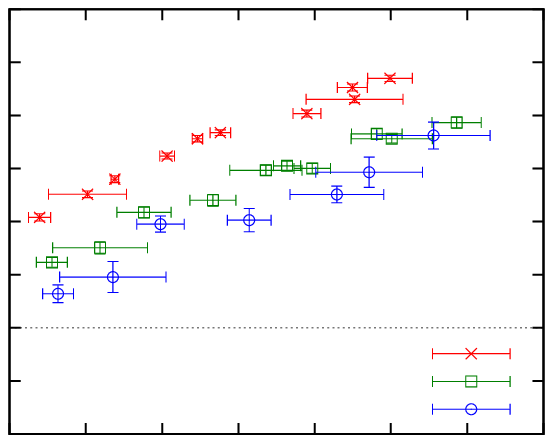}}%
    \gplfronttext
  \end{picture}%
\endgroup

%% file: sigmaWardIdentityEpsilonRegime.tex
\begingroup
\footnotesize
  \makeatletter
  \providecommand\color[2][]{%
    \GenericError{(gnuplot) \space\space\space\@spaces}{%
      Package color not loaded in conjunction with
      terminal option `colourtext'%
    }{See the gnuplot documentation for explanation.%
    }{Either use 'blacktext' in gnuplot or load the package
      color.sty in LaTeX.}%
    \renewcommand\color[2][]{}%
  }%
  \providecommand\includegraphics[2][]{%
    \GenericError{(gnuplot) \space\space\space\@spaces}{%
      Package graphicx or graphics not loaded%
    }{See the gnuplot documentation for explanation.%
    }{The gnuplot epslatex terminal needs graphicx.sty or graphics.sty.}%
    \renewcommand\includegraphics[2][]{}%
  }%
  \providecommand\rotatebox[2]{#2}%
  \@ifundefined{ifGPcolor}{%
    \newif\ifGPcolor
    \GPcolortrue
  }{}%
  \@ifundefined{ifGPblacktext}{%
    \newif\ifGPblacktext
    \GPblacktexttrue
  }{}%
  \let\gplgaddtomacro\g@addto@macro
  \gdef\gplbacktext{}%
  \gdef\gplfronttext{}%
  \makeatother
  \ifGPblacktext
    \def\colorrgb#1{}%
    \def\colorgray#1{}%
  \else
    \ifGPcolor
      \def\colorrgb#1{\color[rgb]{#1}}%
      \def\colorgray#1{\color[gray]{#1}}%
      \expandafter\def\csname LTw\endcsname{\color{white}}%
      \expandafter\def\csname LTb\endcsname{\color{black}}%
      \expandafter\def\csname LTa\endcsname{\color{black}}%
      \expandafter\def\csname LT0\endcsname{\color[rgb]{1,0,0}}%
      \expandafter\def\csname LT1\endcsname{\color[rgb]{0,1,0}}%
      \expandafter\def\csname LT2\endcsname{\color[rgb]{0,0,1}}%
      \expandafter\def\csname LT3\endcsname{\color[rgb]{1,0,1}}%
      \expandafter\def\csname LT4\endcsname{\color[rgb]{0,1,1}}%
      \expandafter\def\csname LT5\endcsname{\color[rgb]{1,1,0}}%
      \expandafter\def\csname LT6\endcsname{\color[rgb]{0,0,0}}%
      \expandafter\def\csname LT7\endcsname{\color[rgb]{1,0.3,0}}%
      \expandafter\def\csname LT8\endcsname{\color[rgb]{0.5,0.5,0.5}}%
    \else
      \def\colorrgb#1{\color{black}}%
      \def\colorgray#1{\color[gray]{#1}}%
      \expandafter\def\csname LTw\endcsname{\color{white}}%
      \expandafter\def\csname LTb\endcsname{\color{black}}%
      \expandafter\def\csname LTa\endcsname{\color{black}}%
      \expandafter\def\csname LT0\endcsname{\color{black}}%
      \expandafter\def\csname LT1\endcsname{\color{black}}%
      \expandafter\def\csname LT2\endcsname{\color{black}}%
      \expandafter\def\csname LT3\endcsname{\color{black}}%
      \expandafter\def\csname LT4\endcsname{\color{black}}%
      \expandafter\def\csname LT5\endcsname{\color{black}}%
      \expandafter\def\csname LT6\endcsname{\color{black}}%
      \expandafter\def\csname LT7\endcsname{\color{black}}%
      \expandafter\def\csname LT8\endcsname{\color{black}}%
    \fi
  \fi
  \setlength{\unitlength}{0.0500bp}%
  \begin{picture}(3968.00,2976.00)%
    \gplgaddtomacro\gplbacktext{%
      \csname LTb\endcsname%
      \put(768,480){\makebox(0,0)[r]{\strut{}$0.8$}}%
      \put(768,969){\makebox(0,0)[r]{\strut{}$0.9$}}%
      \put(768,1459){\makebox(0,0)[r]{\strut{}$1.0$}}%
      \put(768,1948){\makebox(0,0)[r]{\strut{}$1.1$}}%
      \put(768,2438){\makebox(0,0)[r]{\strut{}$1.2$}}%
      \put(768,2927){\makebox(0,0)[r]{\strut{}$1.3$}}%
      \put(1374,320){\makebox(0,0){\strut{}$1$}}%
      \put(2657,320){\makebox(0,0){\strut{}$10$}}%
      \put(3939,320){\makebox(0,0){\strut{}$100$}}%
      \put(208,1703){\makebox(0,0){\strut{}$\frac{2\vev{S_\text{B}}}{3N}$}}%
      \put(2401,80){\makebox(0,0){\strut{}$g^{-2}$}}%
    }%
    \gplgaddtomacro\gplfronttext{%
      \csname LTb\endcsname%
      \put(3204,783){\makebox(0,0)[r]{\strut{}SLAC}}%
      \csname LTb\endcsname%
      \put(3204,623){\makebox(0,0)[r]{\strut{}Wilson}}%
    }%
    \gplbacktext
    \put(0,0){\includegraphics{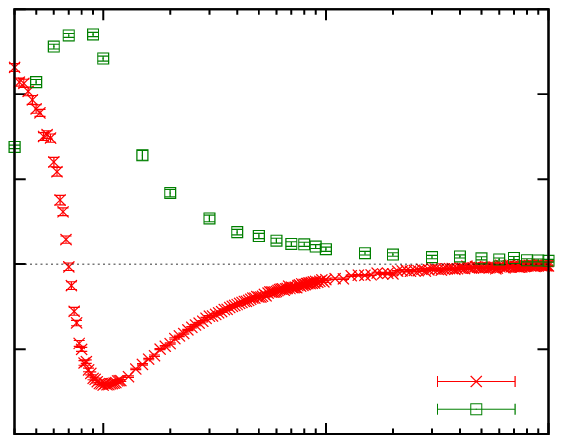}}%
    \gplfronttext
  \end{picture}%
\endgroup

%% file: stereo_bosonicAction.tex
\begingroup
\footnotesize
  \makeatletter
  \providecommand\color[2][]{%
    \GenericError{(gnuplot) \space\space\space\@spaces}{%
      Package color not loaded in conjunction with
      terminal option `colourtext'%
    }{See the gnuplot documentation for explanation.%
    }{Either use 'blacktext' in gnuplot or load the package
      color.sty in LaTeX.}%
    \renewcommand\color[2][]{}%
  }%
  \providecommand\includegraphics[2][]{%
    \GenericError{(gnuplot) \space\space\space\@spaces}{%
      Package graphicx or graphics not loaded%
    }{See the gnuplot documentation for explanation.%
    }{The gnuplot epslatex terminal needs graphicx.sty or graphics.sty.}%
    \renewcommand\includegraphics[2][]{}%
  }%
  \providecommand\rotatebox[2]{#2}%
  \@ifundefined{ifGPcolor}{%
    \newif\ifGPcolor
    \GPcolortrue
  }{}%
  \@ifundefined{ifGPblacktext}{%
    \newif\ifGPblacktext
    \GPblacktexttrue
  }{}%
  \let\gplgaddtomacro\g@addto@macro
  \gdef\gplbacktext{}%
  \gdef\gplfronttext{}%
  \makeatother
  \ifGPblacktext
    \def\colorrgb#1{}%
    \def\colorgray#1{}%
  \else
    \ifGPcolor
      \def\colorrgb#1{\color[rgb]{#1}}%
      \def\colorgray#1{\color[gray]{#1}}%
      \expandafter\def\csname LTw\endcsname{\color{white}}%
      \expandafter\def\csname LTb\endcsname{\color{black}}%
      \expandafter\def\csname LTa\endcsname{\color{black}}%
      \expandafter\def\csname LT0\endcsname{\color[rgb]{1,0,0}}%
      \expandafter\def\csname LT1\endcsname{\color[rgb]{0,1,0}}%
      \expandafter\def\csname LT2\endcsname{\color[rgb]{0,0,1}}%
      \expandafter\def\csname LT3\endcsname{\color[rgb]{1,0,1}}%
      \expandafter\def\csname LT4\endcsname{\color[rgb]{0,1,1}}%
      \expandafter\def\csname LT5\endcsname{\color[rgb]{1,1,0}}%
      \expandafter\def\csname LT6\endcsname{\color[rgb]{0,0,0}}%
      \expandafter\def\csname LT7\endcsname{\color[rgb]{1,0.3,0}}%
      \expandafter\def\csname LT8\endcsname{\color[rgb]{0.5,0.5,0.5}}%
    \else
      \def\colorrgb#1{\color{black}}%
      \def\colorgray#1{\color[gray]{#1}}%
      \expandafter\def\csname LTw\endcsname{\color{white}}%
      \expandafter\def\csname LTb\endcsname{\color{black}}%
      \expandafter\def\csname LTa\endcsname{\color{black}}%
      \expandafter\def\csname LT0\endcsname{\color{black}}%
      \expandafter\def\csname LT1\endcsname{\color{black}}%
      \expandafter\def\csname LT2\endcsname{\color{black}}%
      \expandafter\def\csname LT3\endcsname{\color{black}}%
      \expandafter\def\csname LT4\endcsname{\color{black}}%
      \expandafter\def\csname LT5\endcsname{\color{black}}%
      \expandafter\def\csname LT6\endcsname{\color{black}}%
      \expandafter\def\csname LT7\endcsname{\color{black}}%
      \expandafter\def\csname LT8\endcsname{\color{black}}%
    \fi
  \fi
  \setlength{\unitlength}{0.0500bp}%
  \begin{picture}(3968.00,2976.00)%
    \gplgaddtomacro\gplbacktext{%
      \csname LTb\endcsname%
      \put(768,480){\makebox(0,0)[r]{\strut{}$1.00$}}%
      \put(768,786){\makebox(0,0)[r]{\strut{}$1.05$}}%
      \put(768,1092){\makebox(0,0)[r]{\strut{}$1.10$}}%
      \put(768,1398){\makebox(0,0)[r]{\strut{}$1.15$}}%
      \put(768,1704){\makebox(0,0)[r]{\strut{}$1.20$}}%
      \put(768,2009){\makebox(0,0)[r]{\strut{}$1.25$}}%
      \put(768,2315){\makebox(0,0)[r]{\strut{}$1.30$}}%
      \put(768,2621){\makebox(0,0)[r]{\strut{}$1.35$}}%
      \put(768,2927){\makebox(0,0)[r]{\strut{}$1.40$}}%
      \put(1120,320){\makebox(0,0){\strut{}$1.00$}}%
      \put(1633,320){\makebox(0,0){\strut{}$1.20$}}%
      \put(2145,320){\makebox(0,0){\strut{}$1.40$}}%
      \put(2658,320){\makebox(0,0){\strut{}$1.60$}}%
      \put(3170,320){\makebox(0,0){\strut{}$1.80$}}%
      \put(3683,320){\makebox(0,0){\strut{}$2.00$}}%
      \put(160,1703){\makebox(0,0){\strut{}$\frac{2\langle S_B\rangle}{3N}$}}%
      \put(2401,80){\makebox(0,0){\strut{}$g^{-2}$}}%
    }%
    \gplgaddtomacro\gplfronttext{%
      \csname LTb\endcsname%
      \put(3204,2704){\makebox(0,0)[r]{\strut{}$N=8^2$}}%
      \csname LTb\endcsname%
      \put(3204,2544){\makebox(0,0)[r]{\strut{}$N=16^2$}}%
      \csname LTb\endcsname%
      \put(3204,2384){\makebox(0,0)[r]{\strut{}$N=24^2$}}%
    }%
    \gplbacktext
    \put(0,0){\includegraphics{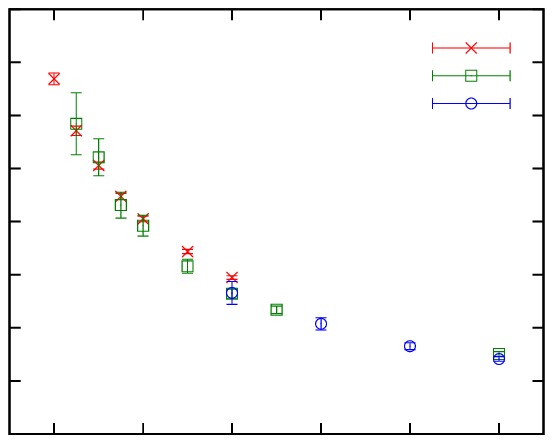}}%
    \gplfronttext
  \end{picture}%
\endgroup

%% file: stereo_bosonicAction_mbL.tex
\begingroup
\footnotesize
  \makeatletter
  \providecommand\color[2][]{%
    \GenericError{(gnuplot) \space\space\space\@spaces}{%
      Package color not loaded in conjunction with
      terminal option `colourtext'%
    }{See the gnuplot documentation for explanation.%
    }{Either use 'blacktext' in gnuplot or load the package
      color.sty in LaTeX.}%
    \renewcommand\color[2][]{}%
  }%
  \providecommand\includegraphics[2][]{%
    \GenericError{(gnuplot) \space\space\space\@spaces}{%
      Package graphicx or graphics not loaded%
    }{See the gnuplot documentation for explanation.%
    }{The gnuplot epslatex terminal needs graphicx.sty or graphics.sty.}%
    \renewcommand\includegraphics[2][]{}%
  }%
  \providecommand\rotatebox[2]{#2}%
  \@ifundefined{ifGPcolor}{%
    \newif\ifGPcolor
    \GPcolortrue
  }{}%
  \@ifundefined{ifGPblacktext}{%
    \newif\ifGPblacktext
    \GPblacktexttrue
  }{}%
  \let\gplgaddtomacro\g@addto@macro
  \gdef\gplbacktext{}%
  \gdef\gplfronttext{}%
  \makeatother
  \ifGPblacktext
    \def\colorrgb#1{}%
    \def\colorgray#1{}%
  \else
    \ifGPcolor
      \def\colorrgb#1{\color[rgb]{#1}}%
      \def\colorgray#1{\color[gray]{#1}}%
      \expandafter\def\csname LTw\endcsname{\color{white}}%
      \expandafter\def\csname LTb\endcsname{\color{black}}%
      \expandafter\def\csname LTa\endcsname{\color{black}}%
      \expandafter\def\csname LT0\endcsname{\color[rgb]{1,0,0}}%
      \expandafter\def\csname LT1\endcsname{\color[rgb]{0,1,0}}%
      \expandafter\def\csname LT2\endcsname{\color[rgb]{0,0,1}}%
      \expandafter\def\csname LT3\endcsname{\color[rgb]{1,0,1}}%
      \expandafter\def\csname LT4\endcsname{\color[rgb]{0,1,1}}%
      \expandafter\def\csname LT5\endcsname{\color[rgb]{1,1,0}}%
      \expandafter\def\csname LT6\endcsname{\color[rgb]{0,0,0}}%
      \expandafter\def\csname LT7\endcsname{\color[rgb]{1,0.3,0}}%
      \expandafter\def\csname LT8\endcsname{\color[rgb]{0.5,0.5,0.5}}%
    \else
      \def\colorrgb#1{\color{black}}%
      \def\colorgray#1{\color[gray]{#1}}%
      \expandafter\def\csname LTw\endcsname{\color{white}}%
      \expandafter\def\csname LTb\endcsname{\color{black}}%
      \expandafter\def\csname LTa\endcsname{\color{black}}%
      \expandafter\def\csname LT0\endcsname{\color{black}}%
      \expandafter\def\csname LT1\endcsname{\color{black}}%
      \expandafter\def\csname LT2\endcsname{\color{black}}%
      \expandafter\def\csname LT3\endcsname{\color{black}}%
      \expandafter\def\csname LT4\endcsname{\color{black}}%
      \expandafter\def\csname LT5\endcsname{\color{black}}%
      \expandafter\def\csname LT6\endcsname{\color{black}}%
      \expandafter\def\csname LT7\endcsname{\color{black}}%
      \expandafter\def\csname LT8\endcsname{\color{black}}%
    \fi
  \fi
  \setlength{\unitlength}{0.0500bp}%
  \begin{picture}(3968.00,2976.00)%
    \gplgaddtomacro\gplbacktext{%
      \csname LTb\endcsname%
      \put(768,480){\makebox(0,0)[r]{\strut{}$1.00$}}%
      \put(768,786){\makebox(0,0)[r]{\strut{}$1.05$}}%
      \put(768,1092){\makebox(0,0)[r]{\strut{}$1.10$}}%
      \put(768,1398){\makebox(0,0)[r]{\strut{}$1.15$}}%
      \put(768,1704){\makebox(0,0)[r]{\strut{}$1.20$}}%
      \put(768,2009){\makebox(0,0)[r]{\strut{}$1.25$}}%
      \put(768,2315){\makebox(0,0)[r]{\strut{}$1.30$}}%
      \put(768,2621){\makebox(0,0)[r]{\strut{}$1.35$}}%
      \put(768,2927){\makebox(0,0)[r]{\strut{}$1.40$}}%
      \put(864,320){\makebox(0,0){\strut{}$1.0$}}%
      \put(1633,320){\makebox(0,0){\strut{}$1.5$}}%
      \put(2402,320){\makebox(0,0){\strut{}$2.0$}}%
      \put(3170,320){\makebox(0,0){\strut{}$2.5$}}%
      \put(3939,320){\makebox(0,0){\strut{}$3.0$}}%
      \put(160,1703){\makebox(0,0){\strut{}$\frac{2\langle S_B\rangle}{3N}$}}%
      \put(2401,80){\makebox(0,0){\strut{}$m_BL$}}%
    }%
    \gplgaddtomacro\gplfronttext{%
      \csname LTb\endcsname%
      \put(1728,2704){\makebox(0,0)[r]{\strut{}$N=8^2$}}%
      \csname LTb\endcsname%
      \put(1728,2544){\makebox(0,0)[r]{\strut{}$N=16^2$}}%
      \csname LTb\endcsname%
      \put(1728,2384){\makebox(0,0)[r]{\strut{}$N=24^2$}}%
    }%
    \gplbacktext
    \put(0,0){\includegraphics{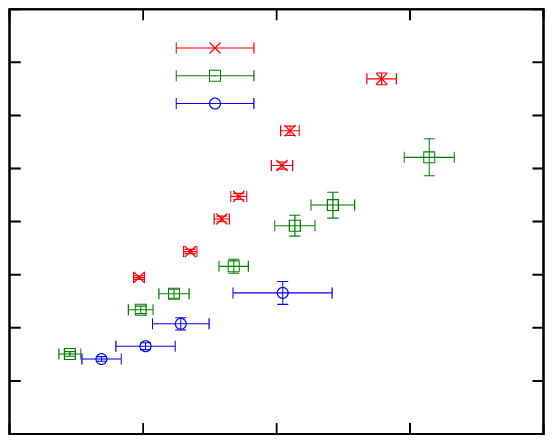}}%
    \gplfronttext
  \end{picture}%
\endgroup

%% file: conclusion.tex
\section{Conclusion}

The purpose of this article was to obtain a lattice formulation of the supersymmetric nonlinear 
O$(3)$ model which maintains the symmetries of the theory, at least in the continuum limit. 
The target manifold of the model is K\"ahler and hence it possesses an $\mathcal{N}=2$ supersymmetry, 
which could provide for a nilpotent supercharge and a discretization prescription based on this which 
maintains an exact supersymmetry on the lattice. We first derived the supersymmetry transformation 
in terms of constrained field variables and used these to analyze the applicability of such an approach
in the nonlinear O$(3)$ model. We realized that there is in fact no way to formulate a discretization of 
the theory which maintains simultanously both the O$(3)$ symmetry and a part of supersymmetry. We explicitly 
verified that the $Q$-exact supersymmetric formulation presented in \cite{Catterall:2006sj} breaks the O$(3)$ symmetry
such that it is not restored even in the continuum limit.
%
%
%
\\
In contrast, we started from a manifestly O$(3)$ symmetric discretization and investigated whether
the supersymmetry is restored in the continuum limit. The spherical geometry of the
target space is treated by two separate approaches, group valued variables on the one hand 
and a stereographic projection on the other hand. The numerical simulations are performed with the SLAC derivative and the
Wilson derivative respectively. The former does not break the chiral $\bbZ_2$ symmetry of the classical
action explicitly, allowing for the evaluation of histograms to verify the symmetric ground state structure 
which is spontaneously broken. The applicability of the SLAC derivative in theories with
curved target space was illustrated by a calculation of the step scaling function in the quenched model.
A disadvantage of the SLAC derivative, however, is the strong sign 
problem that becomes relevant already at comparably small lattices. 
In a sense the SLAC derivative is already too close to the continuum 
limit since it correctly reflects the intrinsic sign problem even on 
moderately sized lattices.
%
In order to test the supersymmetric properties of our lattice models we analyzed the masses 
of fermions and bosons as well as a Ward identity based on the bosonic action. For both derivatives 
the results indicate that the Ward identity is fulfilled in the continuum limit at finite
(large) physical volume. Concerning the expected degeneracy of the masses, no final statement can 
be done based on the small lattice sizes that are accessible by the SLAC derivative. The Wilson derivative
enables us to investigate larger lattices, but breaks chiral symmetry explicitly 
at finite lattice spacing, which leads to the renormalization of relevant operators in 
such a way that one is carried away from the supersymmetric continuum limit. 
Motivated by Super Yang-Mills theory, we fine tuned the fermionic mass in order
to remove the explicit breaking of the chiral symmetry. Furthermore, using the thus tuned ensemble, 
no "run away" of fermionic masses is visible, but rather a degeneracy of boson and fermion masses.
This indicates that a single parameter is sufficient to provide for a supersymmetric continuum limit
and for a cancellation of the encountered divergences. This, however, remains a conjecture until
explicitly checked by means of lattice perturbation theory.\\
We have therefore presented a lattice discretization which incorporates the O$(3)$ symmetry exactly 
at finite lattice spacing and furthermore shows restored supersymmetry as well as the 
spontaneously broken chiral symmetry 
in the continuum limit. The price to pay is a single additional fine tuning parameter.
It remains an open question whether further possibilities for the lattice derivative
like the Neuberger operator provide a discretization which is free from explicit chiral symmetry
breaking and perhaps free from the need for fine tuning, provided that only a mild sign 
problem is encountered at the same time.


%% file: invariance.tex
\section{Conventions and Fierz identities}
\label{app:con}
We choose the Majorana representation 
\begin{equation}
\gamma_0 =\sigma_3,\quad\gamma_1 =-\sigma_1,\quad \gamma_* = \ii\gamma_0 \gamma_1=\sigma_2,
\quad C =-\ii \sigma_2,~~
\label{cf1}
\end{equation}
and the conjugate spinor is defined as $\bar{\chi} = \chi^T C$. 
In the main body of the paper we employ the Fierz relation 
\begin{equation}
\psi\bar{\chi} = - \tfrac{1}{2} \bar{\chi} \psi - \tfrac{1}{2} (\bar{\chi} \gamma^{\mu} \psi) \gamma_{\mu} - \tfrac{1}{2} (\bar{\chi} \gamma_* \psi) \gamma_*.\label{cf3}
\end{equation}
Due to the symmetry properties
\begin{equation}
\bar{\chi}\psi = \bar{\psi} \chi,\quad
\bar{\chi}\gamma^\mu\psi = -\bar{\psi}\gamma^\mu\chi,\quad
\bar\chi\gamma_*\psi=-\bar\psi\gamma_*\chi\, 
\label{cf5}
\end{equation}
the two last terms in (\ref{cf3}) vanish for $\chi=\psi$ such that
\begin{equation}
\psi\bar\psi=-\tfrac{1}{2}\bar\psi\psi\,\id\,.\label{cf7}
\end{equation}

\section{Invariance of the action under the second supersymmetry}

We will prove the invariance of the on-shell action
\begin{equation}
S[\bn,\bpsi] = \int d^2x\left(\partial_{\mu} \bn\cdot\partial^{\mu} \bn + \ii \bar{\bpsi} \slashed{\partial} \bpsi + \tfrac{1}{4} (\bar{\bpsi}\bpsi)^2\right)
\end{equation}
under the second supersymmetry transformations 
(\ref{trafo2}).
The variation of the Lagrangian is\footnote{up to a negligible boundary term $\partial_{\mu}(-\bar{\bpsi}\gamma^{\mu}(\bar{\eps}\bpsi \times \bpsi^{\alpha}))$} 
\begin{equation} \label{varlagr}
\delta \mathcal{L} = 2\ii\partial_{\mu} \bn\cdot \partial^{\mu} (\bn \times \bar{\eps} \bpsi) - 2\ii \bar{\bpsi} \slashed{\partial} (\bn \times \slashed{\partial} \bn \eps) + 2\bar{\bpsi} \slashed{\partial} (\bar{\eps} \bpsi\times \bpsi) - (\bar{\bpsi} \bpsi) ~\bar{\bpsi} (\bn \times \slashed{\partial} \bn \eps)\,.
\end{equation}
The term $\propto \bpsi^5$ vanishes, since $\bpsi$ is a Grassmannian field with only four independent degrees of freedom. We will see that the first and second term cancel each other as well as the third and fourth one. Starting with the first two terms, they can be written as
\begin{equation}
2\ii\partial_{\mu} \bn  (\bn \times \bar{\eps}\partial^{\mu} \bpsi) - 2\ii\bar{\bpsi} (\bn\times \partial^2 \bn \eps) - 2\ii\bar\bpsi \gamma^{\mu} \gamma^{\nu} \eps (\partial_{\mu} \bn \times \partial_{\nu} \bn)\,.
\end{equation}
The last term vanishes since $\partial_\mu\bn\times \partial_\nu \bn$ is parallel to
$\bn$ and hence perpendicular to $\bpsi$. Integrating the second
term by parts one sees that the first and second term cancel
owing to the cyclicity of the triple product.\\
The cancellation of the third and fourth term in (\ref{varlagr}) is a bit
more involved. First, we partially integrate the third term and obtain $- 2 \partial_{\mu} \bar{\bpsi} \gamma^{\mu} (\bar{\eps} \bpsi \times \bpsi)$.
Since $\bn\cdot\bpsi^\alpha=0$ for both spinor components $\alpha$, we
conclude that $\bar\eps\bar\bpsi\times\bpsi$ is parallel to $\bn$
such that
\begin{equation}\label{b.4}
-2 \partial_{\mu} \bar{\bpsi} \gamma^{\mu} (\bar{\eps} \bpsi \times \bpsi) 
=-2 (\partial_{\mu} \bar{\bpsi} \gamma^{\mu} \bn)~ \bn(\bar{\eps} \bpsi \times \bpsi)\,.
\end{equation}
The condition $\bar\bpsi\bn=0$ implies $\partial_\mu\bar\bpsi\bn=
-\bar\bpsi\partial_\mu \bn$. \eqref{b.4} can hence be written as
$2 (\bar{\bpsi} \slashed{\partial}\bn) ~\bn(\bar{\eps} \bpsi \times \bpsi)$. 
To proceed further, we utilize $\bar\psi_1 n_1=-\bar\psi_2 n_2-\bar\psi_3 n_3$ and the Fierz relation
$\bar{\psi}_i \gamma^{\mu} \psi_i = 0$:
\begin{align}
& 2 (\bar{\psi}_1 \slashed{\partial} n_1 + \bar{\psi}_2 \slashed{\partial} n_2 + \bar{\psi}_3 \slashed{\partial} n_3)[n_1 \bar{\eps}\psi_2 \cdot \psi_3 - n_1 \bar{\eps}\psi_3 \cdot \psi_2 +\text{cyclic terms}]\\
&= 2\bar\psi_2\gamma^\mu\psi_3\bigl(\partial_\mu n_2\cdot n_1
-\partial_\mu n_1\cdot  n_2\bigr)\bar\eps\psi_2
  +2\bar\psi_3\gamma^\mu\psi_2\bigl(\partial_\mu n_1\cdot n_3
  -\partial_\mu n_3\cdot  n_1\bigr)\bar\eps\psi_3\nonumber
+\text{cyclic terms\,.}
\end{align}
Finally, we employ the Fierz relation $(\bar\alpha\gamma^\mu\beta)\,\bar\eps\alpha=
\half \bar\alpha\alpha\,(\bar\beta \gamma^\mu\eps)$, which holds for
Majorana spinors, and obtain
$$
(\bar\psi_2 \psi_2) \bar\psi_3\gamma^\mu\eps
(n_1\partial_\mu n_2  - n_2 \partial_\mu n_1)
+(\bar\psi_3 \psi_3) \bar \psi_2\gamma^\mu\eps 
(n_3 \partial_\mu n_1  - n_1\partial_\mu n_3) +\text{cyclic terms}\,.
$$
Finally, using $(\bar\alpha\alpha)\bar\alpha=0$ we obtain for the third term
in (\ref{varlagr}) the simple expression
\begin{equation}
(\bar{\bpsi}\bpsi)~ \bar{\bpsi}(\bn\times \slashed{\partial} \bn \eps)\,.
\end{equation}
As a result, the third and fourth term in (\ref{varlagr}) cancel each other.
This proves that the action is invariant under the second supersymmetry 
transformation (\ref{trafo2}).
\\
The invariance of the constraints can be shown easily:
\begin{eqnarray*}
\delta(\bn^2) &=& 2\ii\bn\cdot(\bn\times \bar{\eps} \bpsi)=0\\
\delta(\bn\cdot \bpsi) &=& \ii(\bn\times \bar{\eps}\bpsi)\cdot\bpsi 
- \bn\cdot (\bn\times \slashed{\partial} \bn \eps) -\ii \bn\cdot (\bar{\eps}\bpsi \times \bpsi) =0\,.
\end{eqnarray*}

%% file: measure.tex
\section{Transformation of the discretized measure} 
The transformation from the constrained fields $(\bo{n},\bo{\psi})$
to the unconstrained fields $(\bu,\blam)$ 
has a non-trivial Jacobian. Since the transformation only relates values 
of the fields on a fixed lattice site it is sufficient to calculate the 
Jacobian for a given site. Denoting values of the fields on this site by
$$
\bn=\begin{pmatrix} n_1\\ \bn_\perp \end{pmatrix},
\quad 
\bpsi=\begin{pmatrix}\psi_1\\ \bpsi_\perp\end{pmatrix}
$$
we are lead to consider
\begin{align*}
\delta(\bn^2-1)\delta(\bn\cdot \bpsi^1)\delta(\bn\cdot\bpsi^2) 
= &\frac{1}{2|n_{1}|} \left[\delta\Big(n_1 - \sqrt{1-\bn_\perp^2}\,\Big) 
+ \delta\Big(n_1 + \sqrt{1-\bn_\perp^2}\,\Big) \right] \\
&\cdot \prod_\alpha n_1 ~\delta\left(\psi_1^\alpha + \frac{\bn_\perp\cdot\bpsi^\alpha_\perp}{n_1}\right) 
\,.
\end{align*}
Consequently, the measure on a given site transforms as
\begin{equation}
\dd \bn\, \dd \bpsi^{1}\,\dd\bpsi^2 ~ \delta(\bn^2-1) 
\delta(\bn\cdot\bpsi^1) \delta(\bn\cdot\bpsi^2)
=\tfrac{1}{2}~\hbox{J}(\bu)\,\dd \bu\,\dd\blam^1\,\dd\blam^2
\end{equation}
with Jacobian
\begin{equation}
\hbox{J}(\bu)=\sqrt{1-\bn_\perp^2(\bu)} 
~ \big|\mathrm{sdet} \{(\bn_\perp, \bpsi_\perp) \rightarrow (\bu,\blam)\}\big|\,.
\label{measuretrans}
\end{equation}
In the super-stereographic projection (\ref{proj}) $\bn$ does not depend on
$\blam^\alpha$, and $\bpsi^{\alpha}$ does not depend on $\blam^\beta$ for 
$\beta \neq \alpha$. The superdeterminant is hence given by
\begin{equation}
\mathrm{sdet} \{(\bn_\perp,\bpsi_\perp) \rightarrow (\bu,\blam)\} 
= \frac{\det(\partial \bn_\perp / \partial \bu)}
{\det(\partial \bpsi_\perp^1 /\partial \blam^1) 
\cdot \det(\partial \bpsi_\perp^2 /\partial \blam^2)}\,.
\end{equation}
In an O$(N)$ model, all three determinants are equal to
$$
(2\rho)^{N-1}\,\frac{1-\bu^2}{1+\bu^2}\quad\hbox{with}\quad \rho=\frac{1}{1+\bu^2}\,.
$$ 
Expressing the square root in (\ref{measuretrans}) in terms of
the new fields,
$$
\sqrt{1-\bn_\perp}=\frac{1-\bu^2}{1+\bu^2},
$$
we end up with the Jacobian
\begin{equation}
\label{eq:app:measure}
\hbox{J}(\bu)=\frac{1}{ (2\rho)^{N-1}}
\propto\left(1+\bu^2\right)^{N-1}\,.
\end{equation}
The functional integral measure for the supersymmetric O$(3)$ model with uncontrained fields is thus
\begin{equation}
\prod_{x} \dd\bu_x\,\dd\blam_x^1\,\dd\blam_x^2\;\left(1+\bu_x^2\right)^2 \;.
\end{equation}
Note that we proved on the way that the Jacobian for the purely bosonic O$(N)$ 
model is
\begin{equation}
{\rm J}_{\rm B}(\bu)\propto \rho^{N-1}=\frac{1}{(1+\bu^2)^{N-1}}\,.
\label{measure_bos}
\end{equation}


%

%% file: derivativeEvaluation.tex
\section{Sign of the fermion determinant}
\FIGURE{
\input{pfaffianSignSigma}\hfill
\input{stereo_pfaffiansign_finetuned}\hfill
\caption{\label{fig:sigma:pfaffianSign} Left Panel: Average sign of the SLAC Pfaffian for
different couplings $g^{-2}$ on lattices sizes ranging from $5\times 5$ to
$11\times 11$. Right Panel: Sign of the Wilson Pfaffian on a $N=16^2$ lattice for different couplings
$g^{-2}$ and normalized finetuning parameter $\kappa/\kappa_c$.}}
\noindent In order to check whether the sign-quenched approximation is applicable in case of the SLAC derivative, 
simulations on lattice sizes ranging from $5\times 5$ to $11\times 11$ have been performed over
a coupling range $g^{-2}\in[0.4,1.2]$. The results that are based on $10^5$
configurations per data point (see Fig.\ref{fig:sigma:pfaffianSign}, left panel) indicate
that the average sign of the Pfaffian is smaller for smaller $g^{-2}$, which is equivalent to
coarser lattices. The problem is that the sign problem worsens for larger
lattice volumes at fixed coupling. In these cases the probability based, i.e.
sign-quenched, Monte-Carlo sampling will \emph{not} correspond to the relevant
configurations in an unquenched ensemble and statistical errors on reweighted 
measurements will become rather large.
Nevertheless, with standard Monte-Carlo techniques simulations are
only possible without taking the sign into account, such that a
reweighting becomes unavoidable. The discretization based on the SLAC derivative becomes thus
unfeasible for larger lattices, which we require in order to study the continuum limit and 
a possible restoration of supersymmetry there. As a consequence, we have to rely for this 
purpose on the Wilson derivative only. The average Pfaffian sign for the 
Wilson derivative is depicted in the right panel of Fig. 
\ref{fig:sigma:pfaffianSign}. We see that for small values of the 
finetuning parameter $\kappa$,
\rFIGURE{\hfill
\input{stereo_pfaffiansign_wilsonslac}\hfill
\caption{\label{fig:sigma:pfaffianSignWilsonSLac} Average Pfaffian sign 
for different box sizes. Wilson sign data is gathered from the finetuned 
ensemble at $\kappa_c$.}}
which correspond to little finetuning, 
changes in the Pfaffian sign are suppressed and it is possible to 
evaluate expectation values directly without the need of reweighting. 
In the vicinity of $\kappa\approx\kappa_c$ however, this behaviour 
changes to a mild correction for large coupling $g^{-2}=2$ up to a 
significant correction for smaller coupling $g^{-2}=1.2$. In the 
finetuned ensemble at $\kappa=\kappa_c$, we regain a sign problem as 
shown in figure \ref{fig:sigma:pfaffianSignWilsonSLac}. However, even 
for lattice volume $N=16^2$ and reasonable box sizes, the average sign 
for Wilson fermions lies considerably above the SLAC data at lattice 
volume $11^2$. Also, the Wilson $N=24^2$ data does not indicate a drastic
decrease of the average sign for larger lattices, as seen in the SLAC 
case. Additionally, applying the Wilson formulation allows for an efficient 
even-odd preconditioning scheme, which provides a sufficient speed-up 
in order to account for the larger configuration numbers that are needed 
for the reweighting procedure.

%% file: pfaffianSignSigma.tex
\begingroup
\footnotesize
  \makeatletter
  \providecommand\color[2][]{%
    \GenericError{(gnuplot) \space\space\space\@spaces}{%
      Package color not loaded in conjunction with
      terminal option `colourtext'%
    }{See the gnuplot documentation for explanation.%
    }{Either use 'blacktext' in gnuplot or load the package
      color.sty in LaTeX.}%
    \renewcommand\color[2][]{}%
  }%
  \providecommand\includegraphics[2][]{%
    \GenericError{(gnuplot) \space\space\space\@spaces}{%
      Package graphicx or graphics not loaded%
    }{See the gnuplot documentation for explanation.%
    }{The gnuplot epslatex terminal needs graphicx.sty or graphics.sty.}%
    \renewcommand\includegraphics[2][]{}%
  }%
  \providecommand\rotatebox[2]{#2}%
  \@ifundefined{ifGPcolor}{%
    \newif\ifGPcolor
    \GPcolortrue
  }{}%
  \@ifundefined{ifGPblacktext}{%
    \newif\ifGPblacktext
    \GPblacktexttrue
  }{}%
  \let\gplgaddtomacro\g@addto@macro
  \gdef\gplbacktext{}%
  \gdef\gplfronttext{}%
  \makeatother
  \ifGPblacktext
    \def\colorrgb#1{}%
    \def\colorgray#1{}%
  \else
    \ifGPcolor
      \def\colorrgb#1{\color[rgb]{#1}}%
      \def\colorgray#1{\color[gray]{#1}}%
      \expandafter\def\csname LTw\endcsname{\color{white}}%
      \expandafter\def\csname LTb\endcsname{\color{black}}%
      \expandafter\def\csname LTa\endcsname{\color{black}}%
      \expandafter\def\csname LT0\endcsname{\color[rgb]{1,0,0}}%
      \expandafter\def\csname LT1\endcsname{\color[rgb]{0,1,0}}%
      \expandafter\def\csname LT2\endcsname{\color[rgb]{0,0,1}}%
      \expandafter\def\csname LT3\endcsname{\color[rgb]{1,0,1}}%
      \expandafter\def\csname LT4\endcsname{\color[rgb]{0,1,1}}%
      \expandafter\def\csname LT5\endcsname{\color[rgb]{1,1,0}}%
      \expandafter\def\csname LT6\endcsname{\color[rgb]{0,0,0}}%
      \expandafter\def\csname LT7\endcsname{\color[rgb]{1,0.3,0}}%
      \expandafter\def\csname LT8\endcsname{\color[rgb]{0.5,0.5,0.5}}%
    \else
      \def\colorrgb#1{\color{black}}%
      \def\colorgray#1{\color[gray]{#1}}%
      \expandafter\def\csname LTw\endcsname{\color{white}}%
      \expandafter\def\csname LTb\endcsname{\color{black}}%
      \expandafter\def\csname LTa\endcsname{\color{black}}%
      \expandafter\def\csname LT0\endcsname{\color{black}}%
      \expandafter\def\csname LT1\endcsname{\color{black}}%
      \expandafter\def\csname LT2\endcsname{\color{black}}%
      \expandafter\def\csname LT3\endcsname{\color{black}}%
      \expandafter\def\csname LT4\endcsname{\color{black}}%
      \expandafter\def\csname LT5\endcsname{\color{black}}%
      \expandafter\def\csname LT6\endcsname{\color{black}}%
      \expandafter\def\csname LT7\endcsname{\color{black}}%
      \expandafter\def\csname LT8\endcsname{\color{black}}%
    \fi
  \fi
  \setlength{\unitlength}{0.0500bp}%
  \begin{picture}(3968.00,2976.00)%
    \gplgaddtomacro\gplbacktext{%
      \csname LTb\endcsname%
      \put(768,480){\makebox(0,0)[r]{\strut{}$0.0$}}%
      \put(768,969){\makebox(0,0)[r]{\strut{}$0.2$}}%
      \put(768,1459){\makebox(0,0)[r]{\strut{}$0.4$}}%
      \put(768,1948){\makebox(0,0)[r]{\strut{}$0.6$}}%
      \put(768,2438){\makebox(0,0)[r]{\strut{}$0.8$}}%
      \put(768,2927){\makebox(0,0)[r]{\strut{}$1.0$}}%
      \put(864,320){\makebox(0,0){\strut{}$0.4$}}%
      \put(1248,320){\makebox(0,0){\strut{}$0.5$}}%
      \put(1633,320){\makebox(0,0){\strut{}$0.6$}}%
      \put(2017,320){\makebox(0,0){\strut{}$0.7$}}%
      \put(2402,320){\makebox(0,0){\strut{}$0.8$}}%
      \put(2786,320){\makebox(0,0){\strut{}$0.9$}}%
      \put(3170,320){\makebox(0,0){\strut{}$1.0$}}%
      \put(3555,320){\makebox(0,0){\strut{}$1.1$}}%
      \put(3939,320){\makebox(0,0){\strut{}$1.2$}}%
      \put(256,1703){\rotatebox{90}{\makebox(0,0){\strut{}$\vev{\sgn\Pf Q}$}}}%
      \put(2401,80){\makebox(0,0){\strut{}$g^{-2}$}}%
    }%
    \gplgaddtomacro\gplfronttext{%
      \csname LTb\endcsname%
      \put(3204,1103){\makebox(0,0)[r]{\strut{}$N=5^2$}}%
      \csname LTb\endcsname%
      \put(3204,943){\makebox(0,0)[r]{\strut{}$N=7^2$}}%
      \csname LTb\endcsname%
      \put(3204,783){\makebox(0,0)[r]{\strut{}$N=9^2$}}%
      \csname LTb\endcsname%
      \put(3204,623){\makebox(0,0)[r]{\strut{}$N=11^2$}}%
    }%
    \gplbacktext
    \put(0,0){\includegraphics{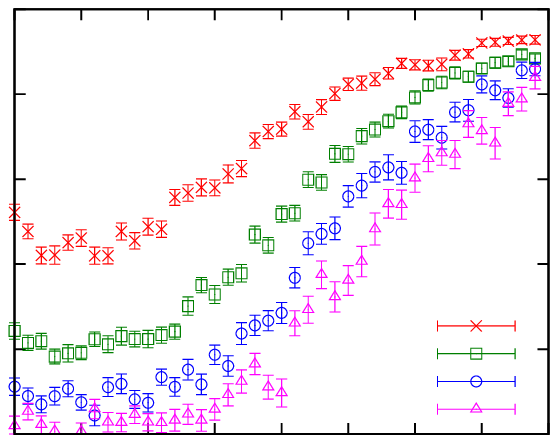}}%
    \gplfronttext
  \end{picture}%
\endgroup

%% file: stereo_pfaffiansign_finetuned.tex
\begingroup
\footnotesize
  \makeatletter
  \providecommand\color[2][]{%
    \GenericError{(gnuplot) \space\space\space\@spaces}{%
      Package color not loaded in conjunction with
      terminal option `colourtext'%
    }{See the gnuplot documentation for explanation.%
    }{Either use 'blacktext' in gnuplot or load the package
      color.sty in LaTeX.}%
    \renewcommand\color[2][]{}%
  }%
  \providecommand\includegraphics[2][]{%
    \GenericError{(gnuplot) \space\space\space\@spaces}{%
      Package graphicx or graphics not loaded%
    }{See the gnuplot documentation for explanation.%
    }{The gnuplot epslatex terminal needs graphicx.sty or graphics.sty.}%
    \renewcommand\includegraphics[2][]{}%
  }%
  \providecommand\rotatebox[2]{#2}%
  \@ifundefined{ifGPcolor}{%
    \newif\ifGPcolor
    \GPcolortrue
  }{}%
  \@ifundefined{ifGPblacktext}{%
    \newif\ifGPblacktext
    \GPblacktexttrue
  }{}%
  \let\gplgaddtomacro\g@addto@macro
  \gdef\gplbacktext{}%
  \gdef\gplfronttext{}%
  \makeatother
  \ifGPblacktext
    \def\colorrgb#1{}%
    \def\colorgray#1{}%
  \else
    \ifGPcolor
      \def\colorrgb#1{\color[rgb]{#1}}%
      \def\colorgray#1{\color[gray]{#1}}%
      \expandafter\def\csname LTw\endcsname{\color{white}}%
      \expandafter\def\csname LTb\endcsname{\color{black}}%
      \expandafter\def\csname LTa\endcsname{\color{black}}%
      \expandafter\def\csname LT0\endcsname{\color[rgb]{1,0,0}}%
      \expandafter\def\csname LT1\endcsname{\color[rgb]{0,1,0}}%
      \expandafter\def\csname LT2\endcsname{\color[rgb]{0,0,1}}%
      \expandafter\def\csname LT3\endcsname{\color[rgb]{1,0,1}}%
      \expandafter\def\csname LT4\endcsname{\color[rgb]{0,1,1}}%
      \expandafter\def\csname LT5\endcsname{\color[rgb]{1,1,0}}%
      \expandafter\def\csname LT6\endcsname{\color[rgb]{0,0,0}}%
      \expandafter\def\csname LT7\endcsname{\color[rgb]{1,0.3,0}}%
      \expandafter\def\csname LT8\endcsname{\color[rgb]{0.5,0.5,0.5}}%
    \else
      \def\colorrgb#1{\color{black}}%
      \def\colorgray#1{\color[gray]{#1}}%
      \expandafter\def\csname LTw\endcsname{\color{white}}%
      \expandafter\def\csname LTb\endcsname{\color{black}}%
      \expandafter\def\csname LTa\endcsname{\color{black}}%
      \expandafter\def\csname LT0\endcsname{\color{black}}%
      \expandafter\def\csname LT1\endcsname{\color{black}}%
      \expandafter\def\csname LT2\endcsname{\color{black}}%
      \expandafter\def\csname LT3\endcsname{\color{black}}%
      \expandafter\def\csname LT4\endcsname{\color{black}}%
      \expandafter\def\csname LT5\endcsname{\color{black}}%
      \expandafter\def\csname LT6\endcsname{\color{black}}%
      \expandafter\def\csname LT7\endcsname{\color{black}}%
      \expandafter\def\csname LT8\endcsname{\color{black}}%
    \fi
  \fi
  \setlength{\unitlength}{0.0500bp}%
  \begin{picture}(3968.00,2976.00)%
    \gplgaddtomacro\gplbacktext{%
      \csname LTb\endcsname%
      \put(768,480){\makebox(0,0)[r]{\strut{}$0.60$}}%
      \put(768,1092){\makebox(0,0)[r]{\strut{}$0.70$}}%
      \put(768,1703){\makebox(0,0)[r]{\strut{}$0.80$}}%
      \put(768,2315){\makebox(0,0)[r]{\strut{}$0.90$}}%
      \put(768,2927){\makebox(0,0)[r]{\strut{}$1.00$}}%
      \put(864,320){\makebox(0,0){\strut{}$0.80$}}%
      \put(1377,320){\makebox(0,0){\strut{}$0.85$}}%
      \put(1889,320){\makebox(0,0){\strut{}$0.90$}}%
      \put(2402,320){\makebox(0,0){\strut{}$0.95$}}%
      \put(2914,320){\makebox(0,0){\strut{}$1.00$}}%
      \put(3427,320){\makebox(0,0){\strut{}$1.05$}}%
      \put(3939,320){\makebox(0,0){\strut{}$1.10$}}%
      \put(160,1703){\rotatebox{90}{\makebox(0,0){\strut{}$\langle\sgn\Pf Q\rangle$}}}%
      \put(2401,80){\makebox(0,0){\strut{}$\kappa / \kappa_c$}}%
    }%
    \gplgaddtomacro\gplfronttext{%
      \csname LTb\endcsname%
      \put(1728,1103){\makebox(0,0)[r]{\strut{}$g^{-2}=1.2$}}%
      \csname LTb\endcsname%
      \put(1728,943){\makebox(0,0)[r]{\strut{}$g^{-2}=1.3$}}%
      \csname LTb\endcsname%
      \put(1728,783){\makebox(0,0)[r]{\strut{}$g^{-2}=1.4$}}%
      \csname LTb\endcsname%
      \put(1728,623){\makebox(0,0)[r]{\strut{}$g^{-2}=2.0$}}%
    }%
    \gplbacktext
    \put(0,0){\includegraphics{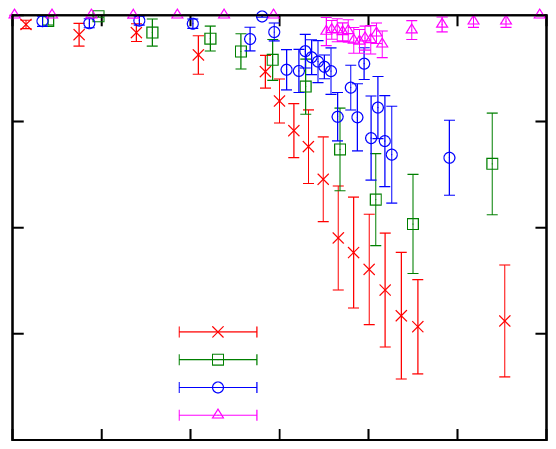}}%
    \gplfronttext
  \end{picture}%
\endgroup

%% file: stereo_pfaffiansign_wilsonslac.tex
\begingroup
\footnotesize
  \makeatletter
  \providecommand\color[2][]{%
    \GenericError{(gnuplot) \space\space\space\@spaces}{%
      Package color not loaded in conjunction with
      terminal option `colourtext'%
    }{See the gnuplot documentation for explanation.%
    }{Either use 'blacktext' in gnuplot or load the package
      color.sty in LaTeX.}%
    \renewcommand\color[2][]{}%
  }%
  \providecommand\includegraphics[2][]{%
    \GenericError{(gnuplot) \space\space\space\@spaces}{%
      Package graphicx or graphics not loaded%
    }{See the gnuplot documentation for explanation.%
    }{The gnuplot epslatex terminal needs graphicx.sty or graphics.sty.}%
    \renewcommand\includegraphics[2][]{}%
  }%
  \providecommand\rotatebox[2]{#2}%
  \@ifundefined{ifGPcolor}{%
    \newif\ifGPcolor
    \GPcolortrue
  }{}%
  \@ifundefined{ifGPblacktext}{%
    \newif\ifGPblacktext
    \GPblacktexttrue
  }{}%
  \let\gplgaddtomacro\g@addto@macro
  \gdef\gplbacktext{}%
  \gdef\gplfronttext{}%
  \makeatother
  \ifGPblacktext
    \def\colorrgb#1{}%
    \def\colorgray#1{}%
  \else
    \ifGPcolor
      \def\colorrgb#1{\color[rgb]{#1}}%
      \def\colorgray#1{\color[gray]{#1}}%
      \expandafter\def\csname LTw\endcsname{\color{white}}%
      \expandafter\def\csname LTb\endcsname{\color{black}}%
      \expandafter\def\csname LTa\endcsname{\color{black}}%
      \expandafter\def\csname LT0\endcsname{\color[rgb]{1,0,0}}%
      \expandafter\def\csname LT1\endcsname{\color[rgb]{0,1,0}}%
      \expandafter\def\csname LT2\endcsname{\color[rgb]{0,0,1}}%
      \expandafter\def\csname LT3\endcsname{\color[rgb]{1,0,1}}%
      \expandafter\def\csname LT4\endcsname{\color[rgb]{0,1,1}}%
      \expandafter\def\csname LT5\endcsname{\color[rgb]{1,1,0}}%
      \expandafter\def\csname LT6\endcsname{\color[rgb]{0,0,0}}%
      \expandafter\def\csname LT7\endcsname{\color[rgb]{1,0.3,0}}%
      \expandafter\def\csname LT8\endcsname{\color[rgb]{0.5,0.5,0.5}}%
    \else
      \def\colorrgb#1{\color{black}}%
      \def\colorgray#1{\color[gray]{#1}}%
      \expandafter\def\csname LTw\endcsname{\color{white}}%
      \expandafter\def\csname LTb\endcsname{\color{black}}%
      \expandafter\def\csname LTa\endcsname{\color{black}}%
      \expandafter\def\csname LT0\endcsname{\color{black}}%
      \expandafter\def\csname LT1\endcsname{\color{black}}%
      \expandafter\def\csname LT2\endcsname{\color{black}}%
      \expandafter\def\csname LT3\endcsname{\color{black}}%
      \expandafter\def\csname LT4\endcsname{\color{black}}%
      \expandafter\def\csname LT5\endcsname{\color{black}}%
      \expandafter\def\csname LT6\endcsname{\color{black}}%
      \expandafter\def\csname LT7\endcsname{\color{black}}%
      \expandafter\def\csname LT8\endcsname{\color{black}}%
    \fi
  \fi
  \setlength{\unitlength}{0.0500bp}%
  \begin{picture}(3968.00,2976.00)%
    \gplgaddtomacro\gplbacktext{%
      \csname LTb\endcsname%
      \put(768,480){\makebox(0,0)[r]{\strut{}$0.00$}}%
      \put(768,725){\makebox(0,0)[r]{\strut{}$0.10$}}%
      \put(768,969){\makebox(0,0)[r]{\strut{}$0.20$}}%
      \put(768,1214){\makebox(0,0)[r]{\strut{}$0.30$}}%
      \put(768,1459){\makebox(0,0)[r]{\strut{}$0.40$}}%
      \put(768,1704){\makebox(0,0)[r]{\strut{}$0.50$}}%
      \put(768,1948){\makebox(0,0)[r]{\strut{}$0.60$}}%
      \put(768,2193){\makebox(0,0)[r]{\strut{}$0.70$}}%
      \put(768,2438){\makebox(0,0)[r]{\strut{}$0.80$}}%
      \put(768,2682){\makebox(0,0)[r]{\strut{}$0.90$}}%
      \put(768,2927){\makebox(0,0)[r]{\strut{}$1.00$}}%
      \put(864,320){\makebox(0,0){\strut{}$1.0$}}%
      \put(1172,320){\makebox(0,0){\strut{}$1.5$}}%
      \put(1479,320){\makebox(0,0){\strut{}$2.0$}}%
      \put(1787,320){\makebox(0,0){\strut{}$2.5$}}%
      \put(2094,320){\makebox(0,0){\strut{}$3.0$}}%
      \put(2402,320){\makebox(0,0){\strut{}$3.5$}}%
      \put(2709,320){\makebox(0,0){\strut{}$4.0$}}%
      \put(3017,320){\makebox(0,0){\strut{}$4.5$}}%
      \put(3324,320){\makebox(0,0){\strut{}$5.0$}}%
      \put(3632,320){\makebox(0,0){\strut{}$5.5$}}%
      \put(3939,320){\makebox(0,0){\strut{}$6.0$}}%
      \put(160,1703){\rotatebox{90}{\makebox(0,0){\strut{}$\langle\sgn\Pf Q\rangle$}}}%
      \put(2401,80){\makebox(0,0){\strut{}$mL$}}%
    }%
    \gplgaddtomacro\gplfronttext{%
      \csname LTb\endcsname%
      \put(3204,2784){\makebox(0,0)[r]{\strut{}$\text{Wilson }N=16^2$}}%
      \csname LTb\endcsname%
      \put(3204,2624){\makebox(0,0)[r]{\strut{}$\text{Wilson }N=24^2$}}%
      \csname LTb\endcsname%
      \put(3204,2464){\makebox(0,0)[r]{\strut{}$\text{SLAC }N=9^2$}}%
      \csname LTb\endcsname%
      \put(3204,2304){\makebox(0,0)[r]{\strut{}$\text{SLAC }N=11^2$}}%
    }%
    \gplbacktext
    \put(0,0){\includegraphics{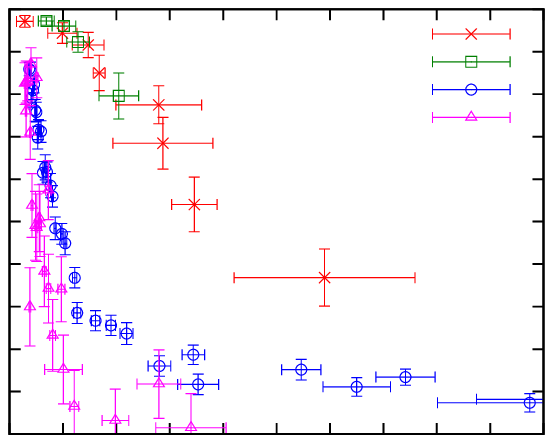}}%
    \gplfronttext
  \end{picture}%
\endgroup

%% file: algorithms.tex
\section{Algorithmic aspects}
We argued in section \ref{ch:o3symmetry} that it is of considerable
importance to tune the many technical parameters of advanced algorithms like
Rational Hybrid Monte Carlo (RHMC) in the right way in order to obtain an appropriate
algorithm that parses configuration space in the correct way. With the
sign problem and large condition numbers lurking, a Hybrid Monte Carlo algorithm
with exact evaluation of the fermion determinant and the inverse fermion matrix
using efficient LAPACK routines was used at small lattice volumes up to $16^2$ in order
to provide a solid ground for the implementation of more sophisticated pseudofermion
algorithms, e.g. RHMC. These algorithms become necessary at large volumes due to the poor
scaling behaviour of the LU decomposition used to solve for the inverse fermion
matrix. However, no efficient preconditiong schemes exist for the
nonlocal SLAC derivative. Therefore we concentrated our efforts on the Wilson
derivative which allows for even-odd preconditioning and, moreover, shows less severe sign
problems at finite volume.
\TABLE{
\begin{tabular}{cccc} \hline \hline
	 			  & original & reduced & even-odd \\ \hline
condition number  & $1.6(6)\cdot 10^{19}$ & $1.3(6)\cdot 10^8$ & $1.4(7)\cdot
10^3$ \\ cg solver steps   & $454(10)$        & $152(1)$      & $48(1)$        \\ \hline
\hline
\end{tabular}
\caption{\label{tab:sigma:stereo_algo} Average condition numbers and CG solver
steps for three different choices of fermion matrix formulation.}} 
In Table \ref{tab:sigma:stereo_algo} average condition numbers
$\kappa$ obtained from the exact matrix norm $\kappa = ||Q||\cdot ||Q^{-1}||$ are shown
as well as typical iteration numbers of a conjugate gradient solver for three different choices of
the fermion matrix: 1. the original fermion matrix, 2. the reduced fermion matrix where a factor $\rho$, whose determinant can
be evaluated analytically, is separated on both sides,
\begin{multline}
{Q'}_{xy,ij}^{\alpha\beta} =  4\big( \delta_{ik}-2u_{x,i}u_{x,k}\rho_x
\big)M_{xy}^{\alpha\beta} \big( \delta_{kj}-2u_{y,k}u_{y,j}\rho_y \big) \\+ 16
\rho_xu_{x,i}M_{xy}^{\alpha\beta}u_{y,j}\rho_y
 +4\beta\sigma_x\delta_{xy}\delta_{ij}\delta^{\alpha\beta},
\end{multline}
and 3. the preconditioned reduced fermion matrix using the well-known even-odd 
preconditioning scheme \cite{DeGrand:1990}. The critical step of the iterative cg solver 
is the application of the inverse of $Q^TQ$, which is used in the pseudofermion algorithm, 
to some random (pseudofermion) vector, $Y=(Q^TQ)^{-1}X$. We see a significant improvement 
in the number of solver steps and in the condition numbers for the reduced fermion matrix.
This can directly be verified by looking at the eigenvalue spectrum depicted in figure
\ref{fig:sigma:stereo_ev}.
\FIGURE{\hfill \input{stereo_eigenvalues_12x12_2_exact}\hfill
 \input{stereo_eigenvalues_12x12_2_evenodd}\hfill\hfill\phantom.
\caption{\label{fig:sigma:stereo_ev} Eigenvalue frequency of the
original and reduced fermion matrix (left panel, normalized such that both the
largest eigenvalue and the integrated surface equal one) and for the even-odd
preconditioned matrix (right panel, no normalization, logarithmic scale) from a
sample of $2000$ configurations on a $12^2$ lattice.}} 
Both matrices, the original and the reduced one, are real and antisymmetric, so that all
eigenvalues are purely imaginary and come in complex conjugate pairs.
Whereas the original fermion matrix exhibits a large number of eigenvalues very
close to zero\footnote{This is expected according to the Banks-Casher-relation.}, this is
not the case for the reduced one and the condition numbers hence decrease
drastically. A further improvement is achieved by even-odd preconditioning.
The preconditioned matrix is no longer antisymmetric and we see that the
majority of eigenvalues lie close to $1$. This is what we expected, since 
the goal of the incomplete LU preconditioning \cite{Oyanagi1986,Peardon:2000}, which can be
regarded as a generalization of the even-odd scheme, is to rewrite the fermion matrix in the
form $1-L-U$, where $1$ is the identity matrix and $L$ ($U$) is a lower (upper)
triangular matrix.

%% file: stereo_eigenvalues_12x12_2_exact.tex
\begingroup
\footnotesize
  \makeatletter
  \providecommand\color[2][]{%
    \GenericError{(gnuplot) \space\space\space\@spaces}{%
      Package color not loaded in conjunction with
      terminal option `colourtext'%
    }{See the gnuplot documentation for explanation.%
    }{Either use 'blacktext' in gnuplot or load the package
      color.sty in LaTeX.}%
    \renewcommand\color[2][]{}%
  }%
  \providecommand\includegraphics[2][]{%
    \GenericError{(gnuplot) \space\space\space\@spaces}{%
      Package graphicx or graphics not loaded%
    }{See the gnuplot documentation for explanation.%
    }{The gnuplot epslatex terminal needs graphicx.sty or graphics.sty.}%
    \renewcommand\includegraphics[2][]{}%
  }%
  \providecommand\rotatebox[2]{#2}%
  \@ifundefined{ifGPcolor}{%
    \newif\ifGPcolor
    \GPcolortrue
  }{}%
  \@ifundefined{ifGPblacktext}{%
    \newif\ifGPblacktext
    \GPblacktexttrue
  }{}%
  \let\gplgaddtomacro\g@addto@macro
  \gdef\gplbacktext{}%
  \gdef\gplfronttext{}%
  \makeatother
  \ifGPblacktext
    \def\colorrgb#1{}%
    \def\colorgray#1{}%
  \else
    \ifGPcolor
      \def\colorrgb#1{\color[rgb]{#1}}%
      \def\colorgray#1{\color[gray]{#1}}%
      \expandafter\def\csname LTw\endcsname{\color{white}}%
      \expandafter\def\csname LTb\endcsname{\color{black}}%
      \expandafter\def\csname LTa\endcsname{\color{black}}%
      \expandafter\def\csname LT0\endcsname{\color[rgb]{1,0,0}}%
      \expandafter\def\csname LT1\endcsname{\color[rgb]{0,1,0}}%
      \expandafter\def\csname LT2\endcsname{\color[rgb]{0,0,1}}%
      \expandafter\def\csname LT3\endcsname{\color[rgb]{1,0,1}}%
      \expandafter\def\csname LT4\endcsname{\color[rgb]{0,1,1}}%
      \expandafter\def\csname LT5\endcsname{\color[rgb]{1,1,0}}%
      \expandafter\def\csname LT6\endcsname{\color[rgb]{0,0,0}}%
      \expandafter\def\csname LT7\endcsname{\color[rgb]{1,0.3,0}}%
      \expandafter\def\csname LT8\endcsname{\color[rgb]{0.5,0.5,0.5}}%
    \else
      \def\colorrgb#1{\color{black}}%
      \def\colorgray#1{\color[gray]{#1}}%
      \expandafter\def\csname LTw\endcsname{\color{white}}%
      \expandafter\def\csname LTb\endcsname{\color{black}}%
      \expandafter\def\csname LTa\endcsname{\color{black}}%
      \expandafter\def\csname LT0\endcsname{\color{black}}%
      \expandafter\def\csname LT1\endcsname{\color{black}}%
      \expandafter\def\csname LT2\endcsname{\color{black}}%
      \expandafter\def\csname LT3\endcsname{\color{black}}%
      \expandafter\def\csname LT4\endcsname{\color{black}}%
      \expandafter\def\csname LT5\endcsname{\color{black}}%
      \expandafter\def\csname LT6\endcsname{\color{black}}%
      \expandafter\def\csname LT7\endcsname{\color{black}}%
      \expandafter\def\csname LT8\endcsname{\color{black}}%
    \fi
  \fi
  \setlength{\unitlength}{0.0500bp}%
  \begin{picture}(3968.00,2976.00)%
    \gplgaddtomacro\gplbacktext{%
      \csname LTb\endcsname%
      \put(768,480){\makebox(0,0)[r]{\strut{} 0}}%
      \put(768,888){\makebox(0,0)[r]{\strut{} 0.05}}%
      \put(768,1296){\makebox(0,0)[r]{\strut{} 0.1}}%
      \put(768,1704){\makebox(0,0)[r]{\strut{} 0.15}}%
      \put(768,2111){\makebox(0,0)[r]{\strut{} 0.2}}%
      \put(768,2519){\makebox(0,0)[r]{\strut{} 0.25}}%
      \put(768,2927){\makebox(0,0)[r]{\strut{} 0.3}}%
      \put(864,320){\makebox(0,0){\strut{}-1}}%
      \put(1172,320){\makebox(0,0){\strut{}-0.8}}%
      \put(1479,320){\makebox(0,0){\strut{}-0.6}}%
      \put(1786,320){\makebox(0,0){\strut{}-0.4}}%
      \put(2094,320){\makebox(0,0){\strut{}-0.2}}%
      \put(2402,320){\makebox(0,0){\strut{} 0}}%
      \put(2709,320){\makebox(0,0){\strut{} 0.2}}%
      \put(3017,320){\makebox(0,0){\strut{} 0.4}}%
      \put(3324,320){\makebox(0,0){\strut{} 0.6}}%
      \put(3632,320){\makebox(0,0){\strut{} 0.8}}%
      \put(3939,320){\makebox(0,0){\strut{} 1}}%
      \put(112,1703){\rotatebox{90}{\makebox(0,0){\strut{}eigenvalue frequency}}}%
      \put(2401,80){\makebox(0,0){\strut{}imaginary part}}%
    }%
    \gplgaddtomacro\gplfronttext{%
      \csname LTb\endcsname%
      \put(3204,2784){\makebox(0,0)[r]{\strut{}original}}%
      \csname LTb\endcsname%
      \put(3204,2624){\makebox(0,0)[r]{\strut{}reduced}}%
    }%
    \gplbacktext
    \put(0,0){\includegraphics{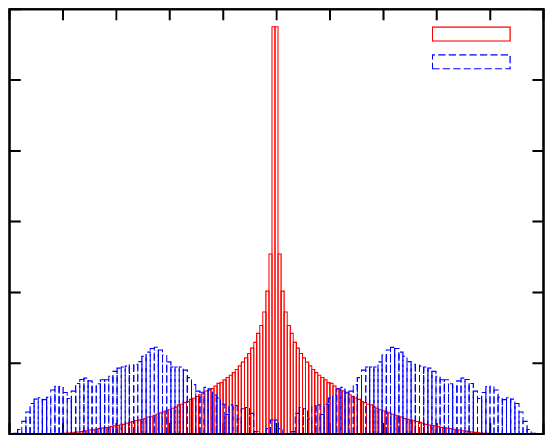}}%
    \gplfronttext
  \end{picture}%
\endgroup

%% file: stereo_eigenvalues_12x12_2_evenodd.tex
\begingroup
\footnotesize
  \makeatletter
  \providecommand\color[2][]{%
    \GenericError{(gnuplot) \space\space\space\@spaces}{%
      Package color not loaded in conjunction with
      terminal option `colourtext'%
    }{See the gnuplot documentation for explanation.%
    }{Either use 'blacktext' in gnuplot or load the package
      color.sty in LaTeX.}%
    \renewcommand\color[2][]{}%
  }%
  \providecommand\includegraphics[2][]{%
    \GenericError{(gnuplot) \space\space\space\@spaces}{%
      Package graphicx or graphics not loaded%
    }{See the gnuplot documentation for explanation.%
    }{The gnuplot epslatex terminal needs graphicx.sty or graphics.sty.}%
    \renewcommand\includegraphics[2][]{}%
  }%
  \providecommand\rotatebox[2]{#2}%
  \@ifundefined{ifGPcolor}{%
    \newif\ifGPcolor
    \GPcolortrue
  }{}%
  \@ifundefined{ifGPblacktext}{%
    \newif\ifGPblacktext
    \GPblacktexttrue
  }{}%
  \let\gplgaddtomacro\g@addto@macro
  \gdef\gplbacktext{}%
  \gdef\gplfronttext{}%
  \makeatother
  \ifGPblacktext
    \def\colorrgb#1{}%
    \def\colorgray#1{}%
  \else
    \ifGPcolor
      \def\colorrgb#1{\color[rgb]{#1}}%
      \def\colorgray#1{\color[gray]{#1}}%
      \expandafter\def\csname LTw\endcsname{\color{white}}%
      \expandafter\def\csname LTb\endcsname{\color{black}}%
      \expandafter\def\csname LTa\endcsname{\color{black}}%
      \expandafter\def\csname LT0\endcsname{\color[rgb]{1,0,0}}%
      \expandafter\def\csname LT1\endcsname{\color[rgb]{0,1,0}}%
      \expandafter\def\csname LT2\endcsname{\color[rgb]{0,0,1}}%
      \expandafter\def\csname LT3\endcsname{\color[rgb]{1,0,1}}%
      \expandafter\def\csname LT4\endcsname{\color[rgb]{0,1,1}}%
      \expandafter\def\csname LT5\endcsname{\color[rgb]{1,1,0}}%
      \expandafter\def\csname LT6\endcsname{\color[rgb]{0,0,0}}%
      \expandafter\def\csname LT7\endcsname{\color[rgb]{1,0.3,0}}%
      \expandafter\def\csname LT8\endcsname{\color[rgb]{0.5,0.5,0.5}}%
    \else
      \def\colorrgb#1{\color{black}}%
      \def\colorgray#1{\color[gray]{#1}}%
      \expandafter\def\csname LTw\endcsname{\color{white}}%
      \expandafter\def\csname LTb\endcsname{\color{black}}%
      \expandafter\def\csname LTa\endcsname{\color{black}}%
      \expandafter\def\csname LT0\endcsname{\color{black}}%
      \expandafter\def\csname LT1\endcsname{\color{black}}%
      \expandafter\def\csname LT2\endcsname{\color{black}}%
      \expandafter\def\csname LT3\endcsname{\color{black}}%
      \expandafter\def\csname LT4\endcsname{\color{black}}%
      \expandafter\def\csname LT5\endcsname{\color{black}}%
      \expandafter\def\csname LT6\endcsname{\color{black}}%
      \expandafter\def\csname LT7\endcsname{\color{black}}%
      \expandafter\def\csname LT8\endcsname{\color{black}}%
    \fi
  \fi
  \setlength{\unitlength}{0.0500bp}%
  \begin{picture}(3968.00,2976.00)%
    \gplgaddtomacro\gplbacktext{%
      \csname LTb\endcsname%
      \put(768,725){\makebox(0,0)[r]{\strut{}-0.6}}%
      \put(768,1051){\makebox(0,0)[r]{\strut{}-0.4}}%
      \put(768,1377){\makebox(0,0)[r]{\strut{}-0.2}}%
      \put(768,1704){\makebox(0,0)[r]{\strut{} 0}}%
      \put(768,2030){\makebox(0,0)[r]{\strut{} 0.2}}%
      \put(768,2356){\makebox(0,0)[r]{\strut{} 0.4}}%
      \put(768,2682){\makebox(0,0)[r]{\strut{} 0.6}}%
      \put(864,320){\makebox(0,0){\strut{} 0}}%
      \put(1184,320){\makebox(0,0){\strut{} 0.2}}%
      \put(1505,320){\makebox(0,0){\strut{} 0.4}}%
      \put(1825,320){\makebox(0,0){\strut{} 0.6}}%
      \put(2145,320){\makebox(0,0){\strut{} 0.8}}%
      \put(2465,320){\makebox(0,0){\strut{} 1}}%
      \put(2786,320){\makebox(0,0){\strut{} 1.2}}%
      \put(3106,320){\makebox(0,0){\strut{} 1.4}}%
      \put(208,1703){\rotatebox{90}{\makebox(0,0){\strut{}imaginary part}}}%
      \put(2065,80){\makebox(0,0){\strut{}real part}}%
    }%
    \gplgaddtomacro\gplfronttext{%
      \csname LTb\endcsname%
      \put(3542,480){\makebox(0,0)[l]{\strut{} 1}}%
      \put(3542,1091){\makebox(0,0)[l]{\strut{} 10}}%
      \put(3542,1703){\makebox(0,0)[l]{\strut{} 100}}%
      \put(3542,2315){\makebox(0,0)[l]{\strut{} 1000}}%
      \put(3542,2927){\makebox(0,0)[l]{\strut{} 10000}}%
    }%
    \gplbacktext
    \put(0,0){\includegraphics{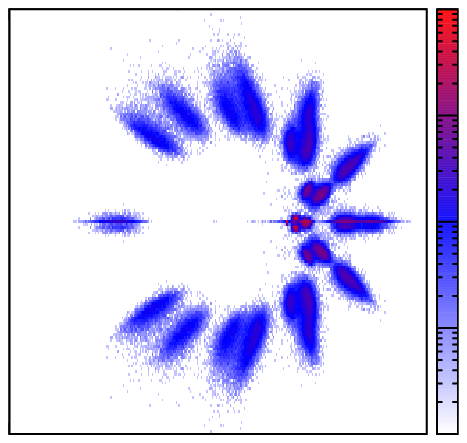}}%
    \gplfronttext
  \end{picture}%
\endgroup

%% file: paper.bbl
\providecommand{\href}[2]{#2}\begingroup\raggedright\begin{thebibliography}{10}

\bibitem{ArkaniHamed2008}
N.~Arkani-Hamed, D.~P. Finkbeiner, T.~R. Slatyer, and N.~Weiner, {\it {A Theory
  of Dark Matter}},  {\em Phys. Rev.} {\bf D79} (2009) 015014,
  [\href{http://xxx.lanl.gov/abs/0810.0713}{{\tt arXiv:0810.0713}}].

\bibitem{RevModPhys.82.557}
J.~E. Kim and G.~Carosi, {\it {Axions and the strong $C\!P$ problem}},  {\em
  Rev. Mod. Phys.} {\bf 82} (Mar, 2010) 557--601,
  [\href{http://xxx.lanl.gov/abs/0807.3125}{{\tt arXiv:0807.3125}}].

\bibitem{Novikov:1984ac}
V.~A. Novikov, M.~A. Shifman, A.~I. Vainshtein, and V.~I. Zakharov, {\it
  {Two-Dimensional Sigma Models: Modeling Nonperturbative Effects of Quantum
  Chromodynamics}},  {\em Phys. Rept.} {\bf 116} (1984) 103.

\bibitem{Witten:1977xn}
E.~Witten, {\it {A Supersymmetric Form of the Nonlinear Sigma Model in Two-
  Dimensions}},  {\em Phys. Rev.} {\bf D16} (1977) 2991.

\bibitem{DiVecchia:1977bs}
P.~Di~Vecchia and S.~Ferrara, {\it {Classical Solutions in Two-Dimensional
  Supersymmetric Field Theories}},  {\em Nucl. Phys.} {\bf B130} (1977) 93.

\bibitem{Shankar1978}
R.~Shankar and E.~Witten, {\it {$S$ matrix of the supersymmetric nonlinear
  $\sigma$ model}},  {\em Phys. Rev. D} {\bf 17} (Apr, 1978) 2134--2143.

\bibitem{Alvarez1978}
O.~Alvarez, {\it Dynamical symmetry breakdown in the supersymmetric nonlinear
  $\sigma{}$ model},  {\em Phys. Rev. D} {\bf 17} (Feb, 1978) 1123--1130.

\bibitem{Evans:1994sy}
J.~M. Evans and T.~J. Hollowood, {\it {The Exact mass gap of the supersymmetric
  {$O(N)$} sigma model}},  {\em Phys. Lett.} {\bf B343} (1995) 189--197,
  [\href{http://xxx.lanl.gov/abs/hep-th/9409141}{{\tt hep-th/9409141}}].

\bibitem{Zumino:1979et}
B.~Zumino, {\it {Supersymmetry and K\"ahler Manifolds}},  {\em Phys. Lett.}
  {\bf B87} (1979) 203.

\bibitem{Dondi:1976tx}
P.~H. Dondi and H.~Nicolai, {\it Lattice supersymmetry},  {\em Nuovo Cim.} {\bf
  A41} (1977) 1.

\bibitem{Montvay:1995rs}
I.~Montvay, {\it {Tuning to N=2 supersymmetry in the SU(2) adjoint Higgs-Yukawa
  model}},  {\em Nucl. Phys.} {\bf B445} (1995) 399--428,
  [\href{http://xxx.lanl.gov/abs/hep-lat/9503009}{{\tt hep-lat/9503009}}].

\bibitem{Catterall:2009it}
S.~Catterall, D.~B. Kaplan, and M.~Unsal, {\it {Exact lattice supersymmetry}},
  {\em Phys. Rept.} {\bf 484} (2009) 71--130,
  [\href{http://xxx.lanl.gov/abs/0903.4881}{{\tt arXiv:0903.4881}}].

\bibitem{Catterall:2003uf}
S.~Catterall and S.~Ghadab, {\it {Lattice sigma models with exact
  supersymmetry}},  {\em JHEP} {\bf 05} (2004) 044,
  [\href{http://xxx.lanl.gov/abs/hep-lat/0311042}{{\tt hep-lat/0311042}}].

\bibitem{Catterall:2006sj}
S.~Catterall and S.~Ghadab, {\it {Twisted supersymmetric sigma model on the
  lattice}},  {\em JHEP} {\bf 10} (2006) 063,
  [\href{http://xxx.lanl.gov/abs/hep-lat/0607010}{{\tt hep-lat/0607010}}].

\bibitem{Kastner:2008zc}
T.~K\"astner, G.~Bergner, S.~Uhlmann, A.~Wipf, and C.~Wozar, {\it
  {Two-Dimensional Wess-Zumino Models at Intermediate Couplings}},  {\em
  Phys.Rev.} {\bf D78} (2008) 095001,
  [\href{http://xxx.lanl.gov/abs/0807.1905}{{\tt arXiv:0807.1905}}].

\bibitem{Kirchberg:2004vm}
A.~Kirchberg, J.~D. Lange, and A.~Wipf, {\it From the {D}irac operator to
  {W}ess-{Z}umino models on spatial lattices},  {\em Ann. Phys.} {\bf 316}
  (2005) 357--392, [\href{http://xxx.lanl.gov/abs/hep-th/0407207}{{\tt
  hep-th/0407207}}].

\bibitem{Hubbard:1959ub}
J.~Hubbard, {\it {Calculation of partition functions}},  {\em Phys. Rev. Lett.}
  {\bf 3} (1959) 77--80.

\bibitem{Drell:1976bq}
S.~D. Drell, M.~Weinstein, and S.~Yankielowicz, {\it Variational approach to
  strong coupling field theory. 1. {$\phi^4$} theory},  {\em Phys. Rev.} {\bf
  D14} (1976) 487.

\bibitem{Bergner:2007pu}
G.~Bergner, T.~K\"astner, S.~Uhlmann, and A.~Wipf, {\it Low-dimensional
  supersymmetric lattice models},  {\em Annals Phys.} {\bf 323} (2008)
  946--988, [\href{http://xxx.lanl.gov/abs/0705.2212}{{\tt arXiv:0705.2212}}].

\bibitem{Wozar:2011gu}
C.~Wozar and A.~Wipf, {\it {Supersymmetry Breaking in Low Dimensional Models}},
   {\em Annals Phys.} {\bf 327} (2012) 774--807,
  [\href{http://xxx.lanl.gov/abs/1107.3324}{{\tt arXiv:1107.3324}}].

\bibitem{Luscher:1991wu}
M.~L\"uscher, P.~Weisz, and U.~Wolff, {\it {A Numerical method to compute the
  running coupling in asymptotically free theories}},  {\em Nucl. Phys.} {\bf
  B359} (1991) 221--243.

\bibitem{Balog:2009np}
J.~Balog, F.~Niedermayer, and P.~Weisz, {\it {The puzzle of apparent linear
  lattice artifacts in the 2d non-linear sigma-model and Symanzik's solution}},
   {\em Nucl. Phys.} {\bf B824} (2010) 563--615,
  [\href{http://xxx.lanl.gov/abs/0905.1730}{{\tt arXiv:0905.1730}}].

\bibitem{Wolff:2009kp}
U.~Wolff, {\it {Simulating the All-Order Strong Coupling Expansion III: O(N)
  sigma/loop models}},  {\em Nucl. Phys.} {\bf B824} (2010) 254--272,
  [\href{http://xxx.lanl.gov/abs/0908.0284}{{\tt arXiv:0908.0284}}].

\bibitem{Balog:2003yr}
J.~Balog and A.~Hegedus, {\it {TBA equations for excited states in the O(3) and
  O(4) nonlinear sigma-model}},  {\em J. Phys.} {\bf A37} (2004) 1881--1901,
  [\href{http://xxx.lanl.gov/abs/hep-th/0309009}{{\tt hep-th/0309009}}].

\bibitem{Hasenbusch:2001ht}
M.~Hasenbusch, P.~Hasenfratz, F.~Niedermayer, B.~Seefeld, and U.~Wolff, {\it
  {Nonstandard cutoff effects in the nonlinear sigma model}},  {\em Nucl. Phys.
  Proc. Suppl.} {\bf 106} (2002) 911--913,
  [\href{http://xxx.lanl.gov/abs/hep-lat/0110202}{{\tt hep-lat/0110202}}].

\bibitem{Bergner:2009vg}
G.~Bergner, {\it {Complete supersymmetry on the lattice and a No-Go theorem: A
  simulation with intact supersymmetries on the lattice}},  {\em JHEP} {\bf 01}
  (2010) 024, [\href{http://xxx.lanl.gov/abs/0909.4791}{{\tt
  arXiv:0909.4791}}].

\bibitem{Mermin:1966fe}
N.~D. Mermin and H.~Wagner, {\it {Absence of ferromagnetism or
  antiferromagnetism in one- dimensional or two-dimensional isotropic
  Heisenberg models}},  {\em Phys. Rev. Lett.} {\bf 17} (1966) 1133--1136.

\bibitem{Sugino:2003yb}
F.~Sugino, {\it {A lattice formulation of super Yang-Mills theories with exact
  supersymmetry}},  {\em JHEP} {\bf 01} (2004) 015,
  [\href{http://xxx.lanl.gov/abs/hep-lat/0311021}{{\tt hep-lat/0311021}}].

\bibitem{Evans:1994sv}
J.~M. Evans and T.~J. Hollowood, {\it {The Exact mass gap of the supersymmetric
  {$CP^{n-1}$} sigma model}},  {\em Phys. Lett.} {\bf B343} (1995) 198--206,
  [\href{http://xxx.lanl.gov/abs/hep-th/9409142}{{\tt hep-th/9409142}}].

\bibitem{Witten:1982df}
E.~Witten, {\it {Constraints on Supersymmetry Breaking}},  {\em Nucl. Phys.}
  {\bf B202} (1982) 253.

\bibitem{Donini:1997}
A.~Donini, M.~Guagnelli, P.~Hernandez, and A.~Vladikas, {\it {Towards N=1
  Super-Yang-Mills on the lattice}},  {\em Nucl.Phys.} {\bf B523} (1998)
  529--552, [\href{http://xxx.lanl.gov/abs/hep-lat/9710065}{{\tt
  hep-lat/9710065}}].

\bibitem{Clark:2006fx}
M.~A. Clark and A.~D. Kennedy, {\it {Accelerating dynamical fermion
  computations using the rational hybrid Monte Carlo (RHMC) algorithm with
  multiple pseudofermion fields}},  {\em Phys. Rev. Lett.} {\bf 98} (2007)
  051601, [\href{http://xxx.lanl.gov/abs/hep-lat/0608015}{{\tt
  hep-lat/0608015}}].

\bibitem{DeGrand:1990}
T.~A. DeGrand and P.~Rossi, {\it Conditioning techniques for dynamical
  fermions},  {\em Comput.Phys.Commun.} {\bf 60} (1990) 211--214.

\bibitem{Oyanagi1986}
Y.~Oyanagi, {\it {An incomplete LDU decomposition of lattice fermions and its
  application to conjugate residual methods}},  {\em Comput.Phys.Commun.} {\bf
  42} (1986), no.~3 333 -- 343.

\bibitem{Peardon:2000}
M.~J. Peardon, {\it {Accelerating the hybrid Monte Carlo algorithm with ILU
  preconditioning}},  \href{http://xxx.lanl.gov/abs/hep-lat/0011080}{{\tt
  hep-lat/0011080}}.

\end{thebibliography}\endgroup
